\documentclass[fleqn,usenatbib,amsmath,amssymb]{emulatemnras}

\def\MyMNRAS#1{#1}
\MyMNRAS{\RequirePackage{rotating}}
\def\MyApJ#1{}

\usepackage{graphicx}
\usepackage{dcolumn}
\usepackage{bm}
\usepackage{epsf}
\usepackage{verbatim}
\usepackage{amsmath}
\usepackage{amssymb}
\usepackage{amsfonts}
\usepackage{ifthen}
\usepackage{epsfig}
\usepackage{verbatim}
\usepackage{dcolumn}
\usepackage{epstopdf}
\usepackage{threeparttable}
\usepackage{booktabs}
\usepackage{calc}
\usepackage{multirow}
\usepackage{rotating}
\usepackage{comment}
\usepackage{dblfloatfix}
\usepackage{appendix}
\usepackage[normalem]{ulem}
\usepackage[percent]{overpic}

\usepackage{footnote}
\makesavenoteenv{tabular}

\MyMNRAS{\usepackage{pdflscape}}

\DeclareSymbolFont{CMletters}{OML}{cmm}{m}{it}
\DeclareMathSymbol{v}{\mathord}{CMletters}{`v}

\def\mybib#1{./#1}

\def\figeps#1{./Figures/#1.eps} 

\newcommand{\myz}{{\tilde{z}}}

\newcommand{\CR}{{CR}}

\newcommand{\mypsi}{{\psi}}

\newcommand{\myn}{{N}}

\newcommand{\myPalpha}{{\widetilde{\alpha}}}

\newcommand{\myi}{{\varrho}}
\newcommand{\myX}{{\mathfrak{r}}}
\newcommand{\myp}{{p}} 
\newcommand{\myrg}{{r_g}}
\newcommand{\myrcr}{{r_{cr}}}

\newcommand{\mach}{{\mathcal{M}}}
\newcommand{\myG}{{G}}
\newcommand{\myg}{{g}}

\newcommand{\mhz}{{\scriptsize{\mbox{ MHz}}}}
\newcommand{\ghz}{{\scriptsize{\mbox{ GHz}}}}
\newcommand{\MJy}{{\mbox{ MJy}}}
\newcommand{\dgr}{^{\circ}}

\MyApJ{\bibliographystyle{apj}}
\MyMNRAS{\bibliographystyle{mnras}}

\defcitealias{Keshet2010}{K10}
\newcommand{\UK}{{\citetalias{Keshet2010}}}

\defcitealias{KeshetLoeb10}{KL10}
\newcommand{\KL}{{\citetalias{KeshetLoeb10}}}

\defcitealias{vanWeerenEtAl19}{vW19}
\newcommand{\VW}{{\citetalias{vanWeerenEtAl19}}}

\newcommand{\MyPlace}{h!}

\newcommand{\ie}{\emph{i.e.} }
\newcommand{\eg}{\emph{e.g.,} }
\newcommand{\cf}{\emph{cf.} }

\newcommand{\be}{\begin{equation}}
\newcommand{\ee}{\end{equation}}
\newcommand{\bea}{\begin{equation*}}
\newcommand{\eea}{\end{equation*}}
\newcommand{\beqr}{\begin{eqnarray} \nonumber}
\newcommand{\eeqr}{\end{eqnarray}}
\newcommand{\beqrb}{\begin{eqnarray}}
\newcommand{\eeqrb}{\nonumber \end{eqnarray}}
\newcommand{\fin}{\mbox{ .}}
\newcommand{\coma}{\mbox{ ,}}

\newcommand{\cm}{\mbox{ cm}}
\newcommand{\sr}{\mbox{ sr}}
\newcommand{\se}{\mbox{ s}}

\newcommand{\Gyr}{\mbox{ Gyr}}
\newcommand{\erg}{\mbox{ erg}}
\newcommand{\Hz}{\mbox{ Hz}}

\newcommand{\MHz}{\mbox{ MHz}}
\newcommand{\GHz}{\mbox{ GHz}}

\newcommand{\km}{\mbox{ km}}

\newcommand{\kpc}{\mbox{ kpc}}
\newcommand{\Mpc}{\mbox{ Mpc}}

\newcommand{\keV}{\mbox{ keV}}
\newcommand{\MeV}{\mbox{ MeV}}
\newcommand{\GeV}{\mbox{ GeV}}
\newcommand{\TeV}{\mbox{ TeV}}

\newcommand{\muG}{\mbox{ $\mu$G}}

\newcommand{\const}{\mbox{const.}}

\newcommand{\gama}{$\gamma$}

\newcommand{\pr}{\partial}

\usepackage{xcolor}
\definecolor{darkgreen}{rgb}{0.0,0.5,0.0}
\MyApJ{\usepackage[colorlinks,citecolor=darkgreen]{hyperref} 
}

\newcommand{\MyTitle}{{Radio halos and relics from extended cosmic-ray ion distributions with strong diffusion in galaxy clusters}}
\newcommand{\MySTitle}{{Hadronic halos and relics}}

\MyMNRAS{
    \title[\MySTitle]{\MyTitle}
	\author[Keshet]{Uri Keshet\thanks{E-mail: ukeshet@bgu.ac.il}  \\
	Physics Department, Ben-Gurion University of the Negev, POB 653, Be'er-Sheva 84105, Israel
}}
	
\MyMNRAS{
    
\begin{document}
    \pubyear{2023}
    \label{firstpage}
    \pagerange{\pageref{firstpage}--\pageref{lastpage}}
    \maketitle
}

\begin{abstract}
A joint hadronic model is shown to quantitatively explain the observations of diffuse radio emission from galaxy clusters in the form of minihalos, giant halos, relics, and their hybrid, transitional stages.
Cosmic-ray diffusion of order $D\sim 10^{31\text{--}32}\text{ cm}^2\text{ s}^{-1}$, inferred independently from relic energies, the spatial variability of giant-halo spectra, and the spectral evolution of relics, reproduces the observed spatio-spectral distributions, explains the recently discovered mega-halos as enhanced peripheral magnetization, and quenches electron (re)acceleration by weak shocks or turbulence.
For instance, the hard-to-soft evolution along secondary-electron diffusion explains both the soft spectra in most halo peripheries and relic downstreams, and the hard spectra in most halo centres and relic edges, where the photon index can reach $\alpha\simeq -0.5$ regardless of the Mach number $\mathcal{M}$ of the coincident shock.
Such spatio-spectral modeling, recent $\gamma$-ray observations, and additional accumulated evidence are thus shown to support a previous claim \citep{Keshet2010} that the seamless transitions among minihalos, giant halos, and relics, their similar energetics, integrated spectra, and delineating discontinuities, the inconsistent $\mathcal{M}$ inferred from radio vs. X-rays in leptonic models, and additional observations, all indicate that these diffuse radio phenomena are manifestations of the same cosmic-ray ion population, with no need to invoke less natural alternatives.
\end{abstract}

\date{Accepted ---. Received ---; in original ---}
	
\MyMNRAS{
}

\section{Introduction}
\label{sec:Introduction}

Diffuse radio sources in the intracluster medium (ICM) of galaxy clusters are broadly classified, according to their location, morphology, and polarization, as minihalos (MHs, or core halos), giant halos (GHs), or relics.
The cool core-spanning MHs and their larger, $\sim$Mpc GH counterparts are in general regular, unpolarized emission around the centre of the cluster, whereas relics are peripheral, typically elongated, and polarized.
This classification is oversimplified, as halos can be irregular or asymmetrically large, relics can bridge to, or merge with, halos, and there are hybrid objects sharing some MH, GH, or relic properties.
Hadronic, and recently mostly leptonic, models were proposed for each of these source types, invoking respectively either secondary cosmic-ray (CR) electrons (CREs) from CR ion (CRI) collisions with ambient nuclei, or primary CREs accelerated or re-accelerated in weak shocks or ICM turbulence.
For reviews, see \citet{FerettiGiovannini96}, \citet{FerrariEtAl08}, \citet[][henceforth {\UK}]{Keshet2010}, and \citet[][henceforth {\VW}]{vanWeerenEtAl19}.

MHs are ubiquitously found in the centres of the more relaxed, cool-core clusters \citep[possibly in $\sim80\%$ of the cores;][]{GiacintucciEtAl17}.
They extend roughly over the cool region \citep{GittiEtAl02}, often engulfing the more compact radio emission from an active galactic nucleus (AGN), and typically truncate at cold fronts \citep[CFs;][]{MazzottaGiacintucci08}: projected tangential discontinuities that confine spiral flows \citep{KeshetEtAl10}, which are observed to strongly magnetize the plasma \citep{ReissKeshet2014, NaorEtAl2020} and appear to regulate the core \citep{ZuHoneEtAl10, Keshet12}.
MHs are typically unpolarized, regular, and spectrally flat, with a typical integrated photon index $-1.0\lesssim\alpha\lesssim-1.2$ (although cases as soft as $\alpha\simeq -1.6$ were reported, as detailed below), showing a fairly universal central ratio $\eta\equiv \nu I_\nu/F_X$ between radio and X-ray surface brightness \citep[][henceforth {\KL}]{KeshetLoeb10}.
Given their high-density environment, MHs can be naturally attributed to secondary CREs \citep[\eg][{\KL}]{PfrommerEnsslin04}, gyrating in the magnetic fields generated by sloshing \citep{MarkevitchVikhlinin07} or a spiral flow \citep{Keshet12, KeshetEtAl22}.
An alternative, leptonic model invokes the re-acceleration of seed electrons \citep{GittiEtAl02} in sloshing-induced turbulence \citep{MazzottaGiacintucci08, ZuHoneEtAl2013}.

GHs are the large, $\sim$Mpc wide counterparts of MHs, found around the centres of some merger clusters that lost their cool cores, preferentially in massive, highly disturbed, X-ray bright clusters (\VW, and references therein).
These sources present a regular morphology which roughly traces the thermal plasma, but sharply truncate sometimes at weak outgoing shocks, in resemblance of the MH truncation at CFs.
In other cases, a radio-bright weak shock is found outside the GH, but still connected to it by a detectable radio protrusion, in which case the emission outside the GH is classified as a radio relic connected to the GH by a radio bridge.
Like MHs, GHs are usually unpolarized, regular, and spectrally flat, with a similar integrated $-1.0\lesssim\alpha\lesssim-1.2$ (although cases with $\alpha$ as soft as $\sim-1.6$ or a high-frequency steepening were reported); their central $\eta$ ratio is also similar to those of MHs (\KL).
The radiating CREs were modelled either as secondaries from CRI collisions \citep{Dennison80, BlasiColafrancesco99, KushnirEtAl09} or as primary CREs (re)accelerated by turbulence \citep{EnsslinEtAl99, BrunettiEtAl01, Petrosian01}.
More recent claims that GHs cannot be predominately hadronic \citep[\VW,][and references therein]{AdamEtAl21}, based mostly on upper limits on their \gama-ray, $\pi^0\to\gamma\gamma$ counterpart, are shown below to be incorrect.

In contrast to halos, relics are peripheral sources, found at radii ranging from a few $100\kpc$ to $r\sim2\Mpc$, often in pairs located at opposite sides of the cluster.
Relics show highly irregular morphologies elongated perpendicular to the radial direction, sometimes present filamentary substructure, and are among the most polarized sources on the sky (reaching up to $\sim70\%$ polarization locally; \VW).
Relics can usually be linked to a recent merger \citep{GiovanniniFeretti04}, and are thought to coincide with weak merger shocks that are seen nearly edge-on, with Mach numbers that are typically $\mach<2$, according to X-ray observations, but can also reach $\mach\gtrsim3$.
Almost all current models attribute the emission to primary CREs (re)accelerated by the shock \citep[][\VW, and references therein]{EnsslinEtAl98}.
However, diffusive shock acceleration (DSA) in such weak shocks is not well tested, and such a model is challenged for example by the fine-tuning needed to explain GH--relic transitions, unrealistic Mach numbers required by DSA (\UK), and an implied large population of shock-accelerated CRIs in possible tension with coincident \gama-ray limits \citep{VazzaBruggen14, VazzaEtAl15}.
A model that alleviates such difficulties attributes relics, too, to secondary CREs injected by CRI collisions with ambient nuclei (\UK), as both the injection rate and the magnetic field are amplified at weak shocks sufficiently to account for relics without invoking any CRE (re)acceleration, naturally reproducing the relic--GH connections and other observed phenomena.

The energetics, multiple similarities, and smooth transitions among MHs, GHs, and relics indicate that at least the majority of cases are simply different manifestations of the same underlying phenomenon, as argued by {\UK} and shown with more evidence below.
Attributing some sources to hadronic processes and others to leptonic processes would thus be unnatural, inflate the number of model parameters, and require fine-tuning. 
A joint model explaining MHs, GHs, relics and their hybrid manifestations, simultaneously, must be hadronic, as no leptonic alternatives operate uniformly and smoothly across such diverse environments, ranging from the high-density, magnetized, relaxed cores (MHs) to the low-density, weakly magnetized, perturbed or recently shocked ICM (GH peripheries and relics).
It should also be noted that electron (re)acceleration in weak shocks or turbulence under ICM conditions, as invoked by the leptonic models, is neither well-understood nor well-constrained elsewhere, and is quenched when CR diffusion exceeds $D\sim 10^{31}\cm^2\se^{-1}$.

Indeed, a simple hadronic model naturally explains MHs, GHs, relics, and their transitions, as arising from the same CRI population, provided that the latter is sufficiently extended (\UK).
In particular, if CRIs are evenly spread out by strong, $D\sim 10^{32}\cm^2\se^{-1}$ diffusion or a comparable combination of diffusion and advection, the observations are recovered in a model with
a single parameter: a mean CRI energy density $u_i\simeq 10^{[-13.3,-12.4]}\erg\cm^{-3}$ (integrated for simplicity over $10^1$--$10^7\GeV$ for a flat, $p=2$ spectral index: equal energy per logarithmic energy bin).
Such a joint model economically explains the observed similarities and transitions among halo and relics, their identical integrated spectra, GH--radio bridges, radio--X-ray correlations, the presence of halo-edge discontinuities, and the disagreement between leptonic radio vs. X-ray relic Mach numbers  (\UK).
Additional observations supporting the model are shown in this work.
For instance, as homogeneous CRIs radio-brighten any magnetized ICM region with no additional particle (re)acceleration, increasingly sensitive observations should reveal an abundance of diffuse emission even in cluster peripheries, such as the recently discovered $\gtrsim$ Mpc scale mega-halos \citep{CucitiEtAl22}.

Even though a single CRI population conservatively accounts for the observations, hadronic models are increasingly dismissed as a viable explanation for halos, mainly in favour of turbulent (re)acceleration models, and are essentially never considered for relics (\VW, and references therein).
The considerable CRI population, which is a persistent and often inevitable counterpart of the radiating CREs, is often disregarded.
Tests of leptonic vs. hadronic models are carried out with unbalanced levels of sophistication, emphasizing the failure of oversimplified hadronic models to match observations.
However, the spatio-spectral distribution and other properties of radio emission from secondary CREs are non-trivial, because they are sensitive to the CRI distribution, the diffusion of secondary CREs, and the spacetime evolution of the magnetic field.
As shown in {\UK} and in more detail in the present work, a hadronic model incorporating these effects does match the observation.

For instance, if one could neglect the evolution and substructure of the magnetic field and the diffusion of the CRs, then CREs would approximately trace the gas distribution and radiate all their energy locally and steadily, giving rise to a cooled, $\alpha\simeq -1$ photon index that directly reflects the typical $p\simeq 2$ CRI index.
The spectral index can differ somewhat from this nominal value, especially at low frequencies $\nu$, where the $\lesssim 100\GeV$ CRI spectrum may deviate from a power-law, and the cross-section for secondary production can no longer be approximated as constant (\eg \KL).
However, the fluid-frame magnetic field is thought to evolve rapidly and non-uniformly in the vicinity of relics, in GH peripheries, and near MH CFs.
Sufficiently fast evolution or pronounced substructure of the magnetic field can strongly modify the spectrum, typically softening it as the field grows, because the synchrotron emission from a steadily Compton-cooling CRE reflects the irregular, intermittent, or time-dependent magnetic field it traverses (\UK).

More importantly, we find increasing evidence that CR mixing in the ICM is indeed strong, equivalent to a diffusion coefficient $D(100\GeV)\simeq 10^{31\text{--}32}\cm^2\se^{-1}$, leading to a fairly homogeneous CRI distribution over $\sim\Mpc$ lengthscales on few Gyr timescales.
Such diffusion, corresponding to a $\sim\kpc$ coherence length, is much stronger than found on galactic scales, but is consistent with the observed relation between $D$ and system size (see \S\ref{sec:Summary}); moreover, this estimate of $D$ may represent a combination of diffusion and mixing processes associated with mergers or spiral flows.
Strong diffusion of CRIs enhances their density in the peripheral regions that harbour GH edges and relics.
We show how strong diffusion of CREs, injected non-homogeneously, modifies the spectrum even for steady-state, uniform magnetic fields, as different regions are dominated by diffusing CREs that experienced different levels of radiative cooling.

This paper examines the joint hadronic model for MHs, GHs, and relics, tests it against observations, and addresses the concerns raised in the literature against hadronic models.
The diffusion coefficient $D$ is estimated independently using spectral variations across GHs and the softening downstream of relics, found to be consistent with each other and with the {\UK}, energetics-based estimate.
The properties of radio emission in a hadronic model with strong diffusion are derived, and shown to reproduce a wide range of observations, resolving problems encountered by alternative models.
Recent observations, such as a diffuse central \gama-ray excess in the Coma cluster \citep{XiEtAl18_Coma, AdamEtAl21, BaghmanyanEtAl22} and the steep-spectrum mega-halos, which we refer to as super-halos (SHs, to avoid confusion with MHs), are also shown to be consistent with the hadronic model.

The paper is organized as follows.
In \S\ref{sec:Overview}, we outline the joint hadronic model, showing that it is supported by an abundance of evidence.
We study CR diffusion more carefully in steady-state systems in \S\ref{sec:Halos}, showing that it facilitates the spatial and spectral properties of MHs and GHs.
In \S\ref{sec:Relics}, we incorporate diffusion and advection in the time-dependent shock environment, showing that the model reproduces the spatio-spectral properties of relics.
Finally, in \S\ref{sec:SH} we outline and model the recent SH observations.
Our results are summarized and discussed in \S\ref{sec:Summary}, where we also discuss the nature of diffusion, the role of the virial shock in accelerating cluster CRs, and additional implications.
Correlations between radio and other signals are discussed in Supporting Information Appendix \S\ref{subsec:Relations}, and exact solutions to the diffusion-loss equation are provided in Appendices \S\ref{app:PLSolutions} (scale-free), \S\ref{app:GreenSolutions} (spherical symmetry), and \S\ref{app:GreenSolutionsPlanar} (shock downstream).
Our method of inferring $D$ from maps of the radio spectral index is detailed in Appendix \S\ref{app:PowerSpectrum}.

We generally follow the notations of {\UK}.
We adopt a flat $\Lambda$CDM model with an $H_0=70\km\se^{-1}\Mpc^{-1}$ Hubble constant, an $f_b\simeq 0.17$ baryon mass fraction, and a $\chi=0.76$ hydrogen mass fraction, so the number densities $n$ of particles and $n_e$ of electrons are related by $n_e\simeq 0.52n$, and $\mu m_p\simeq 0.59m_p$ is the mean particle mass, where $m_p$ is the proton mass.
When discussing X-rays, we exclusively refer to the \emph{ROSAT}, $0.1$--$2.4\keV$ band.
When evaluating volume-integrated signals, we use $r<R_{500}$ unless otherwise specified, where $R_{500}$ is a radius enclosing a mean density $500$ times the critical density of the Universe.
Error bars designate 68\% containment projected for a single parameter.

\section{Joint hadronic model: overview}
\label{sec:Overview}

Here, we outline the joint hadronic model, attributing MHs, GHs, and relics to secondary CREs produced by the same, cluster-wide CRI population.
General arguments for a joint and hadronic origin are given in \S\ref{subsec:Motivation}, leading to a simple, homogeneous CRI model.
The ostensibly strongest argument raised against the hadronic model, namely the weak \gama-ray emission from cluster centres, is shown in \S\ref{subsec:GammaRays} to be incorrect, as the recent \gama-ray excess reported in the centre of Coma actually supports our hadronic model.
The spatio-spectral distribution of radio emission is qualitatively discussed in \S\ref{subsec:CosmicRayDiffusionDdvection}, deferring a quantitative analysis to \S\ref{sec:Halos}--\S\ref{sec:SH}.
Finally, \S\ref{subsec:Evidence} provides a point-by-point summary of the accumulated evidence for a joint hadronic model.
A discussion of the local relations between radio and other signals is deferred to Appendix \S\ref{subsec:Relations}, emphasizing the case of homogeneous CRIs.

\subsection{Motivation and homogeneous CRI limit}
\label{subsec:Motivation}

\subsubsection{Rationale}

A MH is found in the vast majority of relaxed clusters \citep[\eg][]{GiacintucciEtAl17}, but is disrupted and usually replaced by a larger, GH version of itself once the core is destroyed by a sufficiently strong merger.
Some clusters are observed during such a transition, like A2319 (with a CF and subcluster; see {\KL} and references therein), A2256 \citep[where CFs are noticeable in the radio emission of the GH; see][]{RajpurohitEtAl22A, RajpurohitEtAl22B}, CL1821+643 \citep{BonafedeEtAl14}, A2261 \citep{SaviniEtAl19}, A2142 \citep{VenturiEtAl17}, and possibly PSZ1 G139.61+24.20 \citep{SaviniEtAl18UltraSteep} and RX J1720.1+2638 \citep{SaviniEtAl19}.
Such observations challenge the MH--GH classification \citep[\UK; ][]{StormEtAl15}.
Furthermore, the characteristics of radio emission from the centre of the cluster do not change after the MH is replaced by a GH (\eg \KL).

Hence, it would be quite unnatural to attribute MHs to one (say, hadronic) process and GHs to a radically different (leptonic) process, especially considering their many quantitative similarities demonstrated below.
Likewise, attributing both MHs and GHs to leptonic processes but invoking different types of turbulent re-acceleration in each respective environment (\VW, and references therein) would be similarly unnatural and fine-tuned.

An analogous argument ties GHs and relics, which are observed in different stages in different clusters: merged (when the shock is at the edge of the GH), beginning to separate (with a bright radio bridge still connecting the GH and relic as the shock moves outwards), and at large separations (with a faint or undetectable bridge).
Moreover, some MHs and GHs show relic characteristics, such as an irregular or filamentary morphology \citep[RXC J2003.5-2323, A2255, and A2319; see][]{GiacintucciEtAl09, MurgiaEtAl09} or polarization \citep[A2255, MACS J0717.5+3745, A2390, and A523; see][]{GovoniEtAl05, BonafedeEtAl09, BacchiEtAl03, GirardiEtAl16}.
The very defining property of relics --- a coincident weak shock --- is shared by many GHs, challenging leptonic models.
The smooth transitions and quantitative similarities between halos and relics indicate a common underlying origin; here too, enforcing a smooth blend of different types of particle acceleration (\VW, and references therein) would be unnatural and fine-tuned.

Leptonic models invoked in the literature for each type of source face their own, separate challenges (\eg \VW, and additional issues outlined below).
Jointly, requiring primary (re)acceleration to reproduce quantitatively similar radio sources that smoothly blend into each other across diverse environments including relaxed cores (MHs), the turbulent post-shock ICM at large (GHs) and very large (SH) radii, and a range of weak to mild peripheral shocks (distant relics), would place unreasonable and fine-tuned constraints on the accelerator/s.
This argument alone disfavours the popular blend of different leptonic models, pointing at some alternative, common mechanism.
The conclusion is supported by the similarities between sources, like the $-1.2\lesssim\alpha\lesssim-1.0$ integrated spectra they all tend to show.

A common mechanism driving MHs, GHs and relics must be hadronic, as no engine can uniformly (re)accelerate the fast-cooling primary CREs across the entire cluster.
MHs in particular cannot be naturally attributed to primary CREs, as such a leptonic model would need to invoke significant shock or turbulent electron (re)acceleration is most or all cores, including in the most relaxed clusters, against rapid synchrotron cooling, and would struggle to explain observations such as the uniform $\alpha \sim -1$ spectrum.
More importantly, even a conservative lower limit on CRIs in the centres of clusters suffices to explain their GH \citep{KushnirWaxman09} and MH (\KL) brightness, as most of the energy injected into secondary CREs is lost to synchrotron radiation in the strong, $B\gtrsim 3\muG$ central magnetic fields.
We later show that CRE secondaries from the same CRI distribution account for relics, and would overproduce them if particle (re)acceleration were not quenched by strong diffusion.

\subsubsection{Homogeneous CRIs}

The main known sources of CRIs in a cluster are the strong virial shock at its edge, through which a considerable fraction of its baryons were accreted, and supernova remnant (SNR) shocks.
We focus on virial shocks, as their $\xi_e\sim 1\%$ CRE acceleration efficiency was recently measured, thus verifying that CRIs carry the large fraction of downstream energy needed to fuel the diffuse ICM radio sources.
These $\xi_e$ values were inferred from inverse-Compton emission \citep{KeshetEtAl12_Coma, ReissEtAl17, ReissKeshet18, KeshetReiss18}, and from the coincident synchrotron, low-frequency excess \citep{KeshetEtAl12_Coma, HouEtAl22}.
The detection of these shocks was further supported by coincident drops in the Sunyaev-Zel'dovich (SZ) $y$-parameter \citep{KeshetEtAl12_Coma, HurierEtAl19, KeshetAl20} and possibly by an unusual peripheral arc \citep[referred to as an 'accretion relic';][]{BonafedeEtAl22} that precisely coincides with the best-fitting virial ring in \emph{Fermi} data \citep{KeshetReiss18}.
Here, we use a simple, numerically-calibrated analytic model for the virial shock \citep{KeshetEtAl04}.
SNRs may somewhat boost the CRI population, but cannot strongly dominate it.

We begin with the simplest and most conservative case, in which strong diffusion fills the cluster with a homogeneous distribution of CRIs, thus placing a lower-limit on the central CRI density.
Assuming that a fraction $f_s$ of the baryons were processed by strong shocks\footnote{The factor $f_s\sim(\sqrt{2}/5)\dot{M}/(HM)$, where $\dot{M}$ is the mean mass accretion rate, differs from, and is larger than, the factor $f_{\text{acc}}\sim0.12$ calibrated in \citet{KeshetEtAl04}, which quantifies the instantaneous $z=0$ mass accretion rate through $\mach\gtrsim 10$ shocks.} and that the virial shock deposits a fraction $\xi_i$ of the downstream energy in CRIs, the total CRI energy in the cluster is $U_i\simeq \xi_i f_s (3/2) k_B T (f_b M/\mu m_p)$.
Here, we approximated $\int \dot{M}T\,dt\simeq MT$, where $M$ and $T$ are the observed mass and temperature of the cluster, because most of the thermal energy is accreted at late times and, although the cluster gradually heats up, post-accretion compression also raises $U_i$.
The virial shock can be approximated as a sphere of radius $r_s = f_r(2GM/25H^2)^{1/3}\simeq 2h(z)^{-2/3}M_{14}^{2/3}\Mpc$, where $h(z)\equiv H/H_0$, $z$ is the redshift, and the parameter $f_r\simeq 0.9$ was calibrated numerically \citep{KeshetEtAl04}.
Therefore, in this minimal model, the CRI energy density is
\begin{equation}\label{eq:CRI_E1}
  \!\! u_i\!\equiv\! \frac{3U_i}{4\pi r_s^3} \!\simeq\! \frac{225 \xi_i f_s f_b H^2 k_B T}{16 \pi f_r^3 G \mu m_p}\!\simeq\! 7\times 10^{-13} \xi_i f_s T_5\erg\cm^{-3}\! \coma\!\!
\end{equation}
depending only on the $k_B T\equiv 5 T_5 \keV$ cluster temperature.

For comparison, a CRI energy density in the range $u_i\simeq 10^{[-13.3,-12.4]}\erg\cm^{-3}$ accounts for most MHs, GHs, and relics (\UK).
Therefore, as one expects $0.1\lesssim(\xi_i f_s)<1$ values, Eq.~\eqref{eq:CRI_E1} is broadly consistent with the hadronic model.
The model is validated by $\xi_i f_s\sim0.2$, calibrated using \gama-ray observations of Coma in \S\ref{subsec:GammaRays}.
Note that even a much fainter \gama-ray signal would not invalidate the model, given the uncertainties and underlying approximations.
For example, relaxing the assumption of homogeneous CRIs would generally enhance $u_i$ in the cluster core.
Such cores show a particle density $>100$ times higher than the mean density of the cluster, in a volume $\sim 10^{3}$ times smaller than that used to derive Eq.~\eqref{eq:CRI_E1}.
Magnetic structures such as in CFs, shocks, and ICM substructure can hinder the outward diffusion of CRs that accumulate in the core, so a factor $\sim 10$ enhancement of the central $u_i$ is plausible even for overall strong diffusion.

\subsubsection{Halos}

Consider the logarithmic emissivity $\nu j_\nu$ of synchrotron emission from secondary CREs in halo centres, and their radio--X-ray brightness ratio $\eta$.
For simplicity, we assume a flat CRI power-law spectrum of index $p=2$,
as expected of acceleration in strong shocks \citep{AxfordEtAl77, Krymskii77, Bell78, BlandfordOstriker78} if scattering is sufficiently isotropic \citep{keshet2020diffusive}.
For such a spectrum, consistent with present, $p=2.0\pm0.2$ measurements of virial-shock accelerated CREs \citep{KeshetReiss18, HouEtAl22}, the logarithmic energy density depends only logarithmically on the maximal, $\gamma_{\text{max}}\simeq 10^{8.5}$  CRI Lorentz factor.

Approximating \citep[\eg][]{KushnirWaxman09} as $\sigma_{i}\simeq 40\mbox{ mb}$ the cross-section for an inelastic CRI collision with an ambient nucleus, depositing on average $f_e\simeq 5\%$ of the CRI energy in secondary CREs, the radio emissivity becomes
\begin{align}\label{eq:SecondaryEmissivity}
  \nu j_\nu & \simeq \frac{c f_e \sigma_{i} \mu n u_i}{8\pi\ln\gamma_{\text{max}}} \frac{B^2}{B^2+B_{\text{cmb}}^2} \\
& \simeq 5\times 10^{-35}\frac{\xi_i f_s n_{-3}T_{5}}{1+b^{-2}}\erg\se^{-1}\cm^{-3}\sr^{-1} \coma \nonumber
\end{align}
where $n_{-3}\equiv n/10^{-3}\cm^{-3}$, $c$ is the speed of light,
\begin{equation}
B_{\text{cmb}} \equiv \sqrt{8\pi u_{\text{cmb}}} \simeq 3.2 (1+z)^2\muG
\end{equation}
is the magnetic-field equivalent of the cosmic microwave background energy density, $u_{\text{cmb}}$, and $b\equiv B/B_{\text{cmb}}$ is the magnetization parameter.
For a $j_X\simeq 9\times 10^{-25} n_e^2 Z_{0.3}^{0.6}T_{\rm{keV}}^{-0.1}$ X-ray emissivity around $k_B T\equiv 1T_{\rm{keV}}\keV$ temperatures and $Z\equiv 0.3Z_3Z_\odot$ metallicities \citep[\eg][]{KeshetGurwich18}, the ratio between radio and X-ray emissivities becomes
\begin{equation}\label{eq:Eta1}
  \eta_j\equiv \frac{\nu j_\nu}{j_X} \simeq 3\times 10^{-4}\frac{\xi_i f_s T_5^{1.1}}{(1+b^{-2})n_{-3}} \fin
\end{equation}

In a GH, both radio and X-ray signals are dominated by a core of roughly constant, $n\simeq 10^{-3}\cm^{-3}$ density, which is typically also magnetized with $b\gtrsim1$. Hence, for typical parameters and $\xi_i f_s\sim0.2$, the brightness ratio
\begin{equation}\label{eq:etaGHs}
\eta\simeq \eta_j \simeq 10^{-(4.2\text{--}4.5)}T_5^{1.1}
\end{equation}
is consistent with $\eta\simeq 10^{-4.4\pm0.4}$ observed in the centres of both MHs and GHs (\KL) even in the homogeneous CRI limit.
In a MH, both radio and X-ray signals are dominated by a dense cool core, so the higher mean $n$ may lead to a smaller $\eta_j$ in the simple estimate \eqref{eq:Eta1}, despite the higher $b$.
However, as MHs reside in relaxed clusters, CRIs can more easily accumulate in the core and become trapped \eg by strong CF fields; there is indeed evidence for some central $u_i$ enhancement (see \S\ref{subsec:Relations}).
For the $u_i\propto n$ core scaling usually invoked, Eq.~\eqref{eq:Eta1} agrees with the measured $\eta$ also in MHs.

\subsubsection{Relics}

A weak shock driving a radio relic can compress a secondary CRE population by a large factor $r_{\text{cre}}$, but only if cooling may be neglected during compression.
If that were the case, such CREs upstream, already cooled to a $p=3$ spectrum, would be strongly compressed, by $r_{\text{cre}}=16$ in a Mach $\mach=2$ shock or by $r_{\text{cre}}=121$ for $\mach=2.2$, and would undergo even more substantial DSA for $\mach\geq\sqrt{5}$ ({\UK}; see Eq.~\eqref{eq:CR_Compression1} below).
Thus, if particle (re)acceleration were not quenched by cooling, compressed secondary CREs would greatly outshine the freshly injected CREs, over-producing the synchrotron emission from relics even without invoking any additional seed electrons or an alternative (re)acceleration mechanism.
However, as we show in \S\ref{subsec:CosmicRayDiffusionDdvection}, CRE compression and acceleration slow down in the presence of strong diffusion sufficiently to become quenched by radiative cooling.

CRIs, on the other hand, are susceptible to compression strong enough to energize relics (\UK) regardless of diffusion.
For example, the mean compression factor downstream of a Mach $\mach=2.5$ shock is $r_{\text{cri}}\simeq 6$ for $p=2$ CRIs, and $r_{\text{cri}}\simeq 10$ for $p=2.3$ CRIs.
As the shock sweeps up CRIs, they can accumulate in a narrow shell near the shock front and effectively raise $r_{\text{cri}}$, further boosting the signal and changing its spectral signature (see \S\ref{sec:Relics}).
The logarithmic synchrotron power of secondaries produced by such CRIs is given by
\begin{align}\label{eq:RelicPower}
  \nu P_\nu & \simeq \frac{\pi r_{\text{cri}} cf_e\sigma_i L^2 W \mu n u_i}{8(1+b^{-2})\ln(\gamma_{\text{max}})} \\
& \simeq 1.5\times 10^{39}\frac{\xi_i f_s T_5 r_{\text{cri}}L_1^2 W_{0.1}n_{-3}}{1+b^{-2}} \erg \se^{-1}\coma
\nonumber
\end{align}
where $n$ is the downstream density, and $L= 1L_1\Mpc$ and $W=0.1W_{0.1}\Mpc$ are, respectively, the largest linear scale and width of the relic.
Shock magnetization, from $b^2\ll1$ upstream to $b^2\gtrsim 1$ downstream, can further boost the near-downstream signal (\UK).

Equation \eqref{eq:RelicPower} agrees, for plausible $r_{\text{cri}}$ estimates, with observed relics, provided that the immediate downstream is magnetized to $b\gtrsim 1$ levels, as indeed inferred in some relics.
The homogeneous CRI distribution used in Eq.~\eqref{eq:RelicPower} thus accounts, for instance, for the measured $\nu P_\nu\simeq 2\times 10^{39}\erg\se^{-1}$ in A754 \citep{KaleDwarakanath09}, $\sim4\times 10^{39}\erg\se^{-1}$ in Coma \citep{BonafedeEtAl22}, $\sim4\times 10^{40}\erg\se^{-1}$ in A2256 \citep{RajpurohitEtAl22A}, and even $\sim1.2\times 10^{41}\erg\se^{-1}$ in the 'sausage' relic \citep{StroeEtAl16}.

For example, consider the most luminous of these relics.
This $L\simeq 1.9\Mpc$ sausage relic shows a hot, $k_BT\simeq 10\keV$ \citep{OgreanEtAl2014} downstream, where $n\simeq 10^{-3}$ and $B\simeq 5\muG$ \citep{AkamatsuEtAl5}, resulting from the merger of two $\sim 10^{15}M_\odot$ clusters \citep{JeeEtAl14}, so the coincident $u_i$ may easily double our estimate \eqref{eq:CRI_E1}.
The hadronic model thus reproduces the power of this relic with a modest, $r_{\text{cri}}\simeq 10/(f_s\xi_i W_{0.1})$ mean CRI compression in the relic region, consistent with $\mach=2.54_{-0.43}^{+0.64}$ estimated from the temperature profile.
Even the $\mach=1.31_{-0.09}^{+0.12}$ estimate based on surface brightness \citep{OgreanEtAl2014} could explain the relic, provided that CRIs are swept up as it propagates out to its large, $r\simeq 1.5\Mpc$ observed radius.
In either case, the integrated $\alpha=-1.12\pm0.03$ \citep{LoiEtAl20} matches the expected spectrum, as it simply reflects the source CRIs.
Needless to say, leptonic models cannot produce consistent relics in such weak shocks, and certainly not their spectra.

\subsubsection{Criticism and Outlook}

Thus, a simple, minimal hadronic model already accounts for the CREs needed to explain MHs, GHs, and relics, leaving little room for leptonic alternatives, and avoiding the aforementioned difficulties of the latter (\UK).
Nevertheless, hadronic models are increasingly dismissed in favour of leptonic, in particular turbulent (re)acceleration, models for halos (\VW, and references therein), and were are not even considered until now for relics (except in {\UK}) and SHs.

For MHs, both hadronic and leptonic models are often considered, but the former is increasing disfavored, for example because observation do not precisely show its steady-state $-1.1\lesssim\alpha\lesssim-1.0$ spectrum \citep[\eg][concerning $\alpha=-0.95\pm0.10$ in the Phoenix cluster]{TimmermanEtAl21} or by interpreting the MH--CF association as evidence that MHs arise from particle (re)acceleration in sloshing-induced turbulence \citep[][and references therein]{GiacintucciEtAl19Expanding} --- although shear magnetization, anticipated \citep{KeshetEtAl10} and already observed \citep{ReissKeshet2014, NaorEtAl2020} below CFs, suffices to explain the association in a hadronic model.

For GHs, there is an increasing tendency to discount the hadronic model (\eg \VW).
One argument is the disagreement of a simplistic hadronic model with the observed integrated spectral and spatial characteristics and radio--X-ray relations \citep{BrunettiJones14}, the main example being the merged relic--GH system in A521, where the GH is patchy and barely detectable at high frequencies, implying a very soft spectrum ($\alpha\simeq -2.1$ according to \citealt{BrunettiEtAl08}, later corrected to $\alpha\simeq -1.8$ in \citealt{MacarioEtAl13}).
However, this and similar soft GHs are associated with very recent or ongoing mergers \citep[\UK;][]{BourdinEtAl13,YoonEtAl20}, so one cannot neglect transient effects which soften the spectrum (see {\UK} and \S\ref{sec:Halos}--\S\ref{sec:Relics} below).
An allegedly stronger claim is that \gama-ray upper limits and recent detections of the centre of Coma are fainter than the $\pi^0$ counterpart of the secondary CRE-producing $\pi^\pm$ \citep[][\VW]{BrunettiEtAl12, BrunettiEtal17, AdamEtAl21}.

As we show, these objections raised against a hadronic origin are not well-justified, at best pointing out some failure of an oversimplified version of the model.
In particular, many of the possible concerns are alleviated by taking into account CR diffusion, advection, and magnetic evolution, as shown in {\UK} and below.
Future sensitive radio observations of faint halo- or relic-like emission from any magnetized region in the ICM, better constraints on the $\pi^0\to\gamma\gamma$ and secondary inverse-Compton signals, and more sophisticated modeling, would further test and improve the hadronic model.
Next, we address the strongest argument ostensibly made (\eg \VW) against the hadronic model, namely the faint \gama-ray emission from the centre of Coma.

\subsection{Consistent gamma-ray counterpart}
\label{subsec:GammaRays}

A diffuse \gama-ray signal was recently reported near the centre of the Coma cluster \citep{XiEtAl18_Coma}, with a fitted spectral index ranging from $\myp\simeq 2.2$ \citep{BaghmanyanEtAl22} to $\myp\simeq 2.6$ \citep{AdamEtAl21} in energies above $100$ or $200\MeV$.
The \emph{Fermi} point spread function (PSF) radius at $\lesssim200\MeV$ energies is very extended ($>2\dgr$ for $68\%$ acceptance of front-type events), and while the foreground is relatively low near the Galactic pole, it is still more than an order of magnitude above the reported signal; consequently, the inferred properties of the excess are sensitive to modeling and source-removal details.
The remaining excess can be described as a radius $r\simeq 1\dgr$ disk of uniform brightness; the reported fluxes above $200\MeV$ are $F_1\simeq 2\times 10^{-12} \erg\se^{-1}\cm^{-2}$ \citep{XiEtAl18_Coma}, $F_2\simeq 5\times 10^{-13} \erg\se^{-1}\cm^{-2}$ \citep{AdamEtAl21}, and $F_3\simeq 1.6\times 10^{-12}\erg\se^{-1}\cm^{-2}$ \citep{BaghmanyanEtAl22}.
In radio, most of the emission arises within $r\lesssim600\kpc$ (equivalent to $\lesssim 21'$), where the magnetic field is thought to be fairly strong, $b\gtrsim 1$, and the logarithmic flux $\nu F_\nu\simeq 1.4\times 10^{-14}\erg\se^{-1}\cm^{-2}$ is approximately frequency-independent \citep[][and references therein; $F_\nu$ is inferred by integrating their figure 8]{BonafedeEtAl22}.

A CRI deposits approximately twice more energy in neutral pions than in secondary CREs for the relevant frequency ranges \citep[\eg][]{KamaeEtAl06,KushnirWaxman09}.
For the simple case of a flat, $\alpha=-1$ spectrum with no evolution or CR diffusion, the local $\pi^0\to\gamma\gamma$ and synchrotron emissivities are thus related to each other by
\begin{equation}\label{eq:Pi0simple}
  \kappa\equiv \frac{(\nu j_\nu)_{\pi^0}}{(\nu j_\nu)_{\text{syn}}} \simeq 2(1+b^{-2}) \fin
\end{equation}
Attributing the $\gamma$-ray excess to the $r<600\kpc$ region, its spatially-integrated ratio to the radio flux is found to be $\kappa\simeq 20$, $5$, or $14$, if one adopts respectively $F_1$, $F_2$, or $F_3$, with $(20\text{--}50)\%$ statistical uncertainty factors and considerable systematic uncertainties.
Here, we normalized the excess by the $\gtrsim3$ orders of magnitude it spans in photon energy, adopting the $p=2$ CRI spectrum corresponding to the radio $\alpha=-1$.

These \gama-ray-to-radio ratios are comparable to, and even exceed, the $2(1+\langle b^{-2}\rangle) \sim 4$ anticipation for a mean $b\simeq 1$, nicely validating the prediction of the GH hadronic model; for a more detailed analysis, see Kushnir et al. (in preparation).
One may argue for a somewhat smaller ratio of coincident fluxes, especially at high energies, by claiming that the \gama-ray signal is more extended, partly attributed to additional faint sources, or spectrally much softer.
However, given the above results and their uncertainties, such an argument cannot robustly imply that the radio signal is too strong to constitute the $\pi^\pm$ counterpart of the $\pi^0$ signal.

For example, \citet{AdamEtAl21} claim that their estimated \gama-ray excess is too weak to be the counterpart of the radio GH, especially if they attribute $\sim40\%$ of the signal to an uncertain source in the 4FGL catalogue and adopt a soft, $\myp\simeq 2.6$ spectrum, thus allegedly necessitating some additional CRE (re)acceleration processes to explain the GH.
However, such a claim does not seriously challenge the hadronic GH model, given the stronger \gama-ray signals reported by other groups (which used more appropriate photon cuts), the poor photon statistics especially at high energies, the inaccurate \gama-ray localization due to the very extended PSF, the systematic uncertainties associated in particular with the modeling of background and point sources essential for uncovering the diffuse excess, and the highly uncertain magnetic field.
Moreover, more realistic hadronic models correct the naive RHS of Eq.~\eqref{eq:Pi0simple}; in particular, the recent magnetic growth suspected inside the GH can lower the modelled \gama-ray-to-radio ratio throughout the $b>1$ volume (\UK).

Let us estimate the hadronic $\gamma$-ray signal in the simple, homogeneous CRI model of \S\ref{subsec:Motivation}.
Adopting the CRI energy density \eqref{eq:CRI_E1}, a fraction $f_\gamma\simeq0.1$ of the CRI energy radiated through $\pi^0\to\gamma\gamma$ implies a logarithmic \gama-ray luminosity
\begin{equation}\label{eq:Pi0HomogCRI}
  \epsilon L_\epsilon \simeq \frac{225 \xi_i f_s f_b^2 c f_\gamma \sigma_i H^2 M k_B T}{16\pi \mu m_p^2 f_r^3 G \ln\gamma_{\text{max}}}
\simeq 8\times 10^{40}  \xi_i f_s M_{14}T_5\erg\se^{-1} \fin
\end{equation}
For the parameters of Coma --- $M\simeq 10^{15}M_\odot$, $k_BT\simeq 8\keV$, and $z\simeq 0.023$ --- the implied $200\MeV$--$300\GeV$ flux is
\begin{equation}\label{eq:Pi0HomogCRIComa}
  F\simeq \frac{\epsilon L_\epsilon}{4\pi d_L^2}\ln\left(\frac{300}{0.2}\right)
\simeq 8\times 10^{-12}  \xi_i f_s \erg\se^{-1}\cm^{-2} \coma
\end{equation}
where $d_L$ is the luminosity distance.
Comparing this flux to the above measurements confirms that $\xi_i f_s\simeq 0.2$ is a reasonable value.

Finally, note that some of the reported diffuse central \gama-ray excess should be attributed to projected inverse-Compton emission from primary CREs accelerated by the virial shock, although the above studies did not focus on such extended emission and would thus attribute much of it to the background or to removed point sources.
Interestingly, the radially-binned excess reported by \citet[][see their figure 6]{AdamEtAl21} does show a small local enhancement at the virial, $r\simeq 1\dgr$ radius, although it is of low significance as their analysis was not suited for a search for a faint, thin, elliptical ring.
The total $200\MeV$--$300\GeV$ flux from the large, elliptic, virial ring can be estimated as $F\simeq 4\times 10^{-11}\erg\se^{-1}\cm^{-2}\sr^{-1}$ by extrapolating the preliminary $>220\GeV$ VERITAS signal \citep[][with a systematic uncertainty factor of a few and assuming $\myp=2$]{KeshetEtAl12_Coma}, or as $F\simeq 8\times10^{-12}\erg\se^{-1}\cm^{-2}\sr^{-1}$ by extrapolating the 1--100 GeV \emph{Fermi} excess \citep[][with a factor $\sim 2$ systematic uncertainty]{KeshetReiss18}.

\subsection{Cosmic-ray diffusion and advection}
\label{subsec:CosmicRayDiffusionDdvection}

The spatial and spectral properties of radio emission in the hadronic model become much richer in the presence of strong diffusion or advection.
Here, the CRE density $N(t,E,\bm{r})$ per unit volume and unit energy evolves according to
\begin{align} \label{eq:DiffLoss0}
\frac{dN}{dt} & = \frac{\pr \myn}{\pr t} + \pmb{v}\cdot\bm{\nabla}\myn \\
& = \dot{\myn}_+ + \bm{\nabla}\left( D\bm{\nabla} \myn \right) - \frac{2+\phi}{3} N \bm{\nabla}\cdot \pmb{v} - \frac{\pr}{\pr E}\left( \myn \dot{E}_{\text{cool}} \right) \nonumber \coma
\end{align}
where $\pmb{v}(t,\bm{r})$ is the local mean velocity of the CREs,
$\dot{\myn}_+(t,E,\bm{r})$ is the CRE injection rate into the ICM by inelastic CRI collisions,
and diffusion is approximated as isotropic, with a scalar coefficient $D(t,E,\bm{r})$.
The $\bm{\nabla}\cdot \pmb{v}$ term accounts for the adiabatic compression of CREs with a local spectral index
\begin{equation}\label{eq:phiDef}
  \phi(t,E,\bm{r}) \equiv -\frac{\pr\ln N}{\pr\ln E} \fin
\end{equation}
The last term in Eq.~\eqref{eq:DiffLoss0} incorporates the radiative cooling of the CREs, at a rate
\begin{equation}
\label{eq:CRECool}
\dot{E}_{\text{cool}} = -\mypsi E^2 \, ,
\end{equation}
where the cooling parameter
\begin{equation}
\mypsi(t,E,\bm{r}) \!\simeq \!\frac{4\sigma_T}{3m_e^{2}c^{3}} \!\left(\!u_{\text{cmb}}+\frac{B^2}{8\pi}\right)\!
\!\simeq\! 0.83\frac{(1+z)^4(1+b^2)}{\GeV\Gyr}
\end{equation}
is dominated either by Compton scattering off CMB photons or by synchrotron losses, depending on $b$.
Here, $\sigma_T$ is the Thomson cross-section and $m_e$ is the electron mass.

Equation \eqref{eq:DiffLoss0}, the implied CRE population, and the resulting synchrotron signature are discussed here briefly, and studied in more detail below in both steady-state systems (\S\ref{sec:Halos} and \S\ref{sec:SH}) and in the evolving ICM near a shock (\S\ref{sec:Relics}).
The magnetic field $\bm{B}(\bm{r})$ is assumed regular and slowly evolving; time-dependent and irregular magnetic effects, outlined in {\UK} and below, are in general beyond the scope of the present work.

In a hadronic model, one can usually approximate the energy spectrum of CRE injection as spatially uniform, given by a power-law of some fixed index
\begin{equation} \label{eq:pDef}
p\equiv -\frac{\partial\ln \dot{N}_+}{\partial\ln E} \, .
\end{equation}
This approximation, justified by the slow evolution of the CRI spectrum, is particularly good when CRI diffusion is strong.
The upper energy cutoff on CRE injection can be ignored, as it is too high to be relevant for radio emission, and has a negligible effect on the dynamics for $p\geq2$.
When considering a volume-integrated system that evolves slowly and is sufficiently homogeneous, the diffusive term in the transport equation can be omitted, and the problem simplifies considerably.
Here, for a constant $p$
and a spatially uniform, constant cooling rate $\psi$, the integrated CRE population develops an $N\propto \psi^{-1}E^{-(p+1)}$ steady-state distribution, resulting in synchrotron emission of spectral index
\begin{equation}\label{eq:alphaDef0}
\alpha\equiv \frac{\pr\ln j_\nu}{\pr\ln \nu} = \frac{1}{2}+\frac{1}{2}\frac{\pr\ln N}{\pr\ln E} = -\frac{p}{2}\simeq -1\, ;
\end{equation}
here, for consistency with recent literature, $p>0$ and $\alpha<0$.

However, both local and volume-integrated spectra can deviate considerably from Eq.~\eqref{eq:alphaDef0}, even in the weak-diffusion regime, especially if the system evolves rapidly or becomes inhomogeneous on sub-diffusive scales.
For example, a rising level of plasma magnetization can lead to substantial spectral softening even for modest ratios $\mathcal{R}$ of $\psi/\dot{N}_+$ amplification.
Even in a steady state, the combination of magnetic irregularities and CRE streaming and diffusion can further soften the spectrum.
As an example, consider a small magnetic filling factor on scales not much smaller than the CRE Larmor radius.
Here, while CREs Compton-cool steadily, they synchrotron radiate only intermittently, as they cross highly magnetized regions or even become trapped by them.
Correlations between the energy of CREs and the time they spend in more strongly magnetized regions thus modify the spectrum.
In particular, as lower-energy CREs are more easily deflected magnetically, they generally spend longer times in such regions, thus softening the spectrum.
Additional effects can modify the spectrum at very low radio frequencies, where the compressed CRI spectrum may deviate from a pure power law, the cross-section for charged pion production becomes more energy dependent, and CREs can accumulate throughout the life of the cluster.
In addition, when the irregular field contains regions of sufficiently high $B$, the elevated cooling rate in those regions further softens the spectrum.
See {\UK} for further discussion.

More importantly, there is increasing evidence that CR mixing in the ICM is substantial, equivalent to strong, $D(100\GeV)\simeq 10^{31\text{--}32}\cm^2\se^{-1}$ diffusion.
For such a strong diffusion, even a smooth, steady-state ICM shows strong deviations from the nominal spectrum \eqref{eq:alphaDef0}, both locally and when volume-integrated.
In general, secondary CREs diffuse from regions of fast CRE injection or slow cooling to regions of smaller $(\dot{N}_+/\psi)$, gradually cooling in the process.
Consequently, high $(\dot{N}_+/\psi)$ regions harbour a larger fraction of uncooled CREs, which radiate an $\alpha\simeq -(p-1)/2\simeq -1/2$ spectrum; as we show, relic edges can achieve this limit, while the centres of GHs can be as hard as $\alpha\simeq -(p+d-1)/2$ for $D=D_0 E^{d}$ diffusion.
Regions dominated by incoming diffusing CREs show the opposite, softening effect, which is pronounced if these CREs have already cooled substantially; as we show, the resulting spectrum can be arbitrarily soft for sufficiently strong $\dot{N}_+$ gradients.

The typical lengthscale for diffusion-induced variations in synchrotron brightness and spectrum is
\begin{equation}\label{eq:lDef0}
  l(E,\bm{r})
  \equiv \left(\frac{D}{\psi E}\right)^{1/2}
  \propto  \left(\frac{D_0}{1+b^2}\right)^{\frac{1}{2}} \left(\frac{b}{\nu}\right)^{\frac{1-d}{4}} \fin
\end{equation}
This scale is indeed characteristic of the observed variations in $\alpha$, provided that diffusion is strong.
Conversely, the observed variations in GH and relic spectra are separately used below to estimate the diffusion coefficient, indicating in both cases that $D(100\GeV)\simeq 10^{31\text{--}32}\cm^2\se^{-1}$, consistent with the level (\UK) necessary to keep the CRI distribution sufficiently homogeneous.
A detailed analysis is deferred to \S\ref{sec:Halos}--\S\ref{sec:SH} and Appendices \S\ref{app:PLSolutions}--\S\ref{app:GreenSolutions}, which focus on different types of ICM sources and demonstrate that a hadronic model with strong diffusion reproduces, essentially, all observations.

For diffusion as strong as we infer, weak shocks in the ICM cannot efficiently accelerate or re-accelerate electrons to radio-emitting frequencies.
Indeed, the ratio between the cooling time and acceleration time of such CREs,
\begin{equation}\label{eq:AccelerationTimeRatio}
  \frac{t_{\text{cool}}}{t_{\text{acc}}} \simeq \frac{(\psi E)^{-1}} {D/v_s^2}
  \simeq 0.4\frac{\mach_2^2 T_{\rm{keV}}\nu_9^{-1/2}}{(1+z)^4 D_{32}}\frac{2b^{1/2}}{1+b^2} \, ,
\end{equation}
is not much greater than unity.
Here, $v_s$ is the shock velocity, $\mach\equiv 2\mach_2$ is the shock Mach number, $T_{\rm{keV}}$ refers to the upstream plasma, and we defined $D_{32} \equiv D/10^{32}\cm^2\se^{-1}$ and $\nu_9\equiv \nu/\mbox{GHz}$.
An analogous argument disfavours any turbulent (re)acceleration of electrons when diffusion is sufficiently strong.
While the hadronic model does not invoke any such CRE (re)acceleration in weak shocks or turbulence, leptonic models do.
In particular, leptonic models for radio relics all assume some form of CRE (re)acceleration in weak, including $\mach<2$, shocks, unsupported by independent observations or robust theory, and implying radio spectra softer than observed (\UK); we find that such models are inconsistent with strong diffusion, and would over-produce relic energies once secondaries of a sufficiently homogeneous CRI distribution are taken into account.

\subsection{Summary of evidence for a joint hadronic model}
\label{subsec:Evidence}

To conclude this section, we summarize the evidence indicating that
the distinctions between MHs, GHs, and relics are extrinsic, that these systems are manifestations of the same underlying mechanism, and that this mechanism is hadronic.

The main evidence summarized in {\UK} includes:\\
(i) The same $u_i\simeq 10^{-(12.4\text{--}13.3)}\erg\cm^{-3}$ CRI energy density accounts collectively for MHs, GHs, and relics (\UK);\\
(ii) A common origin of MHs and GHs indicated by their identical central $\eta\equiv\nu I_\nu/F_X$ radio--X-ray ratios (\KL);\\
(iii) Radio bridges strongly link GHs and relics as the same phenomenon ({\UK} and references therein);\\
(iv) Moreover, some GHs show a weak shock at their edge: the defining property of relics (\UK, \VW);\\
(v) Clusters with both GH and MH characteristics,
challenging the MH--GH classification \citep[\UK;][]{StormEtAl15}; \\
(vi) GHs and MHs with relic characteristics, such as irregular or filamentary morphology or polarization (see \S\ref{subsec:Motivation});\\
(vii) The hadronic model explains the integrated radio--X-ray relations in GHs \citep{KushnirEtAl09} and MHs (\KL);\\
(viii) Local $I_\nu$--$F_X$ relations in GHs consistent with homogeneous CRIs in both strong and weak field regimes (\UK; \S\ref{subsec:Relations});\\
(ix) The hadronic model naturally yields the integrated $-1.2\lesssim\alpha\lesssim -1.0$ of relaxed MHs, GHs, and relics (\UK); \\
(x) Soft GH spectra are a transient, young-merger effect, as indicated for instance by relic distances (\UK, \eg fig. 28);\\
(xi) Integrated relic spectra are too universal for DSA, and inconsistent with coincident X-ray Mach numbers (\UK);\\
(xii) Spectral softening is anticipated in hadronic models due to magnetic evolution and irregularities, accounting in part for the spectro-spatial properties of GHs and relics (\UK); \\
(xii) Little or no diffuse ICM emission, even around shocks, in galaxy groups and clusters of low mass \citep[\eg in A2146;][]{RussellEtAl11}, as expected in hadronic models;\\
(xiv) It was also argued that invoking DSA in relics requires unreasonably high acceleration efficiencies \citep{KangEtAl07}.\\

Additional evidence accumulated during the past decade has strengthened the case for a joint hadronic model for diffuse ICM radio phenomena, and in part can be considered as predictions of this model.
In particular:\\
(i) Measurements of virial-shock CRE acceleration with $\xi_e\sim1\%$ efficiencies and $p\simeq 2.0\pm0.2$ spectra, supporting the CRI counterpart needed for the hadronic model (see \S\ref{subsec:Motivation});\\
(ii) Detection of the \gama-ray counterpart to the GH in Coma, consistent with the $\pi^0\to\gamma\gamma$ hadronic prediction (see \S\ref{subsec:GammaRays}); \\
(iii) Increasingly irreconcilable radio vs. X-ray Mach numbers in relic leptonic models \citep[\eg][and references therein]{AkamatsuEtal2017, Urdampilleta18, WittorEtAl21, deGasperinEtAl22};\\
(iv) Nearly constant spectra along relic weak shock fronts, consistent with hadronic but not leptonic models (\S\ref{subsec:RelicObs});\\
(v) MHs and GHs show the same, overlapping $P_{\nu}$--$M_{500}$ correlation \citep{YuanEtAl15};\\
(vi) Relics show similar $P_{\nu}$--$L_{500}$ correlations as do GHs and MHs \citep[][and \S\ref{subsec:Relations}]{YuanEtAl15};\\
(vii) The luminosities of neighbouring relics and GHs are approximately equal to each other \citep{PauloEtAl16Presentation};\\
(viii) A tighter, linear $P_\nu\propto Y_{SZ}$ radio--SZ relation emerges when one focuses on the halo region \citep{Basu12}, as expected in the hadronic model with a flat CRI distribution (\UK);\\
(ix) The weak-lensing surface mass density in Coma \citep{BrownRudnick11} correlates nicely with radio brightness, again in agrement with homogenous CRIs; \\
(x) SHs detected at $\gtrsim 1\Mpc$ radii are natural in our hadronic model (see \S\ref{sec:SH}), but challenge leptonic models;\\
(xi) Similarities between neighboring GHs and relics, such as identical integrated spectra \citep[\eg $\alpha=-1.16\pm0.02$ and $\alpha=-1.16\pm0.03$ in the 'toothbrush' relic and halo in 1RXS J0603.3+4214;][]{RajpurohitEtAl20};\\
(xii) Spectral variations within soft GHs correlate with the orientation of nearby relics \citep[\eg in A2256;][]{KaleDwarakanath10}, corroborating the transient GH nature;\\
(xiii) Highly uniform spectral maps in MHs and relaxed GHs \citep[\eg][]{vanWeerenEtAl16}, challenging leptonic models; \\
(xiv) Resolved spectral variations in relics and in some GHs are consistent with the hadronic model (see \S\ref{sec:Halos}--\S\ref{sec:SH}); \\
(xv) Strong CR diffusion measured from spectral variations in GHs and relics is consistent with the hadronic model but not with CRE (re)acceleration (see \S\ref{subsubsec:CoolingDiffusionScale}and \S\ref{subsec:RelicObs}). \\

Next, we supplement the qualitative evidence reviewed above with an analysis of the spatio-spectral distribution of radio emission in the hadronic model.
As we show, incorporating strong diffusion in the model reproduces the detailed spatial and spectral properties of observed MHs and GHs (\S\ref{sec:Halos}), relics (\S\ref{sec:Relics}), and even SHs (\S\ref{sec:SH}).

\section{Halos: steady-state hadronic model}
\label{sec:Halos}

\subsection{Setup}

\begin{bfigure*}
\begin{center}
\includegraphics[width=0.475\linewidth]{\figeps{alpha4_vs_rho}}
\includegraphics[width=0.475\linewidth]{\figeps{alpha5_d05_vs_rho}}
\end{center}
\caption{\label{fig:alphaPLAnalytic}
Self-similar distribution $\alpha(\mathfrak{r})$ of synchrotron spectral index arising from $\dot{N}_+\propto r^{-\myi}E^{-2}$ CRE injection of spatial power-laws $\myi=\{0,1,2,2.5,2.7,2.9\}$ (thin magenta to thick blue curves), computed numerically [using the PDE \eqref{eq:DiffLoss}; curves] and analytically [for integer $\myi$; using the ODE \eqref{eq:ODE}; symbols], for an energy-independent, uniform cooling $\psi$.
The asymptotic limit $\alpha_\delta$ of Eq.~\eqref{eq:Frho3} is also shown (black diamonds).
The results, using the simplified synchrotron function approximation \eqref{eq:syn_j}, are shown for a homogeneous diffusion function $D\propto E^0$ (left panel) and for $D\propto E^{1/2}$ (right panel), as a function of the dimensionless $\mathfrak{r}$ combining $r$, $\psi$, $D$, and $E$ or $\nu$.
}
\end{bfigure*}

Considering the long-term, central diffuse radio emission from a cluster or group of galaxies, we approximate the ICM as spherically symmetric and stationary.
In a hadronic model, the density $N(E,r)$ of CREs per unit volume and unit energy, injected into the ICM at a steady rate $\dot{\myn}_+(E,r)$ by inelastic CRI collisions, then satisfies a steady-state special case of the diffusion-loss partial differential equation (PDE) \eqref{eq:DiffLoss0},
\begin{equation} \label{eq:DiffLoss}
0 = \frac{\pr \myn}{\pr t} = \dot{\myn}_+  + \frac{1}{r^{2}}\frac{\pr}{\pr r}\left(r^2 D \frac{\pr \myn}{\pr r}\right) - \frac{\pr}{\pr E}\left( \myn \dot{E}_{\text{cool}} \right) \coma
\end{equation}
where $r$ is the radius.
The inner boundary condition of PDE \eqref{eq:DiffLoss} is a vanishing diffusive flux at the origin, $r^2 D \partial_r \myn \to 0$ as $r \to 0$.
CRE injection is assumed to vanish as $r$ or $E$ diverge, so the outer boundary conditions forbid particles from reaching infinite energy, $\myn(E\to\infty)=0$, or escaping to spatial infinity, $\myn(r\to\infty)=0$.
CREs are assumed to be injected with a power-law of constant index $p$, giving a spectrum $\alpha=-p/2$ for regular magnetic fields if injection is uniform or diffusion is absent; see Eqs.~\eqref{eq:pDef}--\eqref{eq:alphaDef0} and the accompanying discussion.

\subsection{Self-similar distribution}

We begin with the simple case where $\psi$ and $D$ are both spatially-uniform, so the lengthscale $l(E)$ of Eq.~\eqref{eq:lDef0} is defined globally
for CREs of energy $E$.
Consider the self-similar case where,
in addition, one can approximate as power-laws both the radial dependence of CRE injection,
\begin{equation} \label{eq:PLInj}
\dot{N}_+ = \mathfrak{N} E^{-p} r^{-\myi}   \, ,
\end{equation}
and the energy-dependence of diffusion,
\begin{equation}\label{eq:DiffEOfd}
D = \mathfrak{D} E^{d}  \, ,
\end{equation}
with dimensional coefficients $\mathfrak{N}$ and $\mathfrak{D}$ (using Gothic symbols to denote self-similar variables), and dimensionless constants $\myi\geq0$ and $d<1$.
Dimensional analysis then implies that the steady-state CRE distribution is given by
\begin{equation} \label{eq:PLN}
\myn = \frac{l^{-\myi}\mathfrak{N}}{E^{p+1}\psi} \mathfrak{n}(\mathfrak{r}) \, ,
\end{equation}
where we defined a dimensionless distribution function $\mathfrak{n}(\myX)$ and a dimensionless (energy-dependent) radius
\begin{equation}\label{eq:MathFrakR}
\mathfrak{r}(E)\equiv r/l(E) \, ,
\end{equation}
which monotonically increases with $r$ and, as we assume that $d<1$, also with $E$.

In the more general, non-self-similar case, where the spatial dependence of $\dot{N}_+$ is not a pure power-law, additional lengthscales $r_1,r_2,\ldots$ are introduced to the problem.
Then, the one-dimensional $\mathfrak{n}(\myX)$ is replaced by a multi-dimensional distribution, depending separately on $r$ and $E$ through additional dimensionless parameters $r_1/l, r_2/l,\ldots$.
Such injection profiles, as well as spatial variations in $D(E,r)$ and $\psi(r)$, are addressed in \S\ref{subsec:HaloGeneral} and later in this section.

In the self-similar case, plugging Eqs.~\eqref{eq:PLInj}--\eqref{eq:PLN} into the PDE \eqref{eq:DiffLoss} yields an ordinary differential equation (ODE),
\begin{equation}\label{eq:ODE}
\left(p-1-\frac{1-d}{2}\myi\right)\mathfrak{n}(\myX) - \left(\frac{2}{\myX}+\frac{1-d}{2}\myX \right)\mathfrak{n}'(\myX) -\mathfrak{n}''(\myX) = \myX^{-\myi} \, ,
\end{equation}
for the one-dimensional distribution $\mathfrak{n}$.
For $\myi>0$, the outer boundary condition is a vanishing $\mathfrak{n}(\myX\to \infty)=0$ at an infinite radius or energy.
The inner boundary condition of a vanishing diffusive flux in the $\myX\to0$ centre translates to $\mathfrak{n}(\myX)\myX\to0$.
With these boundary conditions, the ODE \eqref{eq:ODE} admits a formal analytic solution for any radial injection profile $0<\myi<3$; see Appendix \S\ref{app:PLSolutions}.
An asymptotic solution can also be derived for the limiting case $\myi=3$, beyond which the number of particles in the centre diverges non-logarithmically.
The resulting synchrotron spectrum is demonstrated in Fig.~\ref{fig:alphaPLAnalytic}, for the nominal flat, $p=2$ injection and a few choices of $\myi$, both for an energy-independent, $d=0$ diffusion (left panel) and for the $d=1/2$ diffusion corresponding to Kraichnan-like \citep{Kraichnan65, BerezinskiiEtAl90Book} turbulence (right panel).
The analytic solutions to ODE \eqref{eq:ODE} are verified and supplemented by direct numerical solutions to PDE \eqref{eq:DiffLoss}.

For simplicity, in the figure we approximate the specific synchrotron emissivity as 
\begin{align} \label{eq:syn_j}
j_\nu & \simeq \sqrt{3}\, e  B r_e \sin(\myPalpha) \int \myn(E) F_{s} \left[ \frac{\nu}{\nu_s(E)} \right] \,dE \nonumber \\
& \simeq \alpha_e (m_e c^2)^2\sqrt{\frac{b \nu}{3\nu_0}} \myn\left(E=h\sqrt{\frac{\nu_0\nu}{b}}\right) \, ,
\end{align}
where we crudely replaced the synchrotron source function \citep{RybickiLightman86},
\begin{equation}\label{eq:FSynExact}
F_{s}(\myz) \equiv \myz \int_\myz^\infty K_{5/3}(\myz')\,d\myz' \, ,
\end{equation}
by $\delta(\myz-1)$; thus, we retain the exact CRE spectrum and only slightly over-sharpen the photon spectrum, 
as shown for completeness in \S\ref{subsec:HaloGeneral} below.
Here, $e$ is the electron charge, 
$\alpha_e\equiv e^2/(\hbar c)$ is the fine-structure constant, $h\equiv 2\pi\hbar$ is Planck's constant,
$\myPalpha\simeq \pi/4$ is the average of the pitch angle, assumed isotropically distributed,
\begin{equation} \label{eq:nu_s}
\nu_s \equiv
\frac{b}{\nu_0}\left(\frac{E}{h}\right)^2 \simeq 0.9 b \left(\frac{E}{5\GeV}\right)^2 (1+z)^{2} \GHz
\end{equation}
is the synchrotron frequency,
a cooling frequency
\begin{equation} \label{eq:nu0Def}
\nu_0 \equiv \frac{8 \alpha_e^2 m_e c^3}{9\sigma_T e B_{\text{cmb}}\sin\myPalpha} \simeq 1.6\times 10^{39}(1+z)^{-2}\mbox{ Hz}
\end{equation}
was defined ($\nu_0^{-1}\equiv h^2 a B_{\text{cmb}}$ in {\UK} notations), and $K_n(\myz)$ is the modified Bessel function of the second kind.

The approximate Eq.~\eqref{eq:syn_j} provides a simple mapping of the CRE spectrum onto the synchrotron spectrum,
\begin{equation}\label{eq:alphaLocal}
  \alpha \simeq
  \frac{1}{2} + \frac{1}{2}\frac{\partial \ln\myn}{\partial \ln E}\left(E=h\sqrt{\frac{\nu_0\nu}{b}},\,r\right) \, ,
\end{equation}
generalizing Eq.~\eqref{eq:alphaDef0} locally.
In this approximation, the radial dependence $\alpha(r\propto\mathfrak{r})$ of the spectral index can be read directly from Fig.~\ref{fig:alphaPLAnalytic} for any fixed frequency $\nu$.
In the self-similar case considered in this subsection, the spectral index,
\begin{equation}\label{eq:alphaLocalSeS}
  \alpha \simeq
  -\frac{p}{2} + \frac{1-d}{4}\left(\myi+\frac{d \ln\mathfrak{N}}{d \ln \mathfrak{r}} \right) \, ,
\end{equation}
depends on the parameters $\nu$ and $r$ only through their combination
\begin{equation}\label{eq:gorSeS}
  \mathfrak{r}
  \simeq \left(\frac{\psi h}{D}\right)^{\frac{1}{2}}  \left(\frac{\nu_0 \nu}{b}\right)^{\frac{1}{4}} r
  = \left(\frac{\psi}{\mathfrak{D}}\right)^{\frac{1}{2}} h^{\frac{1-d}{2}} \left(\frac{\nu_0 \nu}{b}\right)^{\frac{1-d}{4}} r
   \, ,
\end{equation}
so the full radio spectrum $\alpha[\nu\propto \mathfrak{r}^{4/(1-d)} b]$ can also be read directly from Fig.~\ref{fig:alphaPLAnalytic} for any fixed radius $r$, by stretching the abscissa accordingly.
As $\psi(b)$ was assumed constant, the above holds separately in regions where either $b\gg1$ or $b\ll1$.

\begin{bfigure*}
\begin{center}
\includegraphics[width=0.475\linewidth]{\figeps{alpha4_Rc_vs_rho_rc}}
\includegraphics[width=0.475\linewidth]{\figeps{alpha4D05_Rc_vs_rho_rc}}
\end{center}
\caption{\label{fig:alphaCore}
Same as Fig.~\ref{fig:alphaPLAnalytic}, but for $\dot{N}_+\propto (1+r^2/r_c^2)^{-\myi/2}E^{-2}$ CRE injection with a core of dimensionless radius $\mathfrak{r}_c\equiv r_c/l=1$ (solid curves) and $\myi=\{0,1,\ldots,6\}$ (thin magenta to thick blue curves).
For the case $\myi=4$ (cyan), results are also shown also for smaller $\mathfrak{r}_c=1/3$, $1/10$, and $1/100$ (increasingly longer dashed) and for a larger $\mathfrak{r}_c=2$ (dash-dotted).
The $\alpha_\delta$ limit is again shown (black diamonds).
}
\end{bfigure*}

As the figure shows, the spectrum in a dense ($\myi>0$) centre is harder than the standard $\alpha=-p/2$ spectrum of uniform ($\myi=0$) injection, as CREs have a limited time to soften by cooling before diffusing outward.
This central hardening is offset in part if higher-energy CREs diffuse faster ($d>0$), as their escape from the centre softens it.
The spectrum initially softens outwards from the hard centre, with increasing $r$.
For sufficiently steep, $\myi\gtrsim 1$ injection profiles, the spectrum then softens even beyond $\alpha=-p/2$, reaching a minimal $\alpha$ around $2\lesssim\mathfrak{r}\lesssim5$ before hardening back to $\alpha=-p/2$ at large radii.
The $\alpha$ minimum is shallower and moves outward for diffusion of stronger energy-dependence (\ie a larger $d$); overall, the $\alpha(\mathfrak{r})$ profile smoothes out as $d$ increases.

The spectral index in the $\myi=3$ limit (derived in Appendix \S\ref{app:PLSolutions} and shown in the figure as black diamonds),
\begin{align}\label{eq:LimitingAlpha}
  \alpha_\delta\equiv \alpha(\myi=3) = & -\frac{p-1+d}{2} - \frac{\left(1-d\right)^2}{8}\mathfrak{r}^2  \\
  & -  \frac{2p-3+d}{4} \sqrt{1-d}
  \frac{H_{-\frac{2p-2}{1-d}}\left(\frac{\sqrt{1-d}}{2}\mathfrak{r}\right)}
  {H_{-\frac{2p-3+d}{1-d}}\left(\frac{\sqrt{1-d}}{2}\mathfrak{r}\right)}\mathfrak{r} \, , \nonumber
\end{align}
where $H_n(x)$ is the Hermite polynomial, provides both a tight upper bound on the hard spectrum near the centre, and a lower bound on the soft, $\alpha<-p/2$ spectrum elsewhere.
In particular, sufficiently steep, $\myi\geq2$ injection profiles all asymptote to the same central spectrum,\!\!\!\!
\begin{equation}\label{eq:MaximallyHard}
\alpha(\tau\to 0) = -\frac{p-1+d}{2}  \, .
\end{equation}
The same $\alpha_\delta$ spectral distribution arises from $\dot{N}_+\propto \delta(r)$, or any injection sufficiently peaked near the origin; see Appendix \S\ref{app:GreenSolutions}.

\subsection{General distribution}
\label{subsec:HaloGeneral}

To proceed, it is instructive to solve Eq.~\eqref{eq:DiffLoss} using Green functions, for an arbitrary injection profile.
Consider a spherical shell of CREs injected at a radius $r_0>0$ with an initial energy $E_0$ at time $t=0$,
\begin{equation}
N_0(t=0,E,r)=\frac{\delta(r-r_0)\delta(E-E_0)}{4\pi r_0^2} \,,
\end{equation}
where we normalized $\int N_0 \,d^3r\,dE=1$.
The temporal evolution of this shell follows
\begin{equation} \label{eq:GreenFunction}
\!\!\frac{N_0(t,E,r)}{\delta\left[E-E(t)\right]} \!\equiv\! \myG(t,E,r;r_0) \! = \! \frac{ e^{-\left[\frac{r-r_0}{2r_d}\right]^2}-e^{-\left[\frac{r+r_0}{2r_d}\right]^2}}{8\pi^{3/2} r r_0 r_d} \,,
\end{equation}
solving the homogeneous part (\ie with no injection) of Eq.~\eqref{eq:DiffLoss}.
For a constant $\psi$, the energy evolution
\begin{equation} \label{eq:EnergyEvolve}
E(t)=\frac{E_0}{1+\psi E_0 t}
\end{equation}
solves ODE \eqref{eq:CRECool}, so the diffusion length $r_d$ follows
\begin{equation} \label{eq:DiffLength}
r_d^2(t,E) \!\equiv\! \int_0^t D[E(t')]\, dt' \!=\! \int_0^t D\left[\frac{E}{1-(t-t')\psi E }\right]\, dt' \,,
\end{equation}
and approaches the diffusion--cooling scale over a cooling time: $r_d(t=1/\psi E)=(1-d)^{-1/2}l$.

Next, consider an initial power-law energy spectrum of index $p\geq2$, with $N_0(E)dE = C E^{-p}dE$ particles injected in the energy range $[E,E+dE]$ at time $t=0$, where $C$ is a constant.
The number of particles in this $[E,E+dE]$ energy range at any time $0<t<(\psi E)^{-1}$ then equals $C E^{-p}(1-\psi E t)^{p-2}$.
Hence, the steady-state solution of Eq.~\eqref{eq:DiffLoss} for an ongoing injection $\dot{N}_+=\delta(r-r_0)E^{-p}/(4\pi r_0^2)$ of radius $r_0$ shells is
\begin{equation}\label{eq:N0}
\!N_0(E,r;r_0) \!= \!E^{-p}\!\!\int_{0}^{(\psi E)^{-1}}\!\!\!\!\!\! \myG(t,E,r;r_0) \left(1-\psi E t \right)^{p-2} dt \, ,\!
\end{equation}
accumulating all CREs starting at $r_0$ that managed to reach $r$ with energy $E$ during time $t$.
The steady-state solution for an arbitrary spatial injection $\dot{N}_+(E,r_0)=C(r_0)E^{-p}$ is then
\begin{equation}\label{eq:N}
N(E,r) = \int 4\pi r_0^2 C(r_0) N_0(E,r;r_0) \,dr_0 \, \fin
\end{equation}
Analytic solutions for the kernel $N_0(E,r;r_0)$ with both $d=0$ and $d=1/2$, as well as the distribution $N(E,r)$ arising for select injection profiles (including an exponential distribution), are provided in Appendix \S\ref{app:GreenSolutions}.

\begin{figure}
\begin{center}
\includegraphics[width=0.95\linewidth]{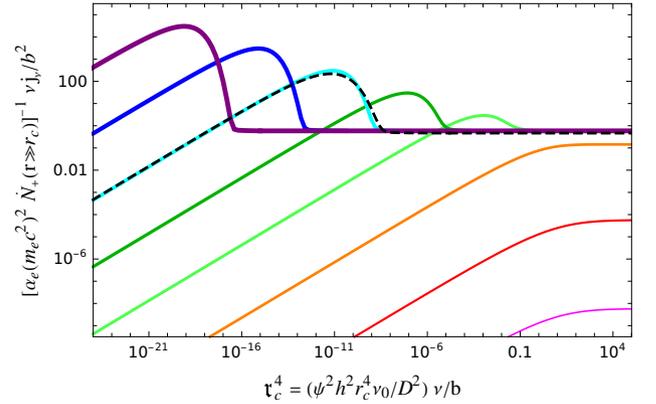}
\end{center}
\caption{\label{fig:alphaCoreJ}
Logarithmic synchrotron emissivity, $\nu j_\nu$, shown normalized as a function of normalized frequency $\mathfrak{r}_c^4\propto\nu$, for $p=2$ CRE injection with a $\myi=4$ core profile and $d=0$ diffusion, at $r/r_c=\{10^{-2},10^{-1},\ldots,10^{5}\}$ (thin magenta to thick purple).
The results (solid curves) are shown, as usual, using the approximation \eqref{eq:syn_j}; the exact $F_s$ of Eq.~\eqref{eq:FSynExact} yields very similar spectra, as demonstrated (dashed black) for $r/r_c=10^{3}$.
}
\end{figure}

\begin{figure}
\begin{center}
\includegraphics[width=0.95\linewidth]{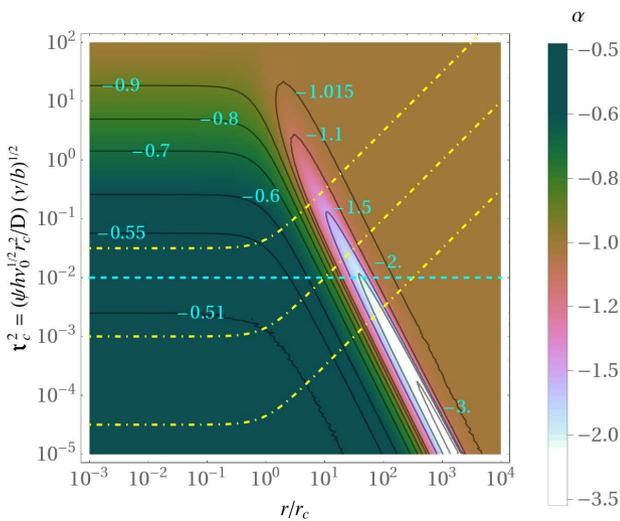}
\vspace{-0.5cm}
\end{center}
\caption{\label{fig:alphaCore2D}
Synchrotron spectral index (shading and labelled solid black contours) as a function of normalized radius $r/r_c$ and frequency $\mathfrak{r}_c^2\propto r_c^2 \nu^{1/2}$ for $p=2$ CRE-injection with a $\myi=4$ core profile and $d=0$ diffusion.
The $\alpha(\nu)$ spectrum at a given $r$ maps onto a vertical cut of this figure.
The $\alpha(r)$ spectrum at a given $\nu$ maps onto a horizontal cut if the magnetic field is uniform (demonstrated by a dashed cyan line, corresponding to the $\mathfrak{r}_c=0.1$ case of Fig.~\ref{fig:alphaCore}), and onto some non-horizontal cut for a variable field (demonstrated by dash-dotted yellow curves for equipartition, $b(r)^2\propto n_\beta$ fields).
}
\end{figure}

Consider a CRE injection profile with a core of radius $r_c$,
\begin{equation}\label{eq:InjectionCore}
  \dot{\myn}_+ = C E^{-p} \left( 1+\frac{r^2}{r_c^2}\right)^{-\myi/2}\, ,
\end{equation}
where $C$ is a constant.
The prescription \eqref{eq:InjectionCore} is proportional to a power of the gas density $n_\beta(r)$ in a $\beta$-model, $\dot{\myn}_+(r)\propto n_\beta(r)^{\myi/(3\beta)}$,
while approaching the $\dot{\myn}_+ \propto r^{-\myi}$ power-law profile of Eq.~\eqref{eq:PLInj} far outside the core.
Figure \ref{fig:alphaCore} demonstrates the resulting spectral index distribution $\alpha(\mathfrak{r})$ for a few choices of index $\myi$ and dimensionless core radius $\mathfrak{r}_c$.
Approximation \eqref{eq:syn_j} is used, so the radial dependence $\alpha(r)$ of the spectral index at a fixed frequency can still be read directly from the figure.
However, as the problem is no longer self-similar, the spectrum $\alpha(\nu)$ at a fixed radius cannot be obtained here simply by stretching the abscissa, for varying $\nu$ changes not only $\mathfrak{r}$ but also $\mathfrak{r}_c$.
The full spectra at specific radii are shown in Fig.~\ref{fig:alphaCoreJ} for $\myi=4$, but here the radial dependence of the spectrum at a given frequency cannot be directly read from the figure.
The two-dimensional distribution of $\alpha$ in $r$--$\nu$ space is shown for $\myi=4$ in Fig.~\ref{fig:alphaCore2D}, providing both the $\alpha(r)$ profile at a given $\nu$ (in a horizontal cut through the figure if $B$ is constant) 
and the $\alpha(\nu)$ profile at a given $r$ (vertical cut).
Figure \ref{fig:alphaCore2D} also provides the former, $\alpha(r)$ distribution for any magnetic $b(r)\ll1$ radial profile, by taking a corresponding non-horizontal cut; such cuts are demonstrated (dash-dotted yellow curves) for equipartition, $b^2\propto n_\beta^1$ fields.

Specifically, Fig.~\ref{fig:alphaCore} demonstrates the spectral index distribution $\alpha(\mathfrak{r})$ in the case of $p=2$ injection, a uniform magnetic field, $d=0$ (left panel) or $d=1/2$ (right panel) diffusion, and a few choices of index $\myi$ (solid curves) with a fixed $\mathfrak{r}_c=r_c/l=1$ core.
For the specific case $\myi=4$, the figure also presents the spectral distribution for a few other choices of $\mathfrak{r}_c$ (dashed and dash-dotted cyan curves).
As the figure confirms, the synchrotron spectrum arising from core injection of index $\myi'$ resembles a mixture of the power-law cases $\myi=\myi'$ and $\myi=0$, approaching the former for $\mathfrak{r}_c\ll1$ and the latter for $\mathfrak{r}_c\gg 1$.
In particular, the spectrum is again found to be hard, $\alpha>-p/2$ in the centre, softens to $\alpha<-p/2$ (if $\myi\geq 1$) at $2\lesssim\mathfrak{r}\lesssim5$, and asymptotes to $\alpha\to-p/2$ at large radii.
The softening at intermediate radii strengthens with an increasing $\myi$, diverging as $\mathfrak{r}_c\to 0$ if $\myi\geq3$.
The case of pure $\myi=3$ power-law injection (black diamonds), equivalent to $\delta(r)$ deposition in the centre, still limits both the hardness of the central region and the softness of the peripheral region.

The full spectrum at a given radius is shown in Fig.~\ref{fig:alphaCoreJ} under the same assumptions of Fig.~\ref{fig:alphaCore}, for the case $\myi=4$ and $d=0$, with a few choices of $r/r_c$.
Instead of the $\alpha(\mathfrak{r})$ profile, this figure shows the spectrum in terms of the emissivity $\nu j_\nu$, normalized in dimensionless form, as a function of the dimensionless frequency $\mathfrak{r}_c^4\propto \nu$.
Shallow injection profiles or large cores (thin, magenta to orange curves) lead to a smooth hard-to-soft transition with an increasing $\nu$, approaching $\alpha\to -p/2$ at high frequencies.
In contrast, for steep injection profiles with a small core (thicker, green to purple curves), the hard-to-soft transition goes through a very soft region where $\nu j_\nu$ rapidly decreases with an increasing $\nu$, in a drop that can be very sharp for large $r/r_c$ when $\myi\geq 3$.

The spectral profiles, even in cases with sharp drops, remain almost unchanged if one uses the exact $F_{s}$ of Eq.~\eqref{eq:FSynExact} or its approximation
\begin{align}\label{eq:FSynApprox}
F_{s}(\myz)  = & \frac{3^{\frac{1}{2}}27\pi }{2^{\frac{2}{3}}160} \frac{\myz^{\frac{11}{3}}\,{_1}F_2\left(\frac{4}{3};\frac{7}{3},\frac{8}{3};\frac{\myz^2}{4}\right)}{\Gamma\left(-1/3\right)}  - \frac{\pi \myz}{\sqrt{3}} \\
& \, + (4\myz)^{\frac{1}{3}} {_1}F_2\left(-\frac{1}{3};-\frac{2}{3},\frac{2}{3};\frac{\myz^2}{4}\right)\Gamma\left(\frac{2}{3}\right)
\simeq c_0 \myz^{c_1} e^{-c_2 \myz}  \, , \nonumber
\end{align}
as demonstrated in the figure (dashed black curve), instead of the crude approximation \eqref{eq:syn_j}.
Here, ${_p}F_q(a;b;z)$ is the generalized hypergeometric function, and the approximation in the last equality of Eq.~\eqref{eq:FSynApprox}, accurate locally within $10\%$ for $c_0=1.83$, $c_1=0.309$, and $c_2=1.03$ ({\UK}), is easier to integrate, while producing results virtually indistinguishable from the exact spectrum.

Figure \ref{fig:alphaCore2D} shows the distribution of the spectral index $\alpha$ under the same assumptions of Fig.~\ref{fig:alphaCoreJ}, but for an arbitrary $r_c$, in the two-dimensional phase space spanned by $r/r_c$ and $\mathfrak{r}_c^2\propto r_c^2(\nu/b)^{1/2}$.
Here, the very soft spectrum in the transition from hard emission at low frequencies to the cooled, $\alpha\to-p/2$ at high-frequencies takes the form of a narrow valley in the $\alpha$ distribution, emerging for $r/r_c\gtrsim 1$ and showing increasingly negative $\alpha$ for larger $r/r_c$.
Qualitatively similar spectral distributions are obtained quite generally when CRE injection is sufficiently concentrated near the centre.
In particular, spectral softening can be very strong in regions where $\dot{N}_+$ diminishes and $N$ is dominated by cooled CREs diffusing from the centre.
These properties are illustrated also in Fig.~\ref{fig:ComaFit} (right panel), showing the CRE distribution as in Fig.~\ref{fig:alphaCore2D}, but for  radially-exponential injection, in which case softening is extreme and extends to arbitrary large radii for any $\mathfrak{r}$.

The above Green-function analysis can be readily generalised for arbitrary spatial distributions of CRE injection and magnetic field, producing hard spectra in maximal-injection regions and soft spectra in low $\dot{N}_+$ regions dominated by incoming cooled CREs.
However, we have assumed that $\psi\propto 1+b^2$ is approximately constant, \ie Compton cooling dominating over synchrotron ($b\ll1$) or a sufficiently uniform magnetic field ($b\simeq \const$), and we did not take into account possible spatial variations in the diffusion function or in the CRI spectral index.
In general, regions of weaker diffusion or a strong ($b\gtrsim 1$) magnetic field would have a shorter characteristic lengthscale $l$, so locally, the spectrum would become closer to its $\alpha\simeq -p/2$ steady state.
A full analysis, incorporating the strong synchrotron cooling in cores, spatial variations in $\psi$ and $D$, and additional effects, generally requires a numerical treatment.

The volume-integrated spectrum depends on the particular volume $V$ in question, as well as on the spacetime distributions of CRE injection, diffusion, and magnetic fields.
These distributions affect the integrated spectrum even in the simplest case where $V$ spans the entire system; in general, CRE diffusion across steady or growing magnetic substructure on scales not much smaller than the CRE Larmor radius tends to soften the spectrum (\UK).
In a steady-state, and even if microscopic substructure can be neglected, any misalignment between CRE injection and magnetic field on diffusive scales generally softens the spectrum, as CREs preferentially synchrotron-radiate after they have already experienced some Compton-cooling.
For the simple, spherical, steady-state, uniformly magnetized systems discussed above, Fig.~\ref{fig:alphaCore2D} illustrates how the choice of $V$ can strongly modify the integrated spectrum.
Here, the spectrum obtained by integrating from the centre out to a given radius is typically a broken power law, with $\alpha(\nu\ll\nu_{br})=-p/2+1/2$ and $\alpha(\nu\gg\nu_{br})=-p/2$, where radiative cooling is balanced at the break frequency $\nu_{br}$ by diffusion out of $V$ or the age of the system.
However, emphasizing the soft regions by effects such as a central mask, strong magnetic fields at large radii, or a particular choice of $V$, would all soften the integrated spectrum, as would a clumpy CRE injection, magnetic substructure, and various contaminating sources in projection, such as the CREs advected or diffused from a nearby relic.
Such spectral softening is typically stronger at high-frequencies, leading to a steepening, concave spectrum.

\subsection{Comparison with observations}
\label{subsec:GH_MH_Obs}

Next, we compare the above results with observed radio MHs and GHs, focusing on the spatio-spectral properties of the radio emission, but also addressing other properties such as the relations to other (X-ray, SZ) signals.
For simplicity, here we avoid microscopic and time-dependent processes, which were outlined in \S\ref{sec:Overview} and discussed in part in {\UK}, although they can significantly modify the spectrum, typically softening it.

When possible, we consider the general spatio-spectral properties inferred from multiple radio systems, rather than focus on individual sources, which can be complicated by rare dynamics, projection effects, and systematic errors; hence, the discussion is not fully inclusive.
Even general trends are susceptible to substantial uncertainties in spectral measurements, associated with a low signal-to-noise, incomplete $uv$-coverage and deconvolution, calibration errors, flux-scale uncertainties, assumptions on map-noise properties, errors induced by contaminant flux subtraction, blurring, source blending, and, at high frequencies, small fields of view \citep[\eg \VW,][]{RiseleyEtAl22}.
The inferred spectra can thus depend on resolution, may show spuriously soft spectra due to undetected low-surface brightness regions, and can be somewhat confused with radio sources such as radio galaxies and AGN lobes especially at low frequencies.
Consequently, different studies have contradicted each other, some reported soft spectra may be unrealistic \citep[\eg][]{RiseleyEtAl22}, and systematic errors should be considered as lower limits on the true uncertainty (\eg \VW).

\subsubsection{Cooling--diffusion scale}
\label{subsubsec:CoolingDiffusionScale}

When diffusion is strong, GHs and possibly MHs may be sufficiently extended to resolve the cooling--diffusion scale
\begin{equation} \label{eq:lEstimate}
l\simeq 280 \left(\frac{b}{\nu_{9}}\right)^{\frac{1}{4}}\! \left[ \frac{D_{32}}{(1+b^2)(1+z)^3}\right]^{\frac{1}{2}}\! \left[\frac{0.003\,\nu_{9}}{(1+z)^2b} \right]^{\frac{d}{4}} \kpc \coma\!
\end{equation}
derived from Eqs.~\eqref{eq:lDef0}, \eqref{eq:DiffEOfd}, \eqref{eq:nu_s}, and \eqref{eq:nu0Def},
so spectral variations are to be expected in the plane of the sky and along the line of sight.
Regions where the injection rate $\dot{N}_+$ is maximal could then show radio emission as hard as the limit \eqref{eq:MaximallyHard}, whereas sufficiently strong spatial gradients of $\dot{N}_+$ could lead to arbitrarily soft regions.
As the spectral index should vary over distances of order $l$, one can use Eq.~\eqref{eq:lEstimate} to directly extract crude estimates of $D$ and $b$ from observations.

\begin{figure}
\begin{center}
\includegraphics[width=1\linewidth]{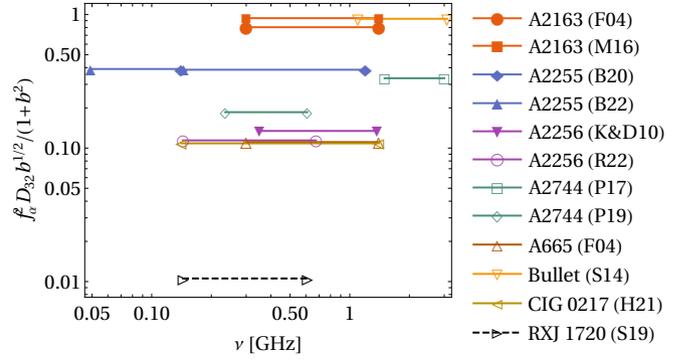}
\end{center}
\caption{\label{fig:DEst}
Diffusion coefficient weighted by magnetic field (ordinate), 
estimated by approximating $l_\alpha\simeq f_\alpha l$, where $f_\alpha$ is a factor of order unity, $l$ is the diffusion--cooling scale of Eq.~\eqref{eq:lEstimate}, and $l_\alpha$ is the characteristic scale of spatial variations in the projected spectral index $\alpha_{\nu_1}^{\nu_2}$, measured (with a factor $\sim2$ uncertainty; see Appendix \S\ref{app:PowerSpectrum}) between frequencies $\nu_1$ and $\nu_2$ (abscissa for symbol pairs) in different GHs (solid lines) and in one MH (dashed line), according to published spectral maps; see text and legend.
Literature references: B20 \citep{BotteonEtAl20}, B22 \citep{BotteonEtAl22}, F04 \citep{FerettiEtAl04a}, H21 \citep[][for ClG 0217+70]{HoangEtAl21}, KD10 \citep{KaleDwarakanath10}, M16 \citep{MhlahloEtAl16}, P17 \citep{PearceEtAl17}, P19 \citep{PaulEtAl19}, R22 \citep{RajpurohitEtAl22B}, S14 \citep{ShimwellEtAl14}, and S19 \citep[][for RXJ1720.1 +2638]{SaviniEtAl19}.
}
\end{figure}

In particular, we use published spectral maps to extract the scale $l_\alpha$ over which the projected spectral index varies, and approximate $l_\alpha\simeq f_\alpha l$ in order to obtain model-independent constraints on $D$ and $b$, as shown in Fig.~\ref{fig:DEst}.
Here, $f_\alpha$ is a dimensionless correction factor of order unity, which depends on details of the spatial distribution and projection.
The results for GHs are consistent with $D\simeq 10^{32}\cm^2\se^{-1}$, within a factor of a few, assuming that $b\lesssim1$ on large scales;
the one MH (in RXJ 1720.1+2638) included in the figure shows a substantially smaller $D$ or larger $b\gg1$.
Our crude estimate of $l_\alpha$, as the main scale found in the circularly averaged periodogram of the $\alpha$ map (in all cases, $l_\alpha$ is found to be larger than the beam size and smaller than the halo size), carries an uncertainty factor of order $\sim2$; our method is outlined in Appendix \S\ref{app:PowerSpectrum}.
This procedure excludes possible small-scale variability in $\alpha$, associated not with diffusion but rather with the filamentary magnetic field in relics and young GHs, as demonstrated by recent observations of A2256 \citep{RajpurohitEtAl22A, RajpurohitEtAl22B}.
In A2744, significant structure in the $\alpha$ distribution was previously reported on $30''$ scales but not on $15''$ scales \citep{PearceEtAl17}, consistent with our $l_\alpha$ which corresponds to $\sim20''$.

\subsubsection{Minihalos}
\label{subsubsec:MHs}

The radial brightness profiles $I_\nu(r)$ of observed MHs were fitted as a combination of a strong central Gaussian and a flatter outer profile that is either exponential or a power-law \citep[\eg][]{MurgiaEtAl09}.
The central Gaussian is probably attributed to unresolved emission from central sources \citep[in particular AGN radio bubbles, which are often masked; \eg][]{IgnestiEtAl20}.
Indeed, secondary CREs would produce an approximately Gaussian $I_\nu(r)$ near the centre of the cluster only for very centrally peaked ($\myi\gtrsim 3$) injection, which is implausible given the shallower $n\propto r^{-1}$ gas profile in the core and the strong CRI diffusion.
It is difficult to distinguish between exponential vs. power-law profiles for the MH itself, after the centre was excluded, as it typically spans only a factor of a few in brightness and a factor of two or so in radius; the same problem applies even to GHs, as discussed in \S\ref{subsubsec:GHs}.
In the presence of strong diffusion, the radial profile should become flatter at lower frequencies, where radiating CREs have a longer time to diffuse outward before cooling.
Such lower-frequency flattening was observed in Ophiuchus \citep{MurgiaEtAl10}, where the $e$-fold lengthscale increases from $\sim100\kpc$ at $1.5\GHz$ to $\gtrsim 400\kpc$ at $\nu\leq240\MHz$.
If these scales are comparable to $l$, one infers $D\sim 10^{32}\cm^2\se^{-1}$ for $b\simeq 1$ and $d=0$, and larger $D$ values for both smaller and larger $b$, as well as for a larger $d$.

MHs extend only over the relatively compact cluster core, of radius comparable to $l$ (within a factor of a few, typically), are strongly mixed by spiral flows or sloshing,
and typically terminate at tangential discontinuities that are highly magnetized and could restrict the escape of CRs.
One thus expects a mixed and confined secondary CRE distribution, radiating a uniform spectrum of index $\alpha\simeq -1$ reflecting the CRI spectrum $p\simeq 2$ (up to small corrections; \KL), except near the MH edge, where cooled CREs escaping the centre could soften the spectrum.
Such a uniform $\alpha\simeq -1$ spectrum, with a hint of softening at the edges, is indeed observed in both resolved MH spectral maps available, in MS 1455.0+2232 \citep{RiseleyEtAl22} and in RX J1720.1+2638 \citep{SaviniEtAl19}, where spectral variations are only $\Delta\alpha_{0.14\ghz}^{0.61\ghz}\sim 0.1$ \citep{BiavaEtAl21}.
Here and below, $\alpha_{\nu_1}^{\nu_2}$ denotes the spectral index fitted between frequencies $\nu_1$ and $\nu_2$.

The integrated spectrum of MHs is typically flat, $\alpha\simeq1$, as expected.
Recent examples include
$\alpha_{0.61\ghz}^{1.15\ghz}=-1.0\pm0.2$ in A2667 \citep{GiacintucciEtAl19Expanding},
$\alpha_{1\ghz}^{12\ghz}=-0.95\pm0.10$ in the Phoenix Cluster \citep{TimmermanEtAl21},
$\alpha_{0.14\ghz}^{0.61\ghz}=-0.93\pm0.10$ in RX J1720.1+2638 \citep{BiavaEtAl21},
and
$\alpha_{0.15\ghz}^{1.28\ghz}=-0.97\pm0.05$ in MS 1455.0+2232 \citep{RiseleyEtAl22}.
There are reports of MHs with softer spectra;
$\alpha_{0.33\ghz}^{1.4\ghz}=-1.02\pm0.10$ steepening to $\alpha_{1.4\ghz}^{4.9\ghz}=-1.41\pm0.13$ in RX J1532.9+3021,
$\alpha_{0.33\ghz}^{1.4\ghz}=-1.21\pm0.05$ in Perseus \citep[][and references therein]{GiacintucciEtAl14New},
$\alpha_{0.14\ghz}^{1.3\ghz}=-1.33\pm0.08$ in PSZ1 G139.61+24.20 \citep{GiacintucciEtAl19Expanding},
and
$\alpha_{0.15\ghz}^{0.24\ghz}=-1.4\pm0.3$ steepening to $\alpha_{0.24\ghz}^{1.48\ghz}=-1.60\pm0.05$ in Ophiuchus \citep{MurgiaEtAl10}.
Such softer spectra, if genuine and truly associated with the MH, can arise if the MH is larger than $l$ and the centre is masked, or if the magnetic field is strengthening,
or if there is magnetic substructure on either microscopic or diffusive scales.
Note that in the presence of strong diffusion, peripheral emission from CREs injected near the masked centre, by the MH or contaminants, would always soften the spectrum to some degree.

The strong magnetic fields typically inferred in MHs \citep[\eg {\KL};][\VW, and references therein]{IgnestiEtAl20} imply subdominant Compton losses and a possible quenching of diffusion especially near the centre, in which case the radio emissivity should trace the local product of CRI and gas densities, $j\propto \dot{N}_+\propto N_i n$.
If CRIs are strongly coupled to the gas, $N_i\propto n$, in such highly magnetized regions, then a linear, $I_\nu\propto F_x$ relation would emerge near the centre; in the periphery, where $b$ declines, $l$ increases, and the radio emission eventually cuts off, the relation could become highly superlinear.
Indeed, one typically finds $I_\nu\propto F_X^{0.8\text{--}1.3}$ in MHs, although rare superlinear relations as strong as $I_\nu\propto F_X^{2}$ were reported \citep{IgnestiEtAl20, BiavaEtAl21, RiseleyEtAl22}.
RXC J1504.1-0248, which is the extreme case presenting an $I_\nu\propto F_X^{2}$ relation, is unusually X-ray bright and shows evidence for particularly strong mixing \citep{IgnestiEtAl20}, consistent with its steep radio--X-ray relation as pointed out in \S\ref{subsec:Relations}.
Note, however, that radio--X-ray correlation studies typically mask or crudely subtract the central emission, and are severely limited by sensitivity and resolution, showing a significant scatter and large uncertainties in the correlation index \citep[\eg][]{RiseleyEtAl22}; moreover, some correlation scatter-plots show pronounced substructure.

\subsubsection{Giant halos}
\label{subsubsec:GHs}

GHs typically present in disturbed clusters, perturbed by merger events that are thought to shock, displace, and magnetize the plasma.
In such an unrelaxed ICM, the CRI distribution is likely modified by the merging component, the magnetic field evolution, and the particular dynamics, driving the distribution towards homogeneity.
Hence, the projected $\dot{N}_+$ distribution may be offset from the X-rays, and in early stages should be irregular and not fully correlated with the magnetic field.

In our model, the disturbed MH in the relaxed cluster core grows into a GH as the core is disrupted and the plasma becomes magnetized with $b\gtrsim 1$ out to several $\sim100\kpc$ radii.
At this early stage, the young GH, which may still present a weak shock at its edge (like the Bullet cluster) or bridge to a shock in a nearby relic (like in A521), would be spectrally soft due to a combination of a few factors:
(i) the elevated magnetic field would cause the existing, cooled CRE population to radiate strongly (\UK), contributing an $\alpha\simeq -1.5$ signal;
(ii) a similar effect is associated with CREs injected by the AGN and other central sources, dispersed by the merger;
(iii) misalignment between CRE injection and magnetization on diffusive scales would soften the spectrum, as discussed in \S\ref{subsec:HaloGeneral};
(iv) a clumpy gas distribution and the advection associated with the merger flows would contribute to this misalignment;
(v) the shock may sweep CRIs into a shell propagating outward, with CREs escaping the shell arriving in the GH already cooled (see \S\ref{sec:Relics});
and
(vi) microscopic magnetic substructure could further soften the spectrum (\UK).

As the GH matures, the CRIs, gas, and magnetic fields relax into more regular distributions, and the integrated spectrum gradually tends to the relaxed $\alpha\simeq -1$.
Gradients in CRE injection and in magnetic field persist, so the local spectrum of the diffusing and cooling CREs is not a pure power-law, resulting is some spectral variability on $\sim l$ lengthscales.
An elevated CRE injection near the centre leads to a general outward-softening trend, with the centre becoming as hard as the $\alpha \simeq -(p-1+d)/2$ of Eq.~\eqref{eq:MaximallyHard} and the periphery becoming arbitrarily soft for sufficiently large $\dot{N}_+$ gradients.
In strongly magnetized, $b\gtrsim1$ regions, spanning the centre and possibly much of the GH, CREs lose most of their energy to synchrotron radiation, so $I_\nu$ directly traces the diffusing CREs and gauges the primary CRI distribution.
In the GH outskirts, $b$ becomes small and modifies the $I_\nu$ profile, rendering it difficult to reconstruct the CRI distribution uniquely.

The radial brightness profile of a GH can usually be approximated as exponential, $I_\nu\propto  e^{-r/r_e}$, with $e$-fold scales $r_e$ of order a few 100 kpc.
Such fits are usually, like in MHs, not unique, as the brightness spans only a factor of a few.
For instance, \citet{MurgiaEtAl09} obtained equally good fits for an exponential $I(r)$ and for $I\propto n^\xi$ with $1.0<\xi<1.2$.
Note that a CRE injection profile equivalent to the latter density power-law 
emerges in a hadronic model with an approximately uniform CRI distribution (\UK).
Nevertheless, there are cases \citep[\eg][]{CucitiEtAl22} where an extended GH fits an exponential $I_\nu(r)$ profile reasonably well over an order of magnitude or so in brightness, corresponding to an approximately exponential emissivity profile $j_\nu(r)$.
Such observed $I_\nu(r)$ profiles suggest an exponential $\dot{N}_+(r)$ profile, or, more likely, emerge in $b\lesssim1$ regions with an approximately exponential radial decline in magnetic field.

\begin{bfigure*}
\begin{center}
\hspace{-0.2cm}\includegraphics[width=0.4\linewidth]{\figeps{ComaJ4}}\hspace{0.5cm}
\includegraphics[width=0.385\linewidth]{\figeps{ComaSpec4}}\\
\hspace{0.1cm} \includegraphics[width=0.39\linewidth]{\figeps{Comaalpha4}}\hspace{0.3cm}
\includegraphics[width=0.41\linewidth]{\figeps{Coma2D3}}
\end{center}
\caption{\label{fig:ComaFit}
Radial distribution of brightness (at $144\MHz$; top-left panel) and spectral index (between $144$ and $342\MHz$; bottom-left panel) from the NE (green diamonds) and SE (red circles) sectors, and the fully integrated spectrum (top-right panel), of the Coma cluster \citep[data from][brightness fits based on projected exponentials shown by dotted green and red curves]{BonafedeEtAl22}.
Result are shown for hadronic models 1 (dot-dashed purple curves), 2 (dashed black), and 3 (solid blue); see Table \ref{tab:ComaParams}.
The 2D spectral distribution of an exponential $\dot{N}_+$ model is shown (bottom-right panel, same notations as Fig.~\ref{fig:alphaCore2D}) with a trajectory (dashed curve) corresponding to model 2.
}
\end{bfigure*}

The volume-integrated spectrum of most GHs is flat, consistent with a pure, $-1.2\lesssim\alpha\lesssim -1.0$ power-law, as expected in a simple, steady-state hadronic model with smooth and steady magnetization.
Some GHs show a significantly softer spectrum, associated in well-analysed cases with rapid ICM evolution indicated for example by very nearby relics, as expected in a hadronic model with substantial changes in CRI and gas distributions, magnetic growth and bulk flows, as discussed in {\UK} and above.

Examples of soft spectra in transient GHs include
$\alpha_{153\mhz}^{1.4\ghz}=-1.81\pm0.02$ \citep[][well-fitted by a power-law only if the $608\MHz$ measurement is excluded]{MacarioEtAl13} in the bridged GH--relic system in the highly disturbed A521,
$\alpha_{144\mhz}^{1.5\ghz}=-1.63\pm0.03$ \citep{RajpurohitEtAl22B} in the very under-luminous and highly irregular GH of A2256, adjacent to a powerful relic, and
$\alpha_{1.1\ghz}^{3.1\ghz}= -1.50\pm0.04$ \citep{ShimwellEtAl14} in the over-luminous GH of the Bullet cluster, adjacent to a powerful relic and still bounded by a shock.
Some GHs detected only at low frequencies suggest similar soft spectra \citep{DiGennaroEtAl21}; if substantiated, they may reflect a population of evolving GHs.
In some strongly evolving systems, with a nearby shock, volume integration yields non-power-law and even convex spectra \citep[\ie high-frequency hardening; see \eg figures 4 and 7, respectively, of][]{KaleDwarakanath10, ShimwellEtAl14}.
In more mildly evolving systems, concave deviations from a power-law (\ie high-frequency softening) were reported: in A2744, MACS J0717.5+3745, and Coma \citep[][and references therein]{PearceEtAl17, RajpurohitEtAl21, BonafedeEtAl22}.
As emphasized above, spectral measurements are difficult, especially when integrated; some claims for very soft or steepening spectra were later challenged \citep[\eg in Abell S1063; see][]{XieEtAl20, RahamanEtAl21}.

The spatial distribution of the projected spectral index varies among GHs.
Most spectral images show a harder spectrum near, albeit usually not precisely at, the cluster centre, gradually softening outwards by $|\Delta\alpha|\simeq 0.5$ or even more at the periphery.
In particular, an $\alpha\simeq -0.8$ spectrum or slightly harder can be found near the centre, softening to a peripheral $\alpha\simeq -1.3$ spectrum or somewhat softer, as indicated for example in Coma, A2744, A2219, A2163, and A665 \citep{GiovanniniEtAl93, OrruEtAl07, FerettiEtAl04a, MhlahloEtAl16, CucitiEtAl22}.
More disturbed clusters can show an overall softer spectrum, but with a similar outward softening trend \citep[A520, MACS J0717.5+3745; see][]{VaccaEtAl14, RajpurohitEtAl21}.
Interestingly, in cases where emission can be detected beyond the soft halo periphery, the spectrum is observed to harden back towards $\alpha\simeq 1$ --- as expected in a hadronic model --- sometimes, but not always \citep[\eg in the south edge of the 'toothbrush' cluster;][]{RajpurohitEtAl18}, within radio bridges leading to a relic.
Some GHs show a very uniform $\alpha$, for example in 1RXS J0603.3+4214 (containing the 'toothbrush' relic), where $\alpha_{150\mhz}^{1.5\ghz}\simeq 1.16$ with only $\Delta\alpha\simeq 0.04$ variations; however, a peripheral softening may still be seen \citep[to the east of this halo;][]{vanWeerenEtAl16}.
Only rarely, do GHs show an inverted trend, with peripheries harder than their centre, as in CIG0217 \citep{HoangEtAl21} and A2256 \citep{RajpurohitEtAl22B}, both indicating a very recent merger.

Overall, these observations are consistent with the hadronic model, especially if one takes into account the strong diffusion.
As a demonstration, Fig.~\ref{fig:ComaFit} shows the GH of the Coma cluster \citep[using data from][]{BonafedeEtAl22}.
The figure shows the radial profiles of the $144\MHz$ brightness and of the $\alpha_{144\mhz}^{342\mhz}$ spectral index in the northeast and southeast sectors \citep[to avoid the $r\simeq 1\Mpc$ discontinuity observed to the west, see][]{BonafedeEtAl22}, as well as the integrated spectrum.
The spectral index varies with radius around $\alpha\sim-1$, but softens suddenly outside $r\simeq 1\Mpc$ to $\alpha<-2$ values, suggesting that the $r\simeq 1\Mpc$ discontinuity to the west may have an eastern, more subtle counterpart.
The integrated spectrum, based on the $\sim 600\kpc$ diameter integration of \citet{Thierbach03} with recent additions from LOFAR and WSRT, suggests some softening at high frequencies, argued to be partly but not entirely \citep{BrunettiEtAl13} due to SZ.
The figure demonstrates three simple hadronic models for Coma, with parameters provided in Table \ref{tab:ComaParams}.

\begin{table*}
\caption{Simple hadronic models for the GH in Coma.}
\centering
\begin{tabular}{cccc}
Model & Diffusion coefficient & CRE injection & Magnetic field \\
\hline
1 & $D=\begin{cases} 10^{32}\cm^2\se^{-1} & \text{if $r_M<1$ ;} \\ 10^{30.5}\cm^2\se^{-1} & \text{if $r_M>1$ .} \end{cases}$ & $\dot{N}_+(r_M<1)\propto E^{-2}e^{-r/r_\beta}$ & $B=6.7f_\beta^{0.7}\muG$\\
2 & $D=10^{32.3}\cm^2\se^{-1}$ & $\dot{N}_+(r)\propto E^{-2}e^{-r_M/0.06}$ & $B=\begin{cases} 0.6\muG & \text{if $r_M<1.1$ ;} \\ 0.1\muG & \text{if $r_M>1.1$ .} \end{cases}$ \\
3 & $D=10^{31.4}\cm^2\se^{-1}$ & $\dot{N}_+(0.3<r_M<1)\propto E^{-2}e^{-r/r_\beta}$ & $B=3.3e^{-\frac{r_M}{0.3}}\muG+0.3e^{-\left(\frac{r_M-1}{0.3}\right)^2}\Theta(1-r_M)\muG$\\
\hline
\end{tabular}
\label{tab:ComaParams}
\begin{tablenotes}
\item
Here, $\Theta(x)$ is the Heaviside step function. For brevity, we denote $r_M\equiv r/\text{Mpc}$.
As explained in the text, here we take $f_\beta(r)\simeq 1.1 e^{-r/r_\beta}$, $r_\beta\simeq 425\kpc$, and $D$ independent of $E$.
\end{tablenotes}
\end{table*}

Our nominal model 1 is based on the isothermal $\beta$-model fit to Coma.
Here, the electron density follows $n_e=n_{e,0}f_\beta(r)$, where $f_\beta\equiv (1+r^2/r_c^2)^{-3\beta/2}$, $n_{e0}=(3.5\pm0.7)\times 10^{-3}\cm^{-3}$, $r_c=343_{-20}^{+22}\kpc$, and $\beta=0.654_{-0.021}^{+0.019}$ \citep{FukazawaEtAl04, ChenEtAl07}.
This model assumes $D=10^{32}\cm^2\se^{-1}$ diffusion, a $B\propto 6.7f_\beta^{0.7}\muG$ magnetic field, and a homogeneous $p=2$ CRI distribution, such that CRE injection follows $\dot{N}_+\propto n\propto f_\beta$.
In the relevant radial range, this $\beta$-model can be approximately replaced (within $<10\%$ accuracy) by an equivalent exponential function, $f_\beta(r)\simeq 1.1 e^{-r/r_\beta}$, where $r_\beta\simeq 425\kpc$.
For simplicity, we thus take $\dot{N}_+\propto e^{-r/r_\beta}$, so the CRE distribution can be evaluated analytically (see Appendix \S\ref{app:GreenSolutions}; although projection along the line of sight is performed numerically).
In order to reproduce the reported sharp softening at large radii, here we assume that CR diffusion and CRE injection sharply drop beyond a discontinuity at $r\simeq 1\Mpc$ (see Table \ref{tab:ComaParams}).

This simple model matches the $144$--$342\MHz$ observations in Coma fairly well, as shown in the left panels of Fig.~\ref{fig:ComaFit}.
Using $E^2\dot{N}_+\simeq (1/2)cf_e \sigma_{i}n\, du_p/d\ln E_p$,
we infer a constant $du_p/d\ln E_p\simeq (5\pm1) \times 10^{-15}\erg\cm^{-3}$ logarithmic CRI energy, corresponding, for example, to $u_p(10\GeV<E_p<10^{10}\GeV)\simeq (1.0\pm0.2)\times 10^{-13}\erg\cm^{-3}$.
Such values are consistent with the CRI energy densities needed to explain MHs, relics, and other GHs (see {\UK} and \S\ref{subsec:Motivation}), as well as the \gama-ray excess reported in Coma.
In this (somewhat high-$B$) model, the \gama-ray flux emanating from the central $r<1\dgr$ is $F(>200\MeV)\simeq 4\times 10^{-14}\erg\se^{-1}\cm^{-2}$, fainter by a factor $\gtrsim 10$ than observed (see \S\ref{subsec:GammaRays}).

In the second model, we avoid any sharp cutoff on $\dot{N}_+$ and $D$, assuming instead that CRE injection is more compact, given by an exponent with a $60\kpc$ $e$-fold radius.
Here, the brightness profile fits the data even with a constant and modest, $B=0.6\muG$ field, and the spectrum noticeably softens with increasing radius, from $\alpha_{144\mhz}^{342\mhz}\simeq -0.65$ near the centre to $\sim -1.6$ around $1\Mpc$.
To obtain a more substantial softening beyond $r\simeq 1.1\Mpc$, we assume that the magnetic field drops to $B=0.1\muG$ outside the discontinuity.
As this model invokes a purely exponential CRE injection, we also present in the figure (bottom-right panel) the corresponding spectral index distribution in the 2D, dimensionless $r$--$\nu$ space (analogous to Fig.~\ref{fig:alphaCore2D} and using the same notations), showing also the trajectory (dashed curve) traced by model 2.
Owing to its compact injection region and weak $B$, this model yields $F(>200\MeV)\simeq 2\times 10^{-11}\erg\se^{-1}\cm^{-2}$, brighter by a factor $\gtrsim 10$ than observed.

These two models are oversimplified, avoiding for example irregularities in CRE injection, magnetic field, and diffusion, which are among the possible explanations for the spectral softening with increasing $r$ or $\nu$.
For instance, misaligned variations in $\dot{N_+}$ and $B$ can soften the integrated spectrum at high frequencies, due to cooled CREs diffusing into magnetized regions dominating their radio emission.
This effect is demonstrated by the third model, which is similar to model 1, but incorporates a region of strong magnetization and weak CRE injection.
For simplicity, we keep the model radial, thus placing this region in the centre of the cluster.
The resulting softening of the integrated spectrum (upper right panel) could in reality be produced by a combination of such $\dot{N}_+$ and $B$ variations, an evolving magnetic field, and a filamentary magnetic structure of a small filling factor.
To better fit the brightness profile with this modified $\dot{N_+}$, model 3 adopts a lower $D$ and incorporates a local $\sim 1\muG$ magnetic enhancement near the discontinuity, followed by a $90\%$ drop in $B$.
In this model, $F(>200\MeV)\simeq 1\times 10^{-12}\erg\se^{-1}\cm^{-2}$, comparable to observations.

These three variants of the hadronic model demonstrate that it can easily reproduce the observations, including the spatio-spectral distributions, once CR diffusion is incorporated, with fewer free parameters than leptonic alternatives.
Although it is difficult to determine the model parameters uniquely for a disturbed, nonspherical system using limited data of substantial statistical and systematic errors, future studies utilizing additional data (X-rays for the gas, Faraday rotations for the magnetic field, etc.) and numerical simulations could disentangle the system components.

\section{Relics: time-dependent model}
\label{sec:Relics}

\subsection{Oversimplified setup: steady-state planar shock}
\label{subsec:SteadyStateRelics}

It is instructive to begin the study of relics with the effect of a weak shock on an ambient CRI or CRE distribution without sources or sinks, in a somewhat artificial shock-frame steady-state that can be handled analytically.
Let the shock lie at $x=0$, with plasma flowing in the positive $x$-direction with velocities $v=v_s$ upstream ($x<0$; subscript $u$) and $v=v_d\equiv v_s/r_g$ downstream ($x>0$; subscript $d$), where $r_g=4\mach^2/(3+\mach^2)$ is the gas compression factor, $\mach$ is the shock Mach number, and we assumed that the pressure is dominated by an adiabatic index $\Gamma=5/3$ plasma.
Other assumptions of \S\ref{sec:Halos} are presumed here too, including an underlying CRI distribution of spectral index $p$, an isotropic diffusion function $D$, and a rate $\psi$ combination of Compton and synchrotron CRE cooling.

Omitting its injection and cooling terms, the transport equation \eqref{eq:DiffLoss0} may be written in a 1D steady-state as
\begin{equation} \label{eq:DiffLoss2}
\pr_x (\myn v) - \pr_x \left( D\pr_x\myn \right) + \frac{\phi-1}{3} N \pr_x v \simeq 0\fin
\end{equation}
Here, $\myn$ pertains to either CRIs or CREs, with a local spectral index $\phi(x)$ which affects the dynamics only near the shock, where $\pr_x v\neq 0$ as the gas decelerates.
Denoting the far upstream CR density by $N_u(E)\equiv N(E,x\to-\infty)$, integrating Eq.~\eqref{eq:DiffLoss2} gives
\begin{equation} \label{eq:int_CR_PDE}
v N -D\pr_x N  - \frac{(\myrg-1)(\phi-1)}{3\myrg} v_s N_{sh} \Theta(x) \simeq v_s N_u \coma \,\,
\end{equation}
where $\Theta(x)$ is the Heaviside step function: zero upstream and unity downstream.
As there is no energy scale in this problem, the spectrum is everywhere a power-law, $N(E)\propto E^{-\phi}$.
Furthermore, the only bound solution of Eq.~\eqref{eq:int_CR_PDE} downstream is uniform, so $N(E,x\geq0)=N_{sh}(E)$.

One can analyse primary particle acceleration by neglecting particles far upstream, $N_u\to 0$, in which case Eq.~\eqref{eq:int_CR_PDE} yields the standard, $\phi=(r_g+2)/(r_g-1)=2(\mach^2+1)/(\mach^2-1)$ DSA spectrum \citep{Krymskii77}, but then the $N_{sh}$ normalization, \ie the acceleration efficiency, cannot be determined.
In the present case, where upstream CRs are guaranteed, we must retain the $N_u$ term, the solution to Eq.~\eqref{eq:int_CR_PDE} becomes
\begin{equation}\label{eq:CRCompression}
  \myn=\begin{cases}
  N_u + (N_{sh}-N_u)e^{v_s x/D} & \text{if $x < 0$ ;} \\ N_{sh} & \text{if $x > 0$ ,}
  \end{cases}
\end{equation}
and continuity implies a {\CR} compression factor (\UK)
\begin{equation}
\label{eq:CR_Compression1}
\myrcr \!\equiv\! \frac{N_{sh}}{N_u} \!=\! \frac{3\myrg}{3-(\myrg-1)(\phi-1)} \!=\! \frac{4\mach^2}{4-(\mach^2-1)(\phi-2)} \,.\!\!
\end{equation}
In the absence of injection or cooling, the persistent population of $\phi\geq 3$ CREs upstream are compressed in a $\mach^2<5$ shock by a factor $\myrcr \geq 4\mach^2/(5-\mach^2)$; for shocks with $\mach^2\geq5$, which accelerate particles with a spectral index $\phi\leq 3$, the compression of upstream $\phi\geq3$ CREs diverges.
For a flat, $\phi=2$ CRI distribution, $\myrcr =\mach^2$, so secondary CRE injection is amplified at the shock by a factor \begin{equation}\label{eq:InjectionAmplification}
  r_{\text{inj}}=r_g\myrcr = \frac{4\mach^4}{3+\mach^2} \fin
\end{equation}

\begin{figure}
\begin{center}
\includegraphics[width=0.95\linewidth]{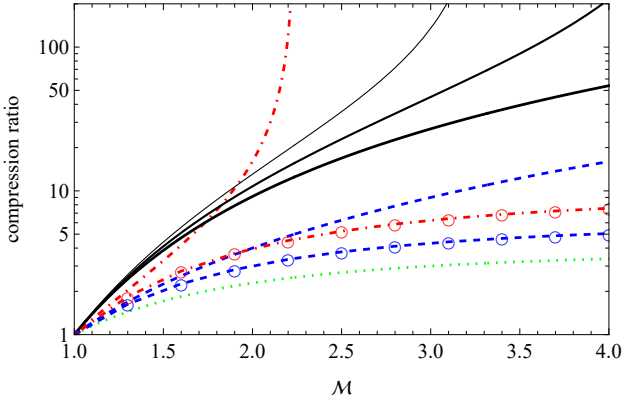}
\end{center}
\caption{\label{fig:Compressions}
Compression factors of $p=2$ CRIs (dashed blue), $\phi=3$ CREs (dash-dotted red), gas (dotted green), and CRE injection (black, thick to thin for CRIs with $p=2.0$, $2.2$, $2.4$) in a Mach $\mach$ shock.
For the CRs, we show both the adiabatic compression associated only with gas and field compression (curves with circles) and the overall compression assuming no cooling (curves without symbols, unrealistic for CREs when diffusion is strong).
}
\end{figure}

Figure \ref{fig:Compressions} illustrates the compression factors of CRIs, CREs, the gas, and CRE injection, as a function of $\mach$.
As the figure shows, secondary CRE injection (black solid curves) due to a $\phi=2$ CRI distribution is significantly enhanced at a weak shock, by a factor $r_{\text{inj}}\sim10$ even for $\mach=2$, increasing with $\mach$ and approaching $r_{\text{inj}}\simeq 4\mach^2$ for strong shocks.
Thus, the compression of CRIs and gas behind weak relic shocks compensates for the lower ambient density, such that even modest magnetic amplification suffices for the relic emissivity to reach or exceed the emissivity of its GH counterpart (\UK).
CRI compression is a strong function of the spectral index $p$, as demonstrated in the figure, so weak shocks preferentially amplify regions where the CRI spectrum is somewhat softer than $p=2.0$.
Note that the CR compression \eqref{eq:CR_Compression1} relies on both second and third terms of Eq.~\eqref{eq:DiffLoss2}, namely, on some form of magnetically-regulated diffusion upstream and on the slowdown and associated compression of CRIs near the shock, mediated by shock-induced electromagnetic fields.

\subsection{Quenched CRE compression/(re)acceleration}
\label{subsec:QuenchedElectrons}

Some cooled CRE population of spectral index $\phi\gtrsim3$ is inevitable upstream, due to inelastic CRI collisions, individual sources such as galaxies, and even slowly-cooling CRE remnants from the virial shock.
While shock-compression of such CREs would be substantial in the absence of cooling, formally diverging as $\mach^2\to5$ for $\phi=3$ (and at smaller $\mach$ for $\phi>3$), compression is quenched when diffusion is strong.
To wit, CR compression and acceleration operate on a timescale $t_{\text{acc}}\simeq D/v_s^2$, over which the fluid crosses the shock precursor given in Eq.~\eqref{eq:CRCompression}.
For strong diffusion, this timescale is not much shorter than the cooling time $t_{\rm cool}$ of radio-emitting CREs, as indicated by the ratio $t_{\text{cool}}/{t_{\text{acc}}}$ in Eq.~\eqref{eq:AccelerationTimeRatio} not greatly exceeding unity for typical ICM conditions.

While strong magnetization by the shock could in principle diminish $D$ and lead to faster compression and acceleration, as inferred for strong shocks, there is no evidence for such strong magnetization upstream of weak shocks.
Hence, for $\mach_2\lesssim 1$ relic and GH shocks, one cannot safely assume that upstream CREs are compressed beyond their adiabatic compression associated with gas and field compression,
\begin{equation}\label{eq:CRE_Compression}
  r_{\text{cre}} \simeq r_g^{(2+\phi)/3} \simeq \left( \frac{4\mach^2}{3+\mach^2}\right)^{5/3} \, ,
\end{equation}
shown (as a red dot-dashed curve with circles) in Fig.~\ref{fig:Compressions}.
The same reasoning applies to electron (re)acceleration in weak shocks and turbulence, which is similarly slowed down by strong diffusion till it becomes quenched by cooling.

Consequently, for strong diffusion, the radio signature of a weak shock is dominated by secondary CREs, injected by inelastic collisions between compressed CRIs and compressed nuclei.
The distribution of these CREs can be derived, as in \S\ref{subsec:HaloGeneral}, using Green functions; see Appendix \S\ref{app:GreenSolutionsPlanar}.
Consider the simple case of a steady-state shock ($v_d=\const$) with energy-independent diffusion ($d=0$) and a uniform ($C=\const$) rate of spectrally-flat ($p=2$) CREs injected downstream.
Integrating the downstream-frame evolution of CREs injected at all earlier times then yield the steady-state distribution
\begin{equation}\label{eq:RelicSteady}
N(E,x_{s}+\Delta x) \propto E^{-2} \int_{0}^{\frac{1}{\psi E}} d\Delta t \int_{v_d \Delta t}^\infty \frac{e^{-\frac{(\Delta x-x_0)^2}{4D \Delta t}}}{\sqrt{4\pi D \Delta t}}dx_0  \coma
\end{equation}
which is readily evaluated analytically.

The corresponding synchrotron brightness rises sharply as one crosses the shock downstream, where it remains fairly constant if $B$ is either strong ($b\gtrsim1$) or uniform (see Fig.~\ref{fig:alphaAnalyticPlanar}).
The spectrum at the front of the shock is hard, approaching $\alpha=-0.5$ at low frequencies regardless of the shock Mach number.
The spectrum gradually softens to its $\alpha=-1$ steady-state farther downstream and at higher frequencies.
Spacetime variations in the magnetic field on sub-diffusive scales can somewhat soften the spectrum downstream (\UK).
However, the above simple setup, introduced in \S\ref{subsec:SteadyStateRelics}, does not naturally reproduce the observed strong downstream decline in brightness, accompanied by substantial spectral softening, as typically observed in relics and demonstrated in Fig.~\ref{fig:Saussage}.

\subsection{CRI shell}
\label{subsec:TimeDependentRelics}

An evolving, weak shock, propagating outward through the merging cluster remnants, may carry with it a shell of elevated CRI density, instead of leaving a precisely constant $u_p$ downstream.
The decline in $u_p$ downstream of the shock, if substantial, would suffice to reproduce the dimming and softening of radio emission observed behind relics.

Indeed, the over-simplified setup in \S\ref{subsec:SteadyStateRelics} with its homogeneous CRIs downstream was based on the poorly-justified assumption of a planar steady-state in a homogeneous medium with sufficiently fast CRI shock-compression.
In practice, CRIs and their secondary CREs can focus near the shock front due to a combination of several factors:
(i) compression by an outward push of gas driving the shock;
(ii) the $t_{\text{acc}}\sim D/v_s^2\simeq 0.3D_{32}M_2^{-2}T_{\text{keV}}\Gyr$ acceleration/compression time being too long for the downstream density to saturate;
(iii) the lateral size of the shock grows as it propagates outwards, with a constant or increasing solid angle with respect to the cluster's centre;
and
(iv) the shock strengthens as it propagates outwards in the cluster periphery, as gas density declines faster than $r^{-2}$;
(v) subsequent adiabatic expansion of the downstream gas may further accentuate the CRI localization in the shell;
(vi) the magnetic field increases with Mach number, suggesting a somewhat smaller $D$ and stronger CR confinement near the shock.

\begin{figure}
\begin{center}
\includegraphics[width=0.9\linewidth]{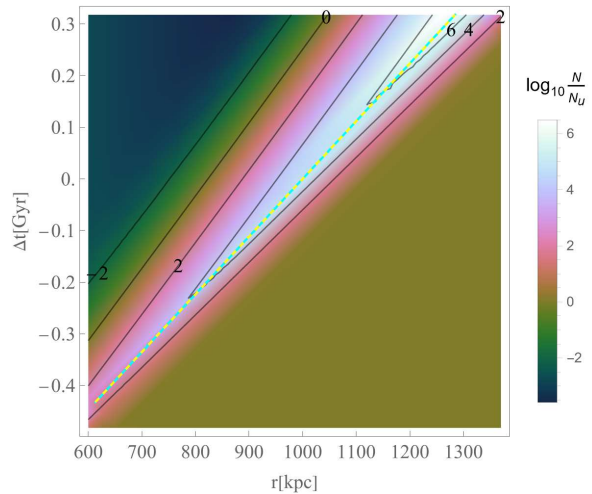}
\end{center}
\caption{\label{fig:Mach2CRIs}
Spacetime distribution of the amplification factor $N/N_u$ (logarithmic contours and colour bar) of an initially homogenous density of $p=2$ CRIs, induced by a Mach $\mach=2$ shock (dashed curve) propagating outward into a stationary, isothermal ICM with a $c_s=10^8\cm\se^{-1}$ speed of sound, assuming $D=10^{30.7}\cm^2\se^{-1}$ diffusion.
}
\end{figure}

Figure \ref{fig:Mach2CRIs} demonstrates the potential of CRI focusing near the shock by incorporating only the first three of these six factors.
Here, a radial shock is assumed to propagate outward (dashed curve), into an isothermal upstream plasma that is at rest in the frame of the cluster, with a constant $\mach=2$ Mach number and a fixed solid angle, reaching $r=1\Mpc$ at time $\Delta t=0$.
The transport equation \eqref{eq:DiffLoss0} is solved numerically for an initially homogeneous distribution of $p=2$ CRIs.
As the figure illustrates, the CRI density declines rapidly downstream, by a factor $>10^3$ within $100\kpc$ for the typical relic parameters used (an upstream $c_s=10^8\cm\se^{-1}$ speed of sound and $D=10^{30.7}\cm^2\se^{-1}$ diffusion).
The amplification factor $N/N_u$ (logarithmic contours and colours in the depicted spacetime) within the shell grows rapidly and exceeds the estimates \eqref{eq:CR_Compression1} and \eqref{eq:InjectionAmplification}, but the effect should saturate as the CRI energy fraction becomes large; a more careful analysis would be needed to determine the outcome.
The three other effects listed above may strengthen the CRI localization.

\begin{figure*}
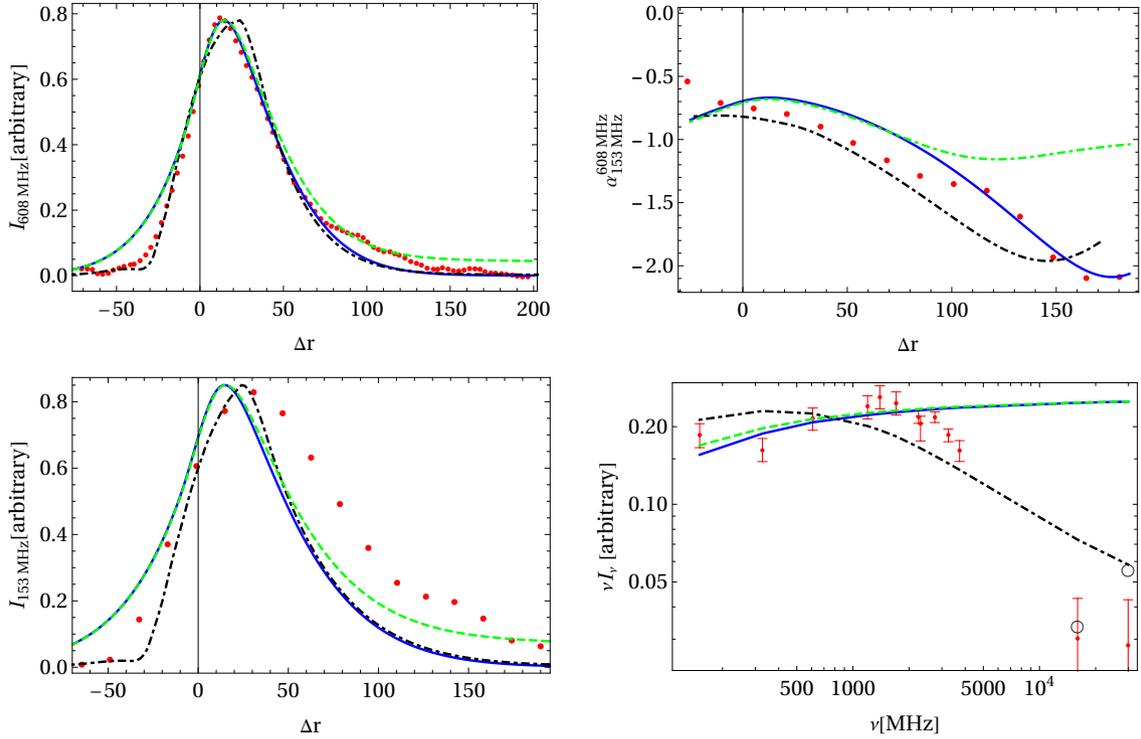

\begin{center}
\includegraphics[width=0.4\linewidth]{\figeps{SaussageV2bI608}}\hspace{0.5cm}
\includegraphics[width=0.4\linewidth]{\figeps{SaussageV2bAlpha1}}\\
\vspace{0.3cm}
\includegraphics[width=0.4\linewidth]{\figeps{SaussageV2bI153}}\hspace{0.5cm}
\includegraphics[width=0.4\linewidth]{\figeps{SaussageV2bSpec1}}
\end{center}
\caption{\label{fig:Saussage}
Projected $608\MHz$ (top left panel) and $153\MHz$ (bottom left) brightness profiles of the 'sausage' relic in CIZA J2242.8+5301, the implied spectral index profile (top right), and the integrated spectrum (bottom right), as observed (symbols; \VW, adopting the same arbitrary brightness normalizations and $10^\circ$ half-opening angle) and modelled (curves) by non-accelerated secondaries injected by shock-compressed CRIs, with uniform $B=1\muG$, $D=10^{30.7}\cm^2\se^{-1}$, and $p=2$.
CRIs are assumed compressed and focused near the shock, with a peak factor $\mathcal{C}=N_{\text{max}}/N_d=10^2$ (dashed green) or $\mathcal{C}=10^4$ (solid blue), with a $\Delta r=12\kpc$ exponential scale and a $u=100\km\se^{-1}$ downstream drift.
For the $\mathcal{C}=10^4$ case, we also show results (dot-dashed black) for a magnetic field that increases linearly from $0.1\muG$ at $r>r_s+35\kpc$ to $2\muG$ at $r<r_s-25\kpc$.
The offset between the observed profile and the shock position is chosen arbitrarily, as there is only a poorly-localized hint of a density discontinuity in X-rays \citep{OgreanEtAl2014}.
Our model neglects the compression of CREs diffusing back upstream (so the upstream brightness is somewhat overestimated), the (unspecified) beam convolution (which could somewhat smooth the profile at lower frequencies), and any softening due to magnetic irregularities and evolution (see \UK).
The integrated spectrum (error bars; from \VW, based on \citet{StroeEtAl16} data) is shown also after correcting for SZ \citep[empty circles; from][]{BasuEtAl16}; it should be noted that a more recent analysis \citep{LoiEtAl20} found a pure $\alpha=-1.12\pm0.03$ power-law integrated spectrum, with no evidence for steepening at least till $18.6\GHz$.
}
\end{figure*}

CRI compression in the vicinity of the shock, combined with the coincident, modest compression of the gas, results in an even stronger local enhancement in secondary CRE injection.
For simplicity, we approximate the downstream decline in CRE injection as exponential, so $\dot{N}_+ \propto 1+\mathcal{C}e^{(r-r_s)/\Delta r}$.
Here, $r_s$ is the shock radius, $\mathcal{C}=N_{\text{max}}/N_d$ is the maximal local enhancement factor of CRE injection with respect to the downstream (larger than $N_{\text{max}}/N_u$, depicted in Fig.~\ref{fig:Mach2CRIs}), and $\Delta r$ is a characteristic shell width.
To further simplify the computation, we also adopt a one-zone approximation, in which the rest frame of these CREs moves downstream, away from the shock, at an effective uniform velocity $u$.
Under these assumptions, one may analytically solve the transport equation \eqref{eq:DiffLoss0} for the CRE evolution.

The resulting, projected solution is demonstrated in Fig.~\ref{fig:Saussage}, where the spectro-spatial properties of the model are also compared to those of the best-studied, 'sausage' relic in the cluster CIZA J2242.8+5301.
The brightness (left panels) rises sharply as one approaches the shock from the upstream, as in the steady-state solution (\cf Fig.~\ref{fig:alphaAnalyticPlanar}).
Here, however, as the CRIs are focused near the shock, the brightness rapidly declines downstream, even if the magnetic field remains uniform (solid blue and dashed green curves).
For a large $\mathcal{C}$ (solid blue curves),
the downstream is dominated by cooled CREs diffusing from the shock, inducing a strong spectral softening (top-right panel).

For a uniform $B$, the volume-integrated spectrum (bottom-right panel) is approximately a power-law with the steady-state index $\alpha\simeq -1$.
However, as high-energy CREs are strongly localized near the CRI shell, $B$ variations that are misaligned with this shell can significantly soften the spectrum at high frequencies, as illustrated in the figure (dot-dashed black curves for the case of $B$ strengthening around shock crossing).
It should be noted, however, that claims for spectral softening in the integrated spectra of observed relics have been largely disputed, as discussed below.

\subsection{Comparison with observations}
\label{subsec:RelicObs}

The spectrum integrated over a relic is typically a power-law of index $\alpha\simeq -1$, spanning in some cases more than three orders of magnitude in frequency.
Reports of substantial deviations from this value or of high frequency steepening were later corrected.
Recent examples include
$\alpha=-1.01\pm0.05$ in ClG 0217+70 \citep{HoangEtAl21},
$\alpha=-1.07\pm0.02$ in A2256 \citep{RajpurohitEtAl22A}, 
$\alpha=-1.08\pm0.05$ and $\alpha=-1.13\pm0.05$ in A1240 \citep{HoangEtAl18},
$\alpha=-1.12\pm0.03$ in the 'sausage' relic in CIZA J2242.8+5301 \citep{LoiEtAl20}, 
$\alpha=-1.12\pm0.03$ in the 'toothbrush' relic in 1RXS J0603.3+4214 \citep{RajpurohitEtAl20}, %
$\alpha=-1.17\pm0.03$ in A2744 \citep{RajpurohitEtAl21_A2744},
and
$\alpha=-1.20\pm0.18$ and $-1.31\pm0.14$ in ZwCl 2341.1+0000 \citep{BensonEtAl17}. 

Integrated spectra so close to $\alpha=-1$ are a robust signature of cooled CRs associated with a very strong shock.
Finding such emission downstream of weak ICM shocks is natural in a hadronic model, but implausible in a leptonic model, where the spectrum should be softer.
Moreover, the very small dispersion of spectral indices among all well-analysed relics is itself natural in a hadronic model, but disagrees with the large dispersion anticipated in leptonic models due to variations in shock Mach numbers (for primary shock acceleration) and environmental properties (for other electron acceleration or reacceleration variants); see {\UK}.

Along the shock front, relics typically show a very uniform spectrum \citep[\eg][]{RajpurohitEtAl18, DiGennaroEtAl18}.
Again, such a spectral uniformity is consistent with a hadronic model, but not with leptonic models, in which inevitable variations in Mach numbers (which are not large) or upstream parameters along the shock front should induce substantial changes in local spectral index.

Perpendicular to the shock front, relic spectra typically evolve from as hard as $\alpha\simeq -0.5$ towards the upstream, to as soft as $\alpha\lesssim-2$ at distances of order a few $100\kpc$ downstream.
Figure~\ref{fig:Saussage} illustrates this behaviour, and shows that it can be easily modeled in a hadronic model, even using only a simple one-zone, analytic model.
Beyond the bright region showing this hard-to-soft downstream trend, some relics show evidence for the spectrum evolving back towards the steady-state $\alpha\simeq-1$, as expected in a hadronic model, both upstream \citep[\eg][]{vanWeerenEtAl16, deGasperinEtAl22} and far downstream \citep[\eg][]{BonafedeEtAl12, vanWeerenEtAl16, HoangEtAl18, deGasperinEtAl22}.

\begin{figure}
\begin{center}
\includegraphics[width=1\linewidth]{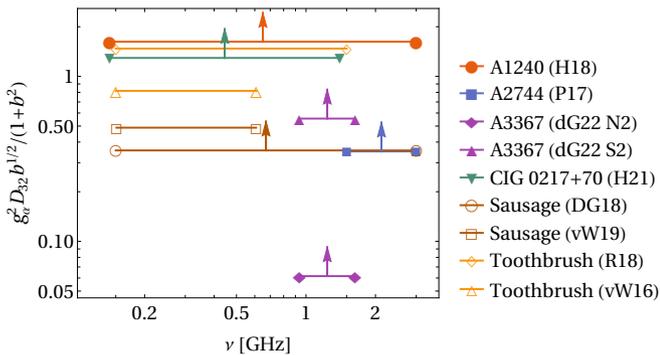}
\end{center}
\caption{\label{fig:DEstRelics}
Diffusion coefficient weighted by magnetic field (ordinate) in relics, 
estimated by approximating $l_\alpha\simeq g_\alpha l$, where $g_\alpha$ is a correction factor of order unity, $l$ is the diffusion--cooling scale, and $l_\alpha$ is the distance between hardest and softest spectrum measured (with $\lesssim 50\%$ systematic uncertainties due mainly to relic orientation) between frequencies $\nu_1$ and $\nu_2$ (abscissa for symbol pairs) according to all useable published spectral profiles; see text and legend.
Arrows designate the (typical) cases where only a lower limit on $l_\alpha$ can be imposed, due to ongoing spectral softening downstream.
Literature references: dG22 \citep[][sectors S2 and N2]{deGasperinEtAl22}, DG18 \citep{DiGennaroEtAl18}, H19 \citep{HoangEtAl18}, H21 \citep[][for ClG 0217+70]{HoangEtAl21}, P17 \citep{PearceEtAl17}, R18 \citep{RajpurohitEtAl18}, and vW16 \citep{vanWeerenEtAl16}.
}
\end{figure}

Within the typical $\Delta r\sim100\Delta r_2\kpc$ extent of spectral softening downstream of a relic, strong diffusion of the CREs dominates over their advection where the respective timescale ratio
$t_{\text{diff}}/t_{\text{adv}}\simeq 0.3D_{32}^{-1} \Delta r_2 (u/10^8\km\se^{-1})$ is small.
Advection further slows if the CREs are partly trapped in the CRI shell moving with the shock, for the same reasons outlined above.
Therefore, as in halos, $l$ again provides a relevant lengthscale for spectral variations, and so can be used to crudely estimate $D$.
Approximating the distance $l_\alpha$ over which the spectrum softens as $g_\alpha l$, where $g_\alpha$ is a dimensionless correction factor of order unity, we typically find $D\simeq 10^{32}g_\alpha^{-2}b^{-1/2}(1+b^2)\cm^2\se^{-1}$, varying by a factor of a few among relics, as shown in Fig.~\ref{fig:DEstRelics}.
These results are comparable to those inferred in Fig.~\ref{fig:DEst} from spatial variations in GH spectra, and with the {\UK} estimate of the diffusion needed to flatten the CRI distribution out to relic radii.

Relics often show a filamentary structure (\VW, and references therein), with filaments too narrow to be attributed to CRE cooling (as sometimes invoked for X-ray filaments in SNR shocks).
Hence, the filamentary structure is usually attributed either to a filamentary magnetic structure or to a corrugated shock surface seen in projection.
The complex structure and hard spectrum of the filaments in A3667 \citep[][]{deGasperinEtAl22} favours a filamentary magnetic field confined near the shock front.

Finally, we caution that the limitations outlined in \S\ref{subsubsec:MHs} in regards to radio observations and modeling apply equally for relics, as pertaining to systematic uncertainties, our focus on general relic properties without accounting for all individual and possibly confused special cases, and the simplifying assumptions made above. In particular, synchrotron cooling was neglected in our relic analysis with respect to Compton cooling, in order to facilitate analytic results; additional effects are associated with highly magnetized relics (see \UK).

\section{Mega/Super halos}
\label{sec:SH}

Recently, highly extended, low-frequency ($44$--$144$ MHz) emission was reported beyond GHs in four very massive clusters \citep{CucitiEtAl22}, in what we refer to as SHs.
These clusters show a clear transition in the radial brightness profile around $0.6\Mpc\lesssim r\lesssim 0.8\Mpc$, from the rapid exponential decline within the GH on smaller scales to a much flatter profile on larger scales, eventually steepening again and becoming undetectable beyond an outer radius $1.2\Mpc\lesssim r\lesssim 1.6\Mpc$.
The spectrum of these SHs is soft, $\alpha_{50\mhz}^{144\mhz}\simeq -1.6$, reflecting partly cooled CREs.
While the GHs are fairly symmetric about the centres of their clusters, all four SHs show substantial deviations from projected circular symmetry, extending towards one side of the cluster.

\subsection{SHs from local peripheral magnetization}

The resolved radial decline in brightness at the outer $(200\text{--}400)\kpc$ of these SHs follows approximately the same exponential decline as in their GH counterparts, namely, with a similar $e$-fold scale $r_e$.
This effect is most pronounced in A665, where the GH and SH each span about an order of magnitude in brightness \citep[][figure 3c]{CucitiEtAl22}.
This similarity suggests that the peripheral drop in SH brightness is dominated by an exponential radial decline in the mean magnetic field, as invoked for some GHs (see model 3 in Table \ref{tab:ComaParams}).
Combining both GHs and SHs, the brightness spans $\sim$two orders of magnitude, corresponding to $\sim$one order of magnitude in $B$.

The flattening of the $I_\nu(r)$ profile, observed in SHs asymmetrically on one side of the cluster, then suggests that the magnetic field is locally enhanced.
We model a cluster harbouring both a GH and a SH by generalizing a simple GH model of core injection \eqref{eq:InjectionCore}, focusing on the most extended case in ZwCl 0634.1+4750 \citep[at $z=0.17$;][]{CucitiEtAl22}.
Here, we adopt $\myi=3$ (equivalently, $\beta=1$), $r_c=400\kpc$, a flat $p=2$ injection spectrum, the nominal $D=10^{32}\cm^2\se^{-1}$, and an exponentially declining magnetic field, $B_0 e^{-r/r_B}$.
To incorporate the SH, we add a local peripheral magnetization, $B_1 e^{-(r-r_1)^2/r_2^2}$, centred on some large radius $r_1$ with an extent $r_2$.
The model is under-constrained, so we fix for simplicity $B_1=1\muG$ ($b_1\simeq0.2$), which is both plausible for a peripheral magnetized region and suffices to explain the SH.

\begin{figure}
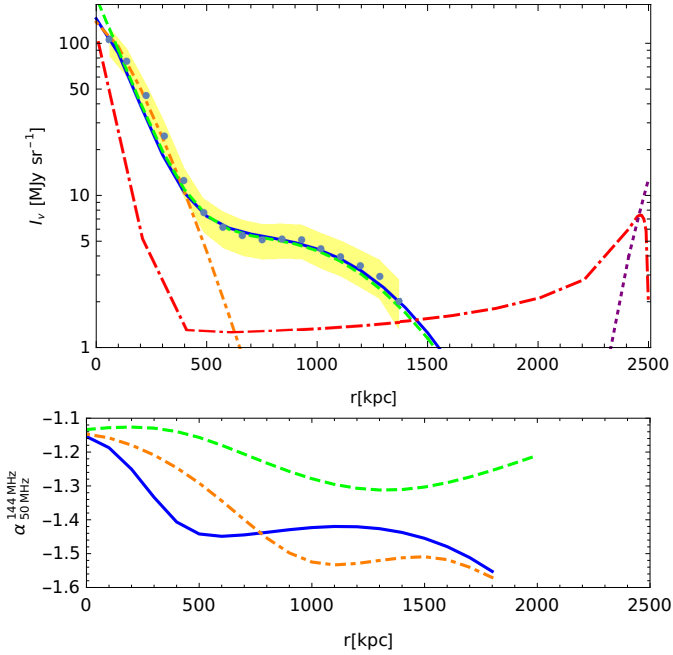

\begin{center}
\includegraphics[width=0.99\linewidth]{\figeps{GH_SH_Fits3}}\\
\hspace{-0.2cm}\includegraphics[width=1.02\linewidth]{\figeps{GH_SH_Fits_alpha3}}
\end{center}
\caption{\label{fig:GH_SH}
Top panel: Radial profile of $144\MHz$ brightness in ZwCl 0634.1+4750 \citep[at $z=0.17$; disks with shaded region for $1\sigma$ uncertainty limits from][]{CucitiEtAl22}, fitted in the hadronic model in the plane of the sky (dashed green, with negligible virial-shock primary CRE contribution in dotted purple) and projected along the line of sight (assuming spherical symmetry) for the full system (solid blue) and for the GH alone (dash-dotted orange, with a possible central contribution from primary CREs at low frequencies in double-dash-dotted red).
Bottom panel: radial profiles of the spectral index $\alpha_{50\mhz}^{144\mhz}$ in the hadronic models.
}
\end{figure}

This oversimplified model already accounts for the observations fairly well, as demonstrated in Fig.~\ref{fig:GH_SH}, both if we assume spherical symmetry and integrate over the line of sight (in which case $B_0\simeq 7\muG$, $r_B\simeq 160\kpc$, $r_1\simeq 1.3\Mpc$, and $r_2\simeq 0.65\Mpc$)
or if we assume that $B$ is enhanced only in the plane of the sky (in which case $B_0\simeq 2.3\muG$, $r_B\simeq 200\kpc$, $r_1\simeq 1.2\Mpc$, and $r_2\simeq 0.7\Mpc$).
For our choice of $\myi=3$, the projected model gives a spectrum $-1.4\lesssim \alpha_{50\mhz}^{144\mhz}\lesssim-1.15$ in the GH and $-1.6\lesssim \alpha_{50\mhz}^{144\mhz}\lesssim-1.4$ in the SH, broadly consistent with observations.
A larger (smaller) $\myi$ produces softer (harder) spectra; additional effects that tend to soften the spectrum of transient systems (see \S\ref{sec:Halos}) are not incorporated.
For simplicity, we neglect synchrotron cooling, so emission in the very centre is slightly overestimated.

\subsection{Contribution of primary, virial-shock CREs}

The brightness of three of the four SHs is of order $I_{\text{SH}}\simeq 8\MJy\sr^{-1}$, and higher by a factor of a few in the SH of A665 \citep[in which only $\sim$ half the cluster area was analysed in][]{CucitiEtAl22}.
Interestingly, these values are similar to the average $I_{\text{VR}}\simeq 7\MJy\sr^{-1}$ low-frequency brightness excess found at the virial radius of stacked clusters \citep{HouEtAl22}, interpreted as synchrotron emission from virial shock-accelerated CREs gyrating in magnetic fields of a few $0.1\muG$.
As primary CREs accelerated in the virial shock propagate inward, through advection and diffusion, $\gtrsim 50\MHz$ emission declines over a few $100\kpc$ scales due to Compton cooling \citep{HouEtAl22} and possibly also the dissipation of the shock-induced magnetic fields.
At smaller radii of order $R_{500}$, as the magnetic field strengthens due to ICM dynamics (compression, shocks, turbulence), lower-energy CREs that only partly cooled can start radiating at $\gtrsim 50\MHz$ frequencies, contributing to both GHs and SHs.

However, these primary CREs are unlikely to generate the observed SHs.
First, for $D\lesssim 10^{32}\cm^2\se^{-1}$ diffusion, unrealistically strong peripheral magnetic fields would be needed to raise the CRE frequency and emissivity sufficiently.
For plausible fields, very strong, $D> 10^{33}\cm^2\se^{-1}$ diffusion would be needed for CREs to reach the SH edges before cooling too much; advection is far too slow for this purpose.
Moreover, if virial-shock CREs would account for SHs, then they would dominate over their secondary counterparts also throughout the GH, endowing it with an implausibly soft spectrum.

Nevertheless, primary virial-shock CREs could enhance the low-frequency emission both at larger, $\gtrsim2\Mpc$ radii, and in the very central, highly-magnetized regions, as illustrated in Fig.~\ref{fig:GH_SH}.
Here, we used the same magnetic fields fitted in the hadronic models, with primary CRE injection at the virial shock constrained by low-frequency stacking \citep{HouEtAl22}.
For $D\simeq 10^{32}\cm^2\se^{-1}$, the effect on the central $\sim100\kpc$ may be noticeable, softening the very low frequency spectrum, but it is unclear if such strong diffusion can be sustained throughout the cluster.

\section{Summary and discussion}
\label{sec:Summary}

We have shown that attributing the different types of diffuse ICM radio sources -- MHs, GHs, relics, and, importantly, their intermediate hybrid states -- to secondary CREs produced by the same, cluster-wide population of CRIs, as argued in {\UK}, provides a more natural, self-consistent, and successful model than the blend of leptonic alternatives dominating the recent literature.
This simpler model not only accounts for essentially all present observations, with fewer free parameters and without invoking poorly constrained (re)acceleration mechanisms in weak shocks or turbulence, but also makes new and in part verified predictions (concerning the radio spectral properties, the \gama-ray counterparts of halos, correlations with other signals, the significance of particle acceleration in virial shocks) and facilitates new measurements (of the extended CRI distribution and spectrum, the strong diffusion coefficient, a more direct handle on magnetization).
Section \ref{sec:Overview} reviews the joint hadronic model, summarizes the evidence accumulated in its favour, demonstrates it using a simple homogeneous CRI distribution, and shows the hadronic model's success in predicting the \gama-ray counterpart of the Coma GH.

We have explored in some detail the spatio-spectral properties of halos (in \S\ref{sec:Halos}) and relics (in \S\ref{sec:Relics}), mainly using Green function and numerical solutions to the transport equation \eqref{eq:DiffLoss0}, and tested them against all observational data presently available.
The data show good agreement with the model in terms of brightness distribution, spectral distribution, integrated spectra, and correlations with additional signals, in both halos (\S\ref{subsec:GH_MH_Obs}) and relics (\S\ref{subsec:RelicObs}).
In particular, we focused on the Coma GH (Fig. \ref{fig:ComaFit}) and the sausage relic (Fig. \ref{fig:Saussage}) as archetypical examples of their classes, finding that even simplified analytic models nicely reproduce the observations.
For a sufficiently homogeneous CRI distribution, any magnetized region in the ICM becomes radio bright, with no need to invoke local primary CRE (re)acceleration.
Hence, a modest peripheral amplification of the magnetic field also accounts for the recently detected, soft, asymmetric extensions of some GHs to large radii, referred to as mega-halos or SHs (\S\ref{sec:SH}).

Resolved spectral maps facilitate a fairly direct measurement of CRE diffusion around $\sim100\GeV$ energies, both in GHs (Fig. \ref{fig:DEst}) and in relics (Fig. \ref{fig:DEstRelics}). The resulting, $D\simeq 10^{31\text{--}32}\cm^2\se^{-1}$ estimates are consistent with each other, as well as with the {\UK} estimate of the diffusion needed to facilitate a CRI distribution sufficiently homogeneous for the joint hadronic model.
Such a strong diffusion homogenizes CRIs across $\sim\Mpc$ scales over the few Gyr age of the cluster, and can marginally homogenize them across the entire cluster volume (especially if $D$ is enhanced in the periphery), while producing a rich radio spectral phenomenology in the hadronic model, allowing for spectra as hard as $\alpha\simeq -(p-1)/2$ in front of relics and $\alpha\simeq -(p+d-1)/2$ in the centres of GHs, along with arbitrarily soft spectra at GH peripheries and downstream of relics.
In addition, such strong diffusion quenches the compression of secondary CREs by weak shocks (which would otherwise over-produce the brightness of relics), and the (re)acceleration of electrons in weak shocks and turbulence in general (see \S\ref{sec:Relics}).
CRE diffusion and growing, irregular magnetic fields also explain why young GHs, characterized by nearby shocks and irregular gas and CRI distributions, show a soft integrated spectrum.

This large diffusion coefficient, inferred on large scales in the ICM, corresponds to a coherence length of order a kpc and is significantly higher than found in galaxies, but is consistent with the observed scaling of $D$ with system size.
Indeed, diffusion measurements from small to large scales include
$D\simeq 10^{23}\cm^2\se^{-1}$ around $10\GeV$ in the solar wind \citep{ChhiberEtAl17,KojimaEtAl18};
$D\simeq 10^{26.5}E_{10}^{0.2\text{--}0.6}\cm^2\se^{-1}$ in molecular clouds \citep{OhiraEtAl11};
$D\simeq 10^{28}E_{100}^{0.9}\cm^2\se^{-1}$ in the central few pc of the Milky Way \citep{ChernyakovaEtAl11};
$D\lesssim 2\times 10^{27} \cm^2\se^{-1}$ at $\sim100\GeV$ perpendicular to galactic disks \citep{DahlemEtAl95};
$D\simeq 10^{27}\text{--}10^{28}\cm^2\se^{-1}$ between $1\GeV$ and $1\TeV$ in starburst galaxies \citep{KrumholzEtAl20};
$D\simeq 10^{28.5}E_{100}^{0.5}\cm^2\se^{-1}$ in other galaxies \citep{HeesenEtAl16};
$D\simeq 10^{28.5}E_{100}^{0.5}\cm^2\se^{-1}$ for the Milky Way \citep[\eg][]{KrumholzEtAl20};
$D\simeq  10^{30}E_{100}^{0.48\pm0.02}\cm^2\se^{-1}$ near the edges of the $\sim10\kpc$ Fermi bubbles \citep{Keshetgurwich17};
$D\gtrsim 10^{30}\cm^2\se^{-1}$ around $100\GeV$ as a possible explanation for uniform, $\gtrsim20\kpc$ radio bubbles in the IGM \citep[\eg][]{MathewsGuo11};
and
$D\simeq  10^{30}\cm^2\se^{-1}$ in MHs, as inferred for an AGN-driven hadronic model \citep{IgnestiEtAl20} and from the spectral map of RX J1720.1+2638 (Fig. \ref{fig:DEst}). Moreover, a combination of somewhat weaker diffusion and additional mixing processes, such as spiral flows in MHs and merger dynamics in GHs, may also explain the data.

Future, more sensitive observations, in particular of faint halo-like or relic-like emission from any magnetized region in the ICM, better constraints on the $\pi^0\to\gamma\gamma$ and secondary inverse-Compton signals, and more sophisticated modeling, would further test and develop the hadronic model.
In the presence of a nearly uniform CRI population, a faint \gama-ray signal should eventually be observed in all clusters, extending out to the virial shock, and any strong ICM magnetization should involve some radio emission even with no coincident CRE sources.
Enhanced radio emission could thus be observed, for example, below the magnetized spiral CFs, and in draped magnetic fields around merger CFs, bubbles, and clumps moving through the ICM.
Radio--\gama-ray anti-correlations are expected, as secondaries lose more energy to synchrotron than to inverse-Compton radiation in highly magnetized regions; such an anti correlation \citep{KeshetEtAl12_Coma} would gauge the magnetic fields and further support the hadronic model.

A point-by-point summary of the accumulated evidence in favour of the joint hadronic model is provided in \S\ref{subsec:Evidence}.

\MyMNRAS{\section*{Acknowledgements}}
I thank D. Kushnir, K.C. Hou, G. Ilani, and D. Ariad for helpful discussions.
This research was supported by the Israel Science Foundation (Grant No. 2126/22), by the IAEC-UPBC joint research foundation (Grant No. 300/18), and by the Ministry of Science, Technology and Space, Israel.

\section*{Data Availability} The data generated from computations are reported in the paper, and any additional data will be made available upon reasonable request to the corresponding author.

\bibliography{\mybib{Clusters}}

\MyMNRAS{\label{lastpage}}

\clearpage
\renewcommand{\appendixname}{Supp.Mat.}
\appendix
\onecolumn

\setcounter{page}{1}
\begin{center}
{\Large \bf{Supporting Information}}
\end{center}
\vspace{-0.5cm}

\section{Relations between radio and other signals}
\label{subsec:Relations}

The joint hadronic model reproduces the observed correlations between radio and other tracers of the ICM, as we show here and in
\S\ref{subsec:GH_MH_Obs} and \S\ref{subsec:RelicObs}.
Local correlations of the radio brightness $I_\nu$ with other signals are particularly informative.
In highly magnetized regions in the cores of clusters, magnetized CRIs follow the gas distribution, $N_i\propto n$, whereas non-magnetized CRIs, in disrupted or weakly magnetized regions, can mix, diffuse, and approach homogeneity, $N_i\propto r^0$.
In $B\gtrsim B_{\text{cmb}}$ regions, Compton losses become subdominant so CREs radiate approximately all their energy within their diffusion radius.
In $N_i\propto n$ regions, the radio brightness $I_\nu$ should then correlate with $n^2$ tracers (X-rays, especially the brightness $F_X$ in the ROSAT band which weakly depends on temperature $T$), whereas in homogeneous CRI regions, $I_\nu$ should correlate better with $n^1$ tracers (SZ, weak lensing).
Indeed, there is evidence for such correlations in MHs, in the centres of GHs, in the peripheries of GHs, and even in radio relics.
Correlations away from GH centres are somewhat complicated by the merger dynamics, and a strong merger can even shift the CRIs with respect to the gas distribution, inducing an offset between X-ray and GH peaks.

Global correlations with the integrated radio power $P_\nu$ are also useful.
The X-ray luminosity (in the ROSAT $0.1\text{--}2.4\keV$ band, within $R_{500}$) shows a similar, $P_\nu\propto L_X^{1.4\text{--}1.6}$ relation in GHs and in relics, and a somewhat steeper, $P_\nu\propto L_X^{1.7\text{--}2.0}$ relation in MHs \citep{SommerBasu14, YuanEtAl15, GiacintucciEtAl19Expanding}.
The results carry substantial systematic uncertainties evident from the large dispersion among fitting variants; within the limited statistic and systematic uncertainties, the correlations are consistent among MHs, GHs and relics.
When the mass $M$ is estimated within $R_{500}$, it shows similar, $P_{\nu}\propto M^{3.4\text{--}4.0}$ relations in GHs \citep{Basu12}, MHs, and relics; GHs and MHs also share the same normalization (\citealt{YuanEtAl15}; but see \citealt{GiacintucciEtAl19Expanding}).
Note that these $P_\nu$--$L_X$ and $P_\nu$--$M$ correlations, jointly, are at tension with the standard, $L_X\propto M^{1.5\text{--}1.7}$ relation in galaxy clusters \citep{PrattEtAl09}.
This tension may be due in part to selection effects, in particular of higher mass clusters, biasing a shallower, $P_{1.4\ghz}\propto M^{2\text{--}3}$ relation \citep{SommerBasu14}.
Otherwise, this inconsistency would suggest that clusters harbouring halos follow an $L_X\propto M^{2.5}$ relation, significantly steeper than in other clusters.
Correlations between $P_\nu$ and the integrated Comptonization $Y_{SZ}$ inside $R_{500}$ are generally similar or slightly steeper than the $P_\nu$--$L_X$ correlation, consistent \citep[\eg][]{SommerBasu14}, within their large systematic and statistic errors, with the approximately linear, \eg $Y_{SZ}\propto L_X^{1.14\pm0.08}$ relation \citep{ArnaudEtAl10}.

MHs are found in the $n\propto r^{-1}$ gas density core, where the $B>B_{\text{cmb}}$ magnetic field is strong, so CREs radiate most of their energy in the radio with an emissivity nearly independent of $B$.
If the magnetic field in the core is sufficiently strong to couple CRIs to the gas and hinder diffusion, then the specific radio emissivity $j_\nu\propto N_i n$ and the ROSAT-band X-ray emissivity $j_X\propto n^2Z^{0.6}T^{-0.1}$ are approximately proportional to each other, as indeed found observationally (\KL) and usually verified by point-to-point correlations (see \S\ref{subsubsec:MHs}).
Here, $Z$ is the metallicity, changing slowly in the core \citep[$Z\propto r^{-0.3}$;][]{SandersonEtAl09}; the $j_X(Z)$ dependence, approximated here for $k_BT\simeq 1\keV$ \citep{KeshetGurwich18}, vanishes at higher, $k_BT\simeq 10\keV$ temperatures.
Mixing of the CRIs would steepen the linear $I_\nu\propto F_X$ relation, approaching $I_\nu\propto F_X^2$ in the limit of homogeneous CRIs.
Assuming that $N_i/n\propto T$, where $T$ is a characteristic temperature in the cluster, and that clusters are self-similar such that the core radius is linear in $R_{500}$, the integrated powers in radio and X-ray are then related by
\begin{equation}\label{eq:IntegratedCorrelationX}
  P_\nu\propto L_X \frac{N_i}{n}Z^{-0.6}T^{0.1}\propto L_X  T^{1.1-0.6\beta_Z} \propto L_X^{1.4\text{--}1.6}\coma
\end{equation}
consistent with observations.
Here, we adopted $L_X\propto T^{2.2\text{--}3.0}$ \citep{PrattEtAl09} and $Z\propto T^{\beta_Z}$ with $-0.35\lesssim \beta_Z\lesssim0$ \citep[more negative indices corresponding to more central regions, see][]{MantzEtAl17, TruongEtAl19}; the uncertainties in both of these scalings have a minor effect on the outcome \eqref{eq:IntegratedCorrelationX}.
The $N_i/n\propto T$ scaling pertains to CRIs accelerated by the virial-shock; alternative CRI sources such as SNe would generally yield different scalings.

In the core of a GH, $n$ is approximately constant, so one expects a fairly uniform, $N_i\propto r^0$ CRI distribution whether diffusion is strong or weak.
As GH emission is typically dominated by the core, like its X-ray counterpart, the same integrated $P_\nu$--$L_X$ relation \eqref{eq:IntegratedCorrelationX} applies to GHs if the core is magnetized, whereas clusters with weak, $b\ll 1$ core magnetization lie below this correlation \citep{KushnirEtAl09}.
Indeed, while GH cores are not as strongly magnetized as MHs, their central  field is still of order $B_{\text{cmb}}$ (\VW, and references therein), so $j_\nu$ depends only weakly on $B$ near the centre.
Therefore, locally, the radio-to-X-ray brightness ratio $\eta\equiv \nu I_\nu/F_X$ is approximately uniform in the centres of GHs, where it is also found to be comparable to its MH value \citep{Keshet2010}.

However, unlike the local correlations in the $n\propto r^{-1}$ core of MHs, a uniform $\eta$ does not imply a linear point-to-point radio--X-ray correlation in the $n\propto r^0$ cores of GHs.
Rather, for $N_i\propto n^\sigma$, $j_\nu\propto N_i n B^\zeta$, and an $n\propto(1+r^2/r_c^2)^{-3\beta/2}$ isothermal $\beta$-model distribution, integration along the line of sight yields
\begin{equation}\label{eq:GH_I_F_relation}
I_\nu\propto F_X^{1+\gamma\delta} \coma
\end{equation}
where $\gamma\equiv \sigma+\zeta-1$ and $\delta\equiv3\beta/(6\beta-1)$ approximately equals $2/3$ for $\beta\simeq 2/3$ (\UK).
Consider the case of homogeneous CRIs ($\sigma=0)$.
Here, Eq.~\eqref{eq:GH_I_F_relation} becomes $I_\nu\propto F_X^{1/3}$ for strong magnetization ($\zeta=0$), and $I_\nu\propto F_X^{2/3}$ for weak ($b\ll1$) equipartition ($\zeta=1/2$) fields.
Indeed, central GH regions often show the former relation \citep[\eg {\UK},][]{BonafedeEtAl22}, whereas more peripheral regions are consistent with the latter \citep[\eg Coma, A2163, A3562, A2256; see][]{GovoniEtAl01, FerettiEtAl01, GiacintucciEtAl05, BrownRudnick11, RajpurohitEtAl22B}.
In Coma, for example, $I_\nu\propto F_X^{0.41\pm0.04}$ in the inner halo and $I_\nu\propto F_X^{0.76\pm0.05}$ in the outer halo \citep{BonafedeEtAl22}.
These correlations are affected to some extent by diffusion or alternative mixing processes; CRI mixing lowers $\sigma$, whereas CRE mixing, mainly by diffusion, has the opposite effect of rendering the $I_\nu(F_X)$ relation steeper.
A few clusters suggest a linear, $I_\nu\propto F_X$ relation \citep[A2255, A2744, and in part of A3562; see][]{GovoniEtAl01, GiacintucciEtAl05}, which may arise from $b\gg1$ regions of CRIs still coupled to the gas ($\sigma=1$) or from very strong CRE mixing.

Under the above assumptions, a local linear tracer $Y$ of density or pressure, such as the $y$-parameter or the surface density inferred from weak lensing, is related to the GH radio brightness as
\begin{equation}\label{eq:GH_I_y_relation}
I_\nu\propto Y^{(1+\gamma \delta)/(1-\delta)} \fin
\end{equation}
For homogeneous CRIs ($\sigma=0$), strongly magnetized regions should thus show a linear, $I_\nu\propto Y$ relation, whereas $b\ll1$ regions would show a steeper, approximately $I_\nu\propto Y^{1+2\zeta}$ correlation.
Indeed, while $P_\nu$ is not linear in the $Y_{SZ}$ integrated inside $R_{500}$, limiting the integration of the $y$-parameter to the halo region yields a tighter, linear $P_\nu\propto Y_{SZ}$ correlation \citep{Basu12}.
Locally, in Coma, an approximately linear, $y\propto I_\nu^{0.92\pm0.04}$ point-to-point correlation was found by the \citet{PlanckComa13}, whereas a steeper, $I_\nu\propto y^{1.76\pm0.08}$ relation was found by \citet{BonafedeEtAl22}, who were able to better remove contaminations at large radii.
Furthermore, at small radii, the radio brightness in Coma correlates well with the surface mass density inferred from weak lensing \citep{BrownRudnick11}, better than it correlates with X-ray emission (S. Brown, private communications, 2012).

\section{Scale-free solution for spherical symmetry}
\label{app:PLSolutions}

The ODE \eqref{eq:ODE} is solved in the range $0<\myi<3$, under the boundary conditions $\mathfrak{n}(\myX\to \infty)=0$ and $\myX\mathfrak{n}(\myX\to 0)=0$, by the $\epsilon\to0$ limit of
\begin{equation}  \label{eq:SolODE}
\mathfrak{n}(\mathfrak{r};\epsilon)= \frac{
{_1}F_1\left(\frac{p-d}{1-d}-\frac{\myi}{2},\frac{1}{2},\frac{1-d}{4}\mathfrak{r}^2\right)\int_x^\infty H\,dr
+
\left[\int_\epsilon^\mathfrak{r} F\,dr -\frac{2^{2\frac{p-d}{1-d}-\myi}\Gamma\left(\frac{1+2p-\myi-3d+\myi d}{2-2d}\right)}{\sqrt{\pi}}\int_\epsilon^\infty H\,dr  \right]
H_{-\frac{2p-\myi-(2-\myi)d}{1-d}}\left(\frac{\sqrt{1-d}}{2}\mathfrak{r}\right)
}{\mathfrak{r} e^{\frac{1-d}{4}\mathfrak{r}^2}}\, ,
\end{equation}
where ${_1}F_1(a;b;c)$ is the Kummer confluent hypergeometric function, $H_n(x)$ is the Hermite polynomial, and we defined
\begin{equation}
\{H,F\} \equiv \frac
{
\frac{2r^{-\myi}\,e^{\frac{1-d}{4}r^2} }{2p-\myi-(2-\myi)d} \times
\left\{
H_{-\frac{2p-\myi-(2-\myi) d}{1-d}}\left(\frac{\sqrt{1-d}}{2}r\right) ,\,
{_1}F_1\left(\frac{p-d}{1-d}-\frac{\myi}{2},\frac{1}{2},\frac{1-d}{4}r^2\right)
\right\}
}
{
\frac{2}{r\sqrt{1-d}}
{_1}F_1\left(\frac{p-d}{1-d}-\frac{\myi}{2},\frac{1}{2},\frac{1-d}{4}r^2\right)
H_{-\frac{1+2p-\myi-(3-\myi) d}{1-d}}\left(\frac{\sqrt{1-d}}{2}r\right)
+
{_1}F_1\left[\frac{2+2p-\myi-(4-\myi)d}{2(1-d)},\frac{3}{2},\frac{1-d}{4}r^2\right]
H_{-\frac{2p-\myi-(2-\myi) d}{1-d}} \left(\frac{\sqrt{1-d}}{2}r\right)
} \, .
\end{equation}
This $\mathfrak{n}(\mathfrak{r};\epsilon\to0)$ limiting process is necessary in the regime $2\leq\myi<3$, where the integrals of $F$ and $H$ separately diverge as $r\to0$.
For $\myi>3$, the $\propto \int r^2 \dot{N}_+ dr$ number of injected particles diverges in the centre, and the one-dimensional steady-state $\mathfrak{n}$ becomes non-physical.
In the intermediate limit $\myi=3$, this divergence is only logarithmic and the solution follows
\begin{equation} \label{eq:Frho3}
\mathfrak{n}(\mathfrak{r};\myi=3) \propto
\mathfrak{r}^{-1} e^{-\frac{1-d}{4}\mathfrak{r}^2}H_{-\frac{2p+d-3}{1-d}}\left(\frac{\sqrt{1-d}}{2}\mathfrak{r}\right)\, .
\end{equation}

Equations \eqref{eq:SolODE}--\eqref{eq:Frho3} reduce to closed expressions for specific parameters.
For example, in the nominal, $p=2$ and $d=0$ case, we find
$\mathfrak{r}\mathfrak{n}=1-(8/\sqrt{\pi})e^{-\mathfrak{r}^2/4}H_{-3}(\mathfrak{r}/2)$ for $\myi=1$,
$\mathfrak{n}=(\sqrt{\pi}/4)G_{1\,1}^{2\,1}(\mathfrak{r}^2/4|_{0,0,-1/2}^{0,1})$ for $\myi=2$,
and $\mathfrak{n}=\mathfrak{r}^{-1}e^{-\mathfrak{r}^2/4}H_{-2}(\mathfrak{r}/2)$ for $\myi=3$,
where $G$ is the Meijer G-function.
The synchrotron spectral index derived from the scaling \eqref{eq:PLN} in the approximation \eqref{eq:alphaLocal},
\begin{equation}\label{eq:AlphaSelfSimilar}
  \alpha = -\frac{2p-(1-d)\left(\myi+\frac{d\ln \mathfrak{n}}{d\ln \mathfrak{r}}\right)}{4} \, ,
\end{equation}
is illustrated in Fig.~\ref{fig:alphaPLAnalytic} for the solution \eqref{eq:SolODE} and for the limiting case \eqref{eq:Frho3}, in which case $\alpha$ reduces to Eq.~\eqref{eq:LimitingAlpha}.

\section{Green-function solutions for spherical symmetry}
\label{app:GreenSolutions}

\subsection{Energy-independent diffusion}

For the simple case of energy-independent, $d=0$ diffusion, the diffusion length becomes $r_d=(D t)^{1/2}$.
For fixed cooling, the kernel in Eq.~\eqref{eq:N0} then becomes
\begin{equation}\label{eq:N0_d0}
N_0 =\frac{E^{-(p+1)}}{8\pi l^3 \psi} \left\{
\left(\frac{1}{\mathfrak{r}_0}+\frac{1}{\mathfrak{r}}\right)\Delta F\left(2-p,\frac{3}{2}\right)
-\frac{\Gamma(p-1)\Delta F\left(\frac{3}{2}-p,\frac{1}{2}\right)}{\mathfrak{r}_0\mathfrak{r}\,\Gamma\left(p-\frac{1}{2}\right)}
+ 2\text{min}\left(\frac{1}{\mathfrak{r}_0},\frac{1}{\mathfrak{r}}\right)\,{_1}F_1\left[2-p,\frac{3}{2},-\frac{(\mathfrak{r}-\mathfrak{r}_0)^2}{4}\right]
\right\}\, ,
\end{equation}
where we assumed $\dot{N}_+\propto E^{-p}$ injection and defined
\begin{equation}
\Delta F(a,b) \equiv {_1}F_1\left[a,b,-\frac{(\mathfrak{r}+\mathfrak{r}_0)^2}{4}\right] - {_1}F_1\left[a,b,-\frac{(\mathfrak{r}-\mathfrak{r}_0)^2}{4}\right] \, .
\end{equation}
Note that the spatially-integrated CRE distribution due to a single spherical shell reduces to the expected
\begin{equation}\label{eq:N_int_r}
\int 4\pi r^2 N_0(E,r;r_0)dr = \frac{E^{-(p+1)}}{(p-1)\psi} \, ,
\end{equation}
corresponding to an
\begin{equation}
\alpha = \frac{1}{2}\left(1+\frac{d\ln N}{d\ln E} \right) = -\frac{p}{2}
\end{equation}
spectral index of synchrotron emission, provided that the entire relevant region is included.
The same spectrum arises for the volume-integrated steady-state emission from any superposition of shells, and for any $D(E)$ diffusion function.
The integrated synchrotron index could differ from ($-p/2$) if part of the volume is excluded or if the underlying assumptions (in particular, a steady-state with a uniform $\psi$) are violated.

\begin{figure}
\centerline{
\epsfxsize=8cm \epsfbox{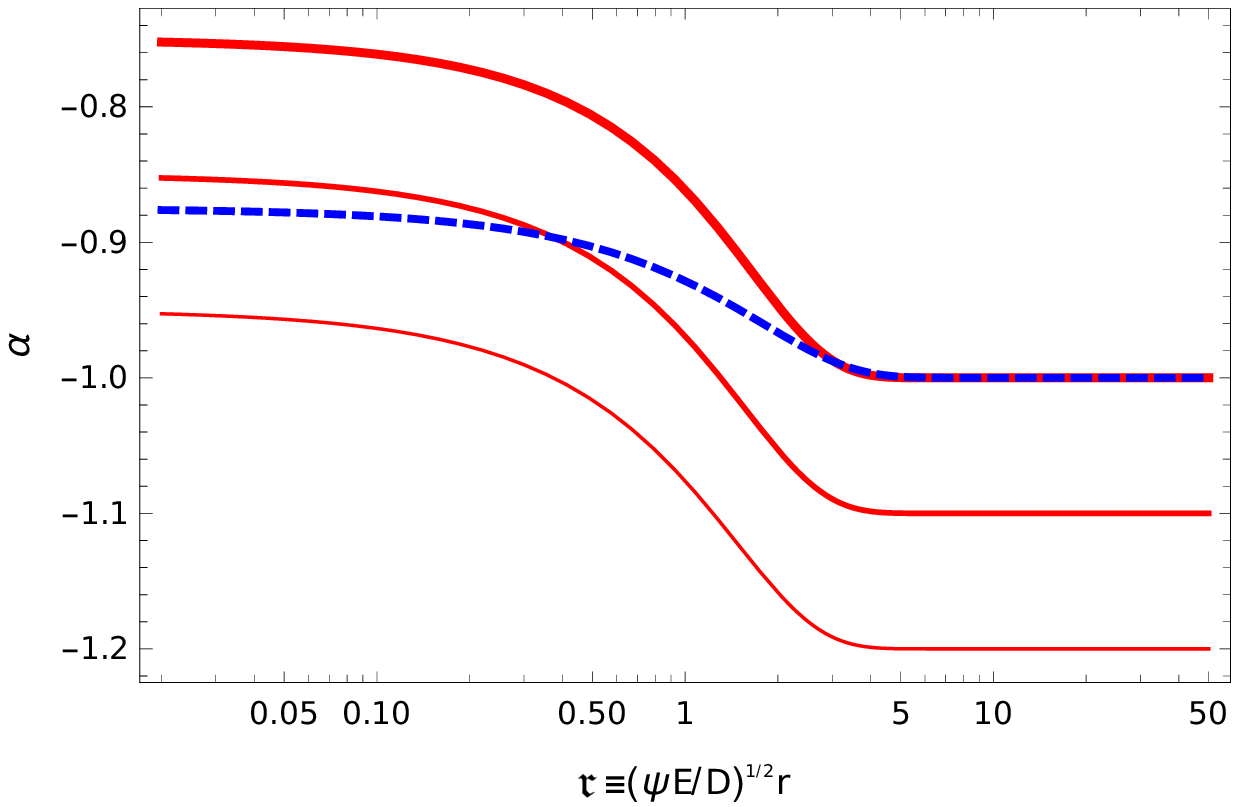}
\hspace{1cm}
\epsfxsize=8cm \epsfbox{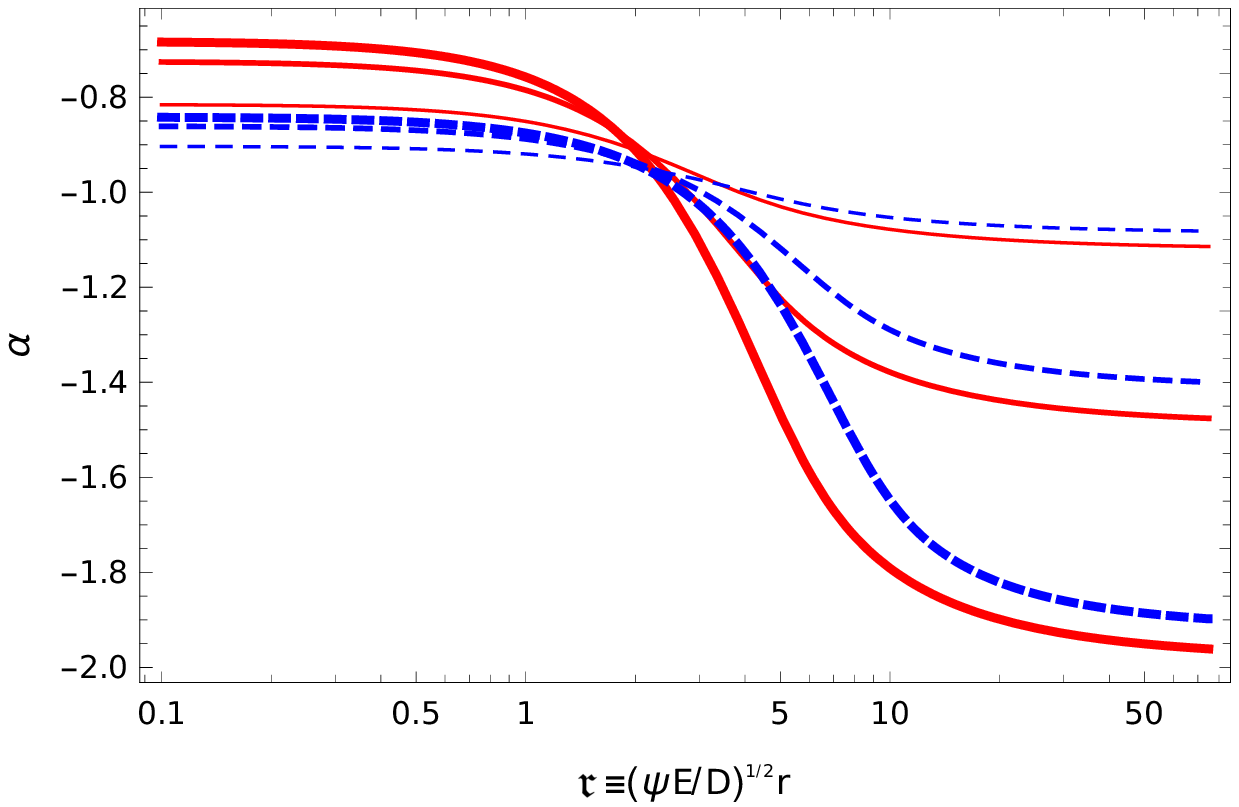}
}
\caption{\label{fig:alphaAnalytic}
Spectral index of synchrotron emission for $\dot{N}_+\propto r_0^{-1}$ power-law (left panel) and $\dot{N}_+\propto e^{-r_0/r_c}$ exponential core (right) CRE injection profiles, shown for $D\propto E^0$ (red solid curves) and $D\propto E^{1/2}$ (dashed blue) diffusion functions.
For power-law injection and energy-independent diffusion, results are shown for $p=2, 2.2, 2.4$ (thick to thin solid red curves).
For exponential core injection, results are shown for $\mathfrak{r}_c=0.6, 0.8, 1.5$ (thick to thin curves).
}
\end{figure}

Next, we demonstrate the steady-state $N(E,r)$ arising from some ongoing $\dot{N}_+=C(r_0)E^{-p}$  injection of a $C(r_0)$ distribution of spherical shells.
The simplest, homogeneous $C=\const$ injection reproduces the $N\propto E^{-(p+1)}$ result as in \eqref{eq:N_int_r} due the $r\leftrightarrow r_0$ symmetry of $N_0$, but here the spectrum $\alpha=-p/2$ applies locally everywhere without volume integration; again, this result holds for any $D(E)$ diffusion function.
For a $C=C_0r_0^{-1}$ power-law injection, plugging Eq.~\eqref{eq:N0_d0} into the integral in Eq.~\eqref{eq:N} yields
\begin{equation}
\frac{N}{C_0} = \frac{E^{-(p+1)}}{(p-1)l\psi } \left[
\frac{\Gamma(p)}{\Gamma\left(p-\frac{1}{2}\right)}
{_1}F_1\left(\frac{3}{2}-p,\frac{3}{2},-\frac{\mathfrak{r}^2}{4}\right)
- \frac{{_1}F_1\left(1-p,\frac{1}{2},-\frac{\mathfrak{r}^2}{4}\right)-1}{\mathfrak{r}}
\right]\, .
\end{equation}
For injection with an exponential, $C=C_0 e^{-r_0/r_c}$ core, the same procedure yields, for the idealized case $p=2$,
\begin{equation}
\frac{N}{C_0} = \frac{E^{-3}\mathfrak{r}_c^{2}}{\psi\chi} \left[
2\text{erfc}\left(\frac{\mathfrak{r}}{2}\right) - \left(2+\chi\right)e^{-\chi}
\pm\left(\frac{1}{\mathfrak{r}_c^2}-1\pm\frac{\chi}{2}\right) e^{\frac{1}{\mathfrak{r}_c^{2}}\pm \chi}\text{erfc}\left(\frac{1}{\mathfrak{r}_c}\pm\frac{\mathfrak{r}}{2}\right)
\right]\, ,
\end{equation}
where both signs $\pm$ are included, $\text{erfc}=1-\text{erf}$ is the complementary error function, and we defined $\chi\equiv r/r_c$ for brevity.
These analytic solutions are illustrated in Fig.~\ref{fig:alphaAnalytic}.
Such analytical solutions may be superimposed, and one can plug Eq.~\eqref{eq:N0_d0} into the integral in Eq.~\eqref{eq:N} and evaluate numerically the energy-independent diffusion of an arbitrary injection profile $C(r_0)$.

In the special case of injection in the origin, $\dot{N}_+\propto\delta(r)$, Eq.~\eqref{eq:N0_d0} is replaced by
\begin{equation}\label{eq:N0_d0_Origin}
N_0 =\frac{E^{-(p+1)}}{4\pi l^3 \psi} \left\{
\frac{{_1}F_1\left[2-p,\frac{1}{2},-\frac{(\mathfrak{r}-\mathfrak{r}_0)^2}{4}\right]}{\mathfrak{r}}
-\frac{\Gamma(p-1)\,{_1}F_1\left[\frac{5}{2}-p,\frac{3}{2},-\frac{(\mathfrak{r}-\mathfrak{r}_0)^2}{4}\right]} {\Gamma\left(p-\frac{1}{2}\right)} \right\} \propto \frac{E^{-(p+1)}}{l^3 \psi} \mathfrak{n}(\mathfrak{r};\myi=3) \propto N(\mathfrak{r};\myi=3) \, .
\end{equation}
so the resulting distribution is equivalent (in the present, $d=0$ case) to the solution \eqref{eq:Frho3} for $\myi=3$ power-law injection.
We verify numerically that Eq.~\eqref{eq:Frho3} is indeed the solution for $\dot{N}_+\propto\delta(r)$ injection for any $0\leq d<1$.

\pagebreak

\subsection{Kraichnan, $D\propto E^{1/2}$ diffusion}

For the interesting case of energy-dependent diffusion with $D(E)=\mathfrak{D} E^{1/2}$, the diffusion length is given by
\begin{equation}
r_d^2 = \frac{2\mathfrak{D}}{\psi \sqrt{E}}\left( 1 - \sqrt{1-\psi E t}\right) \, ,
\end{equation}
where $\mathfrak{D}$ is a constant.
For injection with a $p=2$ spectrum, the kernel in Eq.~\eqref{eq:N0} can then be derived analytically as
\begin{equation}\label{eq:N0_d05}
N_0 =
\frac{-\left(\mathfrak{r}_0\pm\mathfrak{r}\right)\left[6+\frac{(\mathfrak{r}\pm\mathfrak{r}_0)^2}{2}\right]\mbox{erf}\left(\frac{\mathfrak{r}\pm\mathfrak{r}_0}{2^{3/2}}\right)
\pm\sqrt{\frac{2}{\pi}} \left[-8-(\mathfrak{r}\pm\mathfrak{r}_0)^2\right]e^{-\frac{(\mathfrak{r}\pm\mathfrak{r}_0)^2}{8}}
+\Bigl\{\begin{array}{lr}
\mathfrak{r}^3+3\mathfrak{r}_0^2\mathfrak{r}+12\mathfrak{r} & \text{for } r<r_0; \\
\mathfrak{r}_0^3+3\mathfrak{r}^2\mathfrak{r}_0+12\mathfrak{r}_0 & \text{for } r>r_0
\end{array}}
{48\pi l^3\psi \mathfrak{r}\mathfrak{r}_0 E^3}\, .
\end{equation}
Plugging Eq.~\eqref{eq:N0_d05} into the integral in Eq.~\eqref{eq:N} yields, for a $C=C_0r_0^{-1}$ power-law injection,
\begin{equation}
\frac{N}{C_0} = \frac{E^{-3}}{24l\psi} \left[
\frac{24}{\mathfrak{r}}+\sqrt{\frac{2}{\pi}}\left(20+\mathfrak{r}^2\right)e^{-\mathfrak{r}^2/8}
-\left(\frac{24}{\mathfrak{r}}+12\mathfrak{r}+\frac{\mathfrak{r}^3}{2}\right)\text{erfc}\left(\frac{\mathfrak{r}}{2^{3/2}}\right)
\right]\, ,
\end{equation}
whereas the exponential, $C=C_0e^{-r_0/r_c}$ core results in
\begin{equation}
\frac{\psi\chi E^3}{\mathfrak{r}_c^{2}} \frac{N}{C_0} =
2\left(1+\mathfrak{r}_c^2+\frac{\mathfrak{r}^2}{4}\right)\text{erfc}\left(\frac{\mathfrak{r}}{2^{\frac{3}{2}}}\right) - \mathfrak{r}\sqrt{\frac{2}{\pi}}e^{-\frac{\mathfrak{r}^2}{8}}
-2\left(1+\mathfrak{r}_c^2+\frac{\chi}{2}+\frac{\mathfrak{r}_c \mathfrak{r}}{4}\right)e^{-\chi}
+\left(\mathfrak{r}_c^2-1\mp\frac{\mathfrak{r}_c \mathfrak{r}}{4}\right) \left[\text{erfc}\left(\frac{\mathfrak{r}}{2^{\frac{3}{2}}}\pm\frac{\sqrt{2}}{\mathfrak{r}_c}\right)\mp 1\right]
e^{\frac{2}{\mathfrak{r}_c^{2}}\pm\chi}
\, .
\end{equation}
As Fig.~\ref{fig:alphaAnalytic} shows, strengthening the energy-dependence of diffusion to $D\propto E^{1/2}$ renders the spectral index closer to its volume-averaged value $\alpha\to -p/2=-1$.
For instance, the $\{\mathfrak{r},\mathfrak{r}_c\}\to 0 $ limit gives the non-cooled limit $\alpha=-1/2$ for $D\propto E^0$, but a softer $\alpha=-3/4$ for $D\propto E^{1/2}$.
This effect arises because the hardening in the centre, associated with CREs escaping it before they can cool, is partly offset by the softening associated with higher-energy CREs diffusing away faster.

\section{Green-function solutions downstream of a planar shock}
\label{app:GreenSolutionsPlanar}

\begin{figure}
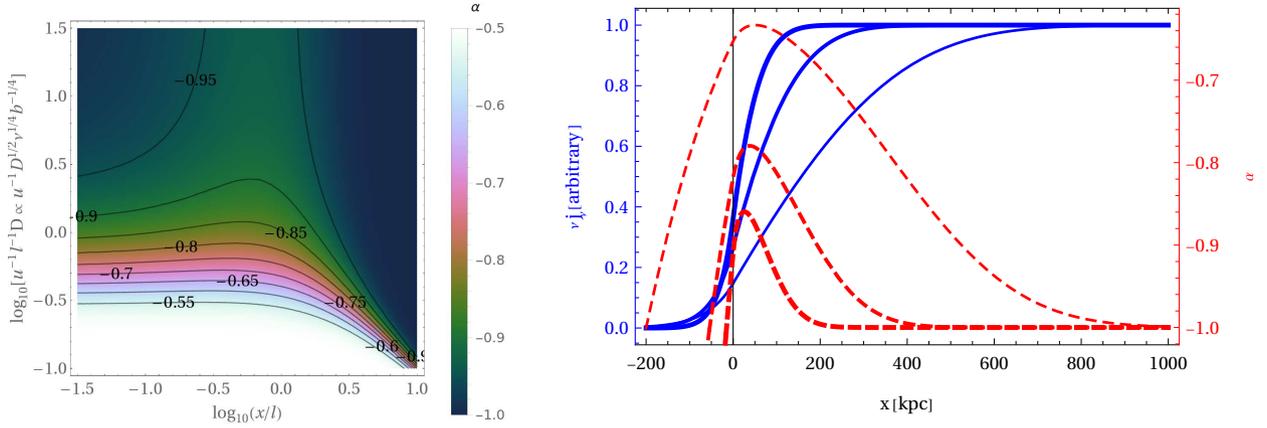

\centerline{
\includegraphics[width=0.37\linewidth]{\figeps{PlanarAlpaDimLess1}}
\hspace{0.5cm}
\raisebox{4mm}[0pt][0pt]{\includegraphics[width=0.51\linewidth,trim={0 0 0 0cm},clip]{\figeps{PlanarProf1}}}
}
\caption{\label{fig:alphaAnalyticPlanar}
Synchrotron emission of secondary CREs in a steady-state planar shock located at $x=0$, for the simple case where $0<u=\const$, $B=\const$, $d=0$, and $p=2$.
Left panel: spectral index (contours and colour bar) in the dimensionless $x$--$\nu$ phase space.
In the downstream, the spectrum hardens towards the shock, approaching $\alpha\to 0.5$ at lower frequencies.
In this simple setup, $\alpha$ softens in the downstream direction, more or less monotonically, towards the steady-state $\alpha=-1$.
Right panel: spatial distribution of emissivity (solid blue curves) and spectral index (dashed red) for parameters $D=10^{31}\cm^2\se^{-1}$, $u=500\km\se^{-1}$, and frequencies $\nu=10^8$, $10^9$, and $10^{10}\Hz$ (thin to thick curves).
}
\end{figure}

Consider a sheet of CREs injected at a distance $x_0>0$ downstream of a planar shock, with an initial energy $E_0$ at time $t=t_0$,\!\!
\begin{equation} \label{eq:PlanarDeltaInject}
N_0(t=t_0,E,x)=\delta(x-x_0)\delta(E-E_0) \,,
\end{equation}
where we normalized $\int N_0 \,dx\,dE=1$.
In the shock frame, the temporal evolution of this sheet follows
\begin{equation} \label{eq:PlanarDeltaEvolve}
\frac{N_0(t>t_0,E,x)}{\delta\left[E-E(\Delta t,E_0)\right]} \equiv \myG(\Delta t\equiv t-t_0,E,x;x_0) = \frac{1}{2\pi^{1/2}r_d}e^{-\left(\frac{x-x_0-\Delta\Phi}{2r_d}\right)^2} \coma
\end{equation}
provided that the downstream magnetic structure moves with respect to the shock at a uniform velocity $u(t)=\Phi'(t)>0$.
Here, the effective downstream velocity $u$ is allowed to evolve only temporally, according to the derivative of some function $\Phi(t)\equiv \Phi(t_0)+\Delta\Phi(t)$; the more physical case of a spatially non-uniform downstream velocity requires, in general, a numerical treatment.
Equivalently, one may work in a uniformly stationary downstream frame, taking $\Delta\Phi=0$ in Eq.~\eqref{eq:PlanarDeltaEvolve} and a shock front moving at a velocity $-u(t)$.
We continue working in the shock frame, where the $N_0$ of Eq.~\eqref{eq:PlanarDeltaEvolve} solves the downstream transport equation
\begin{equation} \label{eq:DiffLossPlanar}
\frac{dN}{dt} = \frac{\pr \myn}{\pr t} + u \partial_x\myn
= \dot{\myn}_+ + D\pr_{x,x}\myn - \frac{\pr}{\pr E}\left( \myn \dot{E}_{\text{cool}} \right) \coma
\end{equation}
with the appropriate injection $N_+$ corresponding to Eq.~\eqref{eq:PlanarDeltaInject}.
In this planar version of Eq.~\eqref{eq:DiffLoss0}, we omitted the adiabatic term, focusing on particles that remain downstream; consequently, the upstream density of particles that manage to diffuse across the shock will be somewhat overestimated below.
Assuming a CRE injection spectrum of power-law index $p$, the energy evolution of Eq.~\eqref{eq:EnergyEvolve}, and the diffusion length of Eq.~\eqref{eq:DiffLength}, we follow the procedure leading to Eq.~\eqref{eq:N}, here implying that
\begin{equation}\label{eq:NPlanarShock}
N(t,E,x) = \int_0^{\infty} dx_0 \int_{0}^{(\psi E)^{-1}} d\Delta t\, C(t-\Delta t, x_0) \myG(\Delta t,E,r;r_0) E^{-p}\left(1-\psi E \Delta t \right)^{p-2} \fin
\end{equation}
The injection amplitude $C$ is a constant for a steady-state shock with a uniform downstream CRI distribution, but the result \eqref{eq:NPlanarShock} would hold for any injection function $C(t,x_0)$.
In the special case where $u$ and $C$ are time-independent, the CRE distribution \eqref{eq:NPlanarShock} is stationary, as depicted in Fig.~\ref{fig:alphaAnalyticPlanar}.

\pagebreak

\section{Inferring the diffusive scale from a spectral index map}
\label{app:PowerSpectrum}

Given the spatial distribution of the spectral radio index $\alpha$ across a halo, one can infer the diffusion-cooling scale $l$ from the radial power spectrum $|\alpha_\lambda|^2$.
To illustrate the expected power spectrum, we superimpose a fixed number of point-like CRE sources, randomly distributed spatially, and compute the resulting two-dimensional spectral map (by combining Eqs. \eqref{eq:PLN},\eqref{eq:MathFrakR}, \eqref{eq:syn_j}, and \eqref{eq:Frho3}; for simplicity, we consider an unprojected 2D slice).
At small scales, the power spectrum of this synthetic map is approximately a power-law, $|\alpha_\lambda|^2\propto \lambda^{\myg}$.
While the index $\myg$ is not sensitive to $l$, the power-law behaviour of $|\alpha_\lambda|$ extends only up to a wavelength $\lambda\equiv l_\alpha$ that is comparable to $l$, as demonstrated in Fig.~\ref{fig:PowerSpecSynth}.

\begin{bfigure*}
\begin{center}
\vspace{0.5cm}
\begin{overpic}[width=0.33\linewidth]{\figeps{P2SpecSynthR20d}}
\put (30,67) {\scriptsize\textcolor{blue}{Synthetic halo with small $l$}}
\put (65,30) {\scriptsize\textcolor{red}{$\myg=2.9\pm1.1$}}
\put (-2,27) {\colorbox{white}{\rotatebox{90}{\scriptsize\textcolor{black}{\tiny{$\lambda^{-\myg}|\alpha_\lambda|^2$}}}}}
\end{overpic}
\begin{overpic}[width=0.33\linewidth]{\figeps{P2SpecSynthR40d}}
\put (25,67) {\scriptsize\textcolor{blue}{Synthetic halo with medium $l$}}
\put (65,30) {\scriptsize\textcolor{red}{$\myg=2.9\pm1.1$}}
\put (-2,27) {\colorbox{white}{\rotatebox{90}{\scriptsize\textcolor{black}{\tiny{$\lambda^{-\myg}|\alpha_\lambda|^2$}}}}}
\end{overpic}
\begin{overpic}[width=0.33\linewidth]{\figeps{P2SpecSynthR60d}}
\put (25,67) {\scriptsize\textcolor{blue}{Synthetic halo with large $l$}}
\put (65,30) {\scriptsize\textcolor{red}{$\myg=2.9\pm1.1$}}
\put (-2,27) {\colorbox{white}{\rotatebox{90}{\scriptsize\textcolor{black}{\tiny{$\lambda^{-\myg}|\alpha_\lambda|^2$}}}}}
\end{overpic}
\end{center}
\caption{\label{fig:PowerSpecSynth}
The radial power-spectra (green disks show, in arbitrary units, the normalized power $\lambda^{-\myg}|\alpha_\lambda|^2$ at wavelength $\lambda\equiv \lambda_1\Mpc$) of the spatial distribution of $\alpha$ in synthetic halo maps, computed for different diffusion-cooling scales $l$ (increasing from left to right panels; vertical solid green lines show $l^{-1}$), are well fit at small scales by a power-law (dashed red, with labeled index $\myg$).
The estimated lengthscale $l_\alpha$ (vertical dot-dashed blue lines show $l_\alpha^{-1}$), above which the power spectrum consistently deviates from its small scale power-law best fit, reproduces the injected $l$ with a factor $\sim2$ uncertainty, as shown by the cumulative $l_\alpha$ distribution inferred from many such synthetic halos (dotted black vertical line with horizontal error bar showing the mean and standard deviation).
}
\end{bfigure*}

The radial power-spectra extracted from published maps of $\alpha$ measured across radio halos are indeed well-fitted by a power-law at small scales, extending out to a maximal wavelength $l_\alpha$ (see Fig.~\ref{fig:PowerSpec} and labels therein). Hence, inasmuch as a determination of $l_\alpha$ gauges $l$, one can measure the diffusion coefficient $D$; the results are summarized in Fig.~\ref{fig:DEst}. We crudely estimate $l_\alpha$ as the scale above which the data consistently deviates from the small-scale best-fit power law.
The combined statistical and systematic uncertainty in $l_\alpha$ can be estimated as $\lesssim 2$; however, calibrating the value of $f_\alpha\equiv l_\alpha/l$ for different halo morphologies and circumstances would require dedicated numerical simulations that are beyond the scope of the present work.
As Fig.~\ref{fig:DEst} shows, the similar results obtained from independent spectral maps of the same halo support the viability of our method.

\begin{bfigure*}
\begin{center}
\vspace{0.5cm}
\begin{overpic}[width=0.33\linewidth]{\figeps{P2SpecA2163F04}}
\put (40,67) {\scriptsize\textcolor{blue}{A2163 (F04)}}
\put (65,60) {\scriptsize\textcolor{red}{$\myg=1.4\pm0.5$}}
\put (-2,27) {\colorbox{white}{\rotatebox{90}{\scriptsize\textcolor{black}{\tiny{$\lambda^{-\myg}|\alpha_\lambda|^2$}}}}}
\end{overpic}
\begin{overpic}[width=0.33\linewidth]{\figeps{P2SpecA2163M16R}}
\put (40,67) {\scriptsize\textcolor{blue}{A2163 (M16)}}
\put (65,60) {\scriptsize\textcolor{red}{$\myg=1.9\pm0.2$}}
\put (-1,27) {\colorbox{white}{\rotatebox{90}{\scriptsize\textcolor{black}{\tiny{$\lambda^{-\myg}|\alpha_\lambda|^2$}}}}}
\end{overpic}
\begin{overpic}[width=0.33\linewidth]{\figeps{P2SpecA2255B20}}
\put (40,67) {\scriptsize\textcolor{blue}{A2255 (B20)}}
\put (65,15) {\scriptsize\textcolor{red}{$\myg=1.7\pm0.3$}}
\put (-1,27) {\colorbox{white}{\rotatebox{90}{\scriptsize\textcolor{black}{\tiny{$\lambda^{-\myg}|\alpha_\lambda|^2$}}}}}
\end{overpic}\\
\vspace{0.3cm}
\begin{overpic}[width=0.33\linewidth]{\figeps{P2SpecA2255B22}}
\put (40,68) {\scriptsize\textcolor{blue}{A2255 (B22)}}
\put (65,15) {\scriptsize\textcolor{red}{$\myg=3.2\pm0.4$}}
\put (-2,27) {\colorbox{white}{\rotatebox{90}{\scriptsize\textcolor{black}{\tiny{$\lambda^{-\myg}|\alpha_\lambda|^2$}}}}}
\end{overpic}
\begin{overpic}[width=0.33\linewidth]{\figeps{P2SpecA2256KD10}}
\put (40,67) {\scriptsize\textcolor{blue}{A2256 (KD10)}}
\put (65,60) {\scriptsize\textcolor{red}{$\myg=2.0\pm0.6$}}
\put (-1,27) {\colorbox{white}{\rotatebox{90}{\scriptsize\textcolor{black}{\tiny{$\lambda^{-\myg}|\alpha_\lambda|^2$}}}}}
\end{overpic}
\begin{overpic}[width=0.33\linewidth]{\figeps{P2SpecA2256R22}}
\put (40,67) {\scriptsize\textcolor{blue}{A2256 (R22)}}
\put (65,60) {\scriptsize\textcolor{red}{$\myg=1.9\pm0.6$}}
\put (-1,27) {\colorbox{white}{\rotatebox{90}{\scriptsize\textcolor{black}{\tiny{$\lambda^{-\myg}|\alpha_\lambda|^2$}}}}}
\end{overpic}\\
\vspace{0.3cm}
\begin{overpic}[width=0.33\linewidth]{\figeps{P2SpecA2744P17}}
\put (40,69) {\scriptsize\textcolor{blue}{A2744 (P17)}}
\put (65,60) {\scriptsize\textcolor{red}{$\myg=2.0\pm0.2$}}
\put (-2,27) {\colorbox{white}{\rotatebox{90}{\scriptsize\textcolor{black}{\tiny{$\lambda^{-\myg}|\alpha_\lambda|^2$}}}}}
\end{overpic}
\begin{overpic}[width=0.33\linewidth]{\figeps{P2SpecA2744P19}}
\put (40,67) {\scriptsize\textcolor{blue}{A2744 (P19)}}
\put (65,60) {\scriptsize\textcolor{red}{$\myg=1.3\pm0.4$}}
\put (-1,27) {\colorbox{white}{\rotatebox{90}{\scriptsize\textcolor{black}{\tiny{$\lambda^{-\myg}|\alpha_\lambda|^2$}}}}}
\end{overpic}
\begin{overpic}[width=0.33\linewidth]{\figeps{P2SpecA665F04}}
\put (40,67) {\scriptsize\textcolor{blue}{A665 (F04)}}
\put (65,60) {\scriptsize\textcolor{red}{$\myg=1.9\pm0.4$}}
\put (-2,27) {\colorbox{white}{\rotatebox{90}{\scriptsize\textcolor{black}{\tiny{$\lambda^{-\myg}|\alpha_\lambda|^2$}}}}}
\end{overpic}\\
\vspace{0.3cm}
\begin{overpic}[width=0.33\linewidth]{\figeps{P2SpecBulletS14}}
\put (40,67) {\scriptsize\textcolor{blue}{Bullet (S14)}}
\put (65,60) {\scriptsize\textcolor{red}{$\myg=1.4\pm0.2$}}
\put (-2,27) {\colorbox{white}{\rotatebox{90}{\scriptsize\textcolor{black}{\tiny{$\lambda^{-\myg}|\alpha_\lambda|^2$}}}}}
\end{overpic}
\begin{overpic}[width=0.33\linewidth]{\figeps{P2SpecCIG217H21}}
\put (40,68) {\scriptsize\textcolor{blue}{CIG 0217 (H21)}}
\put (65,60) {\scriptsize\textcolor{red}{$\myg=1.9\pm0.5$}}
\put (-2,27) {\colorbox{white}{\rotatebox{90}{\scriptsize\textcolor{black}{\tiny{$\lambda^{-\myg}|\alpha_\lambda|^2$}}}}}
\end{overpic}
\begin{overpic}[width=0.33\linewidth]{\figeps{P2SpecRXJ1720S19}}
\put (40,67) {\scriptsize\textcolor{blue}{RXJ 1720 (S19)}}
\put (65,60) {\scriptsize\textcolor{red}{$\myg=2.0\pm0.8$}}
\put (-1,27) {\colorbox{white}{\rotatebox{90}{\scriptsize\textcolor{black}{\tiny{$\lambda^{-\myg}|\alpha_\lambda|^2$}}}}}
\end{overpic}
\end{center}
\caption{\label{fig:PowerSpec}
Same as Fig.~\ref{fig:PowerSpec}, but for published $\alpha$ maps of different radio halos (see panel labels; reference are provided in Fig.~\ref{fig:DEst}).
}
\end{bfigure*}

\clearpage
\twocolumn

\bibliography{\mybib{Clusters}}

\end{document}